\documentclass[reprint,aps,pre]{revtex4-2}
\usepackage{bm}
\usepackage{graphicx} 
\usepackage{amsmath} 
\usepackage{amssymb} 
\usepackage{enumitem}
\usepackage{soul}
\usepackage[mathscr]{eucal}
\usepackage{dsfont}
\usepackage{comment}
\usepackage{physics}
\usepackage[cal=boondoxo]{mathalfa}
\usepackage{hyperref}
\usepackage{placeins}%

\renewcommand{\vec}[1]{{\mathbf #1}}
\newcommand{\be}{\begin{equation}}
\newcommand{\ee}{\end{equation}}
\usepackage{xcolor}

\newcommand{\ls}{\ell_s}
\newcommand{\um}{\,\text{\textmu m}}
\newcommand{\dkm}{\Delta k_{\rm M}}
\newcommand{\dks}{\Delta k_{\rm shell}}
\newcommand{\dksf}{\Delta k_{\rm shell}^{\rm finite}}
\newcommand{\ash}{\alpha_{\rm shell}^{\rm eff}}

\begin{document}

\title{\texorpdfstring{On-shell compression and reconstruction (OSCAR)\\ of monochromatic wave fields in weakly scattering media}{On-shell compression and reconstruction (OSCAR) of monochromatic wave fields in weakly scattering media}}

\author{Pablo Jara$^1$}
\email{pxjdbk@mst.edu}
\author{Alexey Yamilov$^1$}
\email{yamilov@mst.edu}

\affiliation{$^1$Department of Physics, Missouri University of Science and Technology, Rolla, Missouri 65409, USA}

\date{\today}

\begin{abstract}
Advances in computational methods have made full-wave simulations in large disordered media increasingly feasible, but the resulting volume of field data, scaling with the cube of the ratio of system size to wavelength, creates a severe storage and post-processing bottleneck.
Generic compression methods are sample-specific and preclude operations on compressed data. We introduce OSCAR, a physics-based lossy compression scheme for monochromatic wave fields in weakly scattering media.  
OSCAR exploits the universal confinement of the Fourier representation of wave fields to a thin dispersion shell, a direct consequence of wave propagation when the scattering mean free path significantly exceeds the wavelength, and a property of every disorder realization rather than of the ensemble average, so that one mask serves an entire ensemble.
The resulting compression ratio reflects two distinct scale separations: on-shell confinement due to weak scattering, and the excess Fourier-space volume introduced by sub-wavelength discretization of the scatterers.
Crucially, second-order quantities, demonstrated here on sensitivity maps, can be computed via convolution entirely in compressed space while preserving the interference effects.
We validate the method in 2D scalar and 3D vector simulations of electromagnetic waves, spanning quasi-ballistic to diffusive transport.
The reconstruction error is set by the discarded spectral weight.
Demonstrated compression ratios reach ${\sim}810\times$ in 2D and ${\sim}420\times$ in 3D at a $3\%$ error, with cell-averaged second-order observables accurate to an even smaller error.
Established general-purpose scientific compressors perform well for underdeveloped speckle, while becoming less efficient, in comparison, as speckle develops.
Tested from quasi-ballistic to diffusive transport, OSCAR stores the fields in up to $19\times$ less space at a $2.5$--$4.2\times$ lower runtime cost than SZ3.
This lowers the storage barrier to ensemble studies at scales relevant to biomedical optics, seismology, and underwater acoustics.
\end{abstract}

\maketitle
\section{Introduction\label{sec:intro}}

\noindent Increasingly available large-scale computational resources~\cite{2023_Notarus_Computational_EM} enable full-wave simulations in complex scattering media. 
While electromagnetic, acoustic, seismic, and other waves obey different governing equations, sub-wavelength discretization permits direct numerical solution without approximations, capturing interference and other wave phenomena inaccessible to approximate methods. 
In the context of biomedical optics, wavefront shaping (WFS) through tissue requires direct access to spatial wave fields $E(\vec{r})$ rather than ensemble-averaged intensity-like quantities~\cite{2012_Mosk_review,2018_Park_perspective,2022_Gigan_roadmap,2022_Cao_Mosk_Rotter_review}.
Optical coherence tomography (OCT) and digital holographic microscopy (DHM) can also experimentally capture these fields in weakly scattering samples~\cite{2003_Fercher_OCT,2010_Kim_holographic}, but multiple scattering regimes remain achievable primarily through numerical simulation~\cite{2016_Kraszewski_coherent,2021_McCoy_FDTD}. 
Similarly, seismic full-waveform inversion (FWI) requires complete wavefield solutions to reconstruct subsurface velocity structures~\cite{2009_Virieux_FWI,2012_Liu_seismic}, while three-dimensional ocean acoustic propagation modeling demands high computational resources to capture range-dependent environmental effects~\cite{2011_Jensen_ocean,2012_Lin_parabolic}.

Large-scale full-wave simulations~\cite{2012_Graham_book,2026_Mache_domain_decomposition} have to contend with a dual challenge: computational cost itself as well as the data storage / post-processing requirements.
The former can be mitigated through algorithmic improvements and hardware-software integration, whereas the latter presents a separate problem, which is difficult to address in an application-agnostic manner.
The storage burden arises from fundamental scaling.
To capture wave phenomena, three-dimensional fields must be discretized on sub-wavelength scales $\Delta x\!=\!\lambda/n_{\Delta x}$, resulting in $\propto\!(L/\Delta x)^3\!=\!(n_{\Delta x}L/\lambda)^3$ data points, where $L$ is the system size and $\lambda$ the wavelength. 
For finite-difference time-domain (FDTD) electromagnetic simulations~\cite{1966_Yee_FDTD,2005_Taflove_FDTD} in biological tissue at near-infrared wavelengths, this scaling becomes prohibitive for millimeter-scale samples~\cite{2024_Ball_microscopy}. 

Seismic FWI faces analogous challenges, with 4D wavefields (three spatial dimensions plus time) demanding terabyte-scale storage even for modest-sized domains~\cite{2015_Krischer_LASIF,2024_Zhao_Bayesian}. 
Ocean acoustic simulations over kilometer scales similarly encounter severe memory constraints~\cite{2021_Oliveira_shallow_acoustics,1995_Lee_3D_acoustics}. 
Mesoscopic transport phenomena compound these demands. Ensemble averaging over many disorder realizations is required to extract physically meaningful quantities~\cite{2007_Akkermans_Mesoscopic_Physics}, multiplying storage requirements by the number of samples.

Existing compression methods provide substantial data reduction. 
Tucker decomposition and higher-order singular value decomposition (HOSVD) achieve compression ratios exceeding 100:1 for seismic wavefields~\cite{2023_Wang_Tucker_Compression}, while specialized approaches have been developed for specific applications~\cite{2024_Ni_Wavefield_Reconstruction_DAS,2021_Muir_wavefield}. 
Lossy compression of forward wavefields for adjoint-based full-waveform inversion has also been explored~\cite{2016_Boehm_wavefield_compression,2022_Kukreja_lossy_checkpoint}.
In high-performance computing, error-bounded compressors such as SZ3~\cite{sm_sz3, sm_sz3b} and zfp~\cite{sm_zfp} have become the general-purpose tools for floating-point simulation data, and together with the truncated singular value decomposition (SVD)~\cite{sm_svd} they serve as the reference schemes for our benchmarks below.
However, generic compression methods suffer a critical limitation, namely that they are sample-specific.
Each sample of disorder compresses differently, precluding ensemble operations in compressed space. 
Computing disorder-averaged intensities, correlations, or other second-order quantities (even in one sample) requires decompressing all quantities, negating computational benefits and creating a technical barrier to wider adoption of full-wave methods.
Unlike sample-specific approaches such as Tucker decomposition or lossy floating-point compression~\cite{2023_Wang_Tucker_Compression,2022_Kukreja_lossy_checkpoint}, a physics-based compression scheme that exploits universal shell structure can enable combining different fields and ensemble operations on them directly in compressed space.

The linear weak scattering regime ($k\ell_s \gg 1$, where $k = 2\pi/\lambda$ is the wavenumber and $\ell_s$ the scattering mean free path) admits a physics-based solution to the large data set problem. 
In this regime, the spatial Fourier transform of wave fields exhibits a universal structure. Field amplitude concentrates in a characteristic shell of width $\sim 1/\ell_s$ centered at radius $k$~\cite{2007_Akkermans_Mesoscopic_Physics,1978_Ishimaru_Wave_Propagation,2006_Sheng_Wave_Scattering,1996_Lagendijk_vanTiggelen_review}. 
This on-shell concentration arises because propagating waves must satisfy the dispersion relation, while disorder-induced scattering introduces only a finite spread $\propto\ell_s^{-1}$ in wave-vector magnitude. 
Importantly, this shell structure of the field persists across different disorder realizations since it is a property of individual fields, not merely a statistical artifact of ensemble averaging. 
This sample-independent property enables a compression approach described below that reduces storage from $(n_{\Delta x}L/\lambda)^3$ to $(L/\lambda)^2\!\times\!(L/\ell_s)$ in three dimensions leading to compression by a factor $\propto\!k\ell_s\!\times\!n_{\Delta x}^3\!\gg\!1$. 
Furthermore, the universality of the shell structure permits computing disorder-averaged second-order quantities via convolution entirely in compressed $k$-space. 
Since ensemble-averaged observables commonly vary on the scale $\ell_s\!\gg\!\lambda\!>\!\lambda/n_{\Delta x}$, final discretization requires only $(L/\ell_s)^3$ points leading to an additional $(n_{\Delta x}\ell_s/\lambda)^3$ advantage over uncompressed field representations.
We term this approach OSCAR (On-Shell Compression And Reconstruction). 
In media with the reduced scattering coefficients of biological tissue in the near-infrared~\cite{2013_Jacques_optical}, compression ratios of over $400$ are demonstrated even at the coarsest stable discretization, while maintaining fidelity of interference effects and second-order statistics.

This paper makes several distinct claims, each with its own scope:
(i)~\emph{field-storage compression}: the on-shell projection reduces per-realization storage by a factor $C_{\rm OSCAR} \propto k\ell_s \times n_{\Delta x}^d$, validated on plane-wave ensembles in 2D and across 3D systems spanning quasi-ballistic to diffusive transport.
(ii)~\emph{reconstruction fidelity}: the field-level errors are given parameter-free by the discarded spectral weight $P_{\rm out}$.
(iii)~\emph{compressed-space computation}: products of independently compressed fields are formed without decompression and are accurate at the coarse-cell-averaged (envelope) resolution, with errors well below the field-level value.
(iv)~\emph{ensemble averaging in compressed space}: compression commutes with ensemble averaging when all realizations share the same mask, so the average can be performed without decompression.
(v)~\emph{output coarse-graining}, valid where observables are used at envelope resolution and explicitly not claimed for coherent-kernel applications that require wavelength-scale output, such as finite-frequency seismic inversion.

The paper is organized as follows. 
Section~\ref{sec:theory} presents theoretical background, including the on-shell structure and its universality across disorder realizations, the compression criteria, and the correction factors and techniques improving fidelity of the compression and reconstruction.  
Sections~\ref{sec:2d} and~\ref{sec:3d} describe numerical verification of the OSCAR algorithm in 2D and 3D simulations of electromagnetic waves, respectively.
Section~\ref{sec:benchmarks} benchmarks OSCAR against general-purpose scientific compressors at matched error and at matched storage, and reports the measured storage and runtime costs.
We conclude with a discussion of extensions of our approach to time-dependent problems and waves of other types.
Appendix~\ref{app:metrics} derives the exact identities and the statistical error theory behind the metrics used throughout.
The Supplemental Material characterizes every medium, derives the on-shell structure from the radiative transfer equation, and verifies each relationship numerically.

\section{On-shell compression and reconstruction: Theory\label{sec:theory}}
\noindent The compression scheme developed here exploits the spectral structure imposed by the wave equation on fields in weakly scattering media. We consider scalar waves satisfying the Helmholtz equation in the regime $k\ell_s \gg 1$, whereas an extension to vector fields is demonstrated in Sec.~\ref{sec:3d}. We proceed from the idealized case of an infinite medium to the corrections required for practical implementation in finite discrete systems.

\subsection{On-shell structure of wave fields in weakly scattering media\label{sec:onshell}}
In this section we consider monochromatic scalar waves satisfying the Helmholtz equation
\begin{equation}
\nabla^2 E(\vec{r}) + k_0^2 \varepsilon(\vec{r}) E(\vec{r}) = 0,
\label{eq:helmholtz}
\end{equation}
where $k_0 = 2\pi/\lambda_0$ is the free-space wavenumber and $\varepsilon(\vec{r}) = \bar{\varepsilon} + \delta\varepsilon(\vec{r})$ is the spatially varying dielectric permittivity. The mean permittivity renormalizes the wavelength as $\lambda=\lambda_0/\sqrt{\bar{\varepsilon}}$. The spatially varying (random) component $\delta\varepsilon(\vec{r})$ describes the disorder, while its statistical properties determine the scattering mean free path $\ell_s$~\cite{1978_Ishimaru_Wave_Propagation,2007_Akkermans_Mesoscopic_Physics,2006_Sheng_Wave_Scattering,1996_Lagendijk_vanTiggelen_review}.

The Green function of the wave equation, Eq.~\eqref{eq:helmholtz}, can be constructed through the perturbative expansion~\cite{1999_vanRossum_diffuse_waves}
\begin{equation}
G = G_0 + G_0 V G,
\label{eq:dyson_expansion}
\end{equation}
where $V \propto \delta\varepsilon(\vec{r})$ is the scattering potential and $G_0$ is the Green function of the homogeneous-medium equation with $\delta\varepsilon(\vec{r})\equiv 0$. Diagrammatic averaging over disorder realizations, denoted by an overbar, yields the Dyson equation~\cite{2007_Akkermans_Mesoscopic_Physics,1999_vanRossum_diffuse_waves} which in the spatial Fourier space takes the form
\begin{equation}
\bar{G}(\vec{k}^\prime) = \frac{1}{k_0^2\bar{\varepsilon}+\Sigma(\vec{k}^\prime)-k^{\prime 2}}.
\label{eq:dyson}
\end{equation}
Here the self-energy $\Sigma(\vec{k}^\prime)$ encodes the cumulative effect of scattering. The weak scattering regime corresponds to $|\Sigma(\vec{k}^\prime)|\ll k_0^2\bar{\varepsilon}$. The real part, which is not negligible, shifts the shell radius rather than broadening it and is absorbed by measuring $k\equiv 2\pi/\lambda$ as the position of the spectral peak. The imaginary part of $\Sigma(\vec{k}^\prime)$ reduces to $\mathrm{Im}\,\Sigma \approx k/\ell_s$~\cite{1999_vanRossum_diffuse_waves}, which serves as the formal definition of the scattering mean free path $\ell_s$.
This definition implies extinction of the coherent (ballistic) intensity, $|\bar{E}|^2 \propto \exp[-z/\ell_s]$~\cite{1999_vanRossum_diffuse_waves}.
Therefore
\begin{equation}
\bar{G}(\vec{k}^\prime) = \frac{1}{k^2 + ik/\ell_s - k^{\prime 2}}.
\label{eq:green_avg}
\end{equation}
The spectral density, proportional to $\mathrm{Im}\,\bar{G}$, has an approximately Lorentzian envelope as a function of $|\vec{k}^\prime|$:
\begin{equation}
\rho(k') \propto \frac{k/\ell_s}{(k^2 - k'^2)^2 + (k/\ell_s)^2},
\label{eq:lorentzian}
\end{equation}
peaked on the shell $|k'| = k$ with characteristic width $\sim1/\ell_s$. The full width at half maximum of this shell defines the shell width, equal to $\dks = 1/\ls$ in the bulk.
In a finite box the measured width $\dksf$ exceeds $\dks$ and approaches it as the box grows, cf.\ Supplemental Material Sec.~\ref{sec:sm1_estimator}. Here, we have made use of the weak scattering approximation $k\ell_s \gg 1$. The connection of this on-shell structure to the radiative transfer equation is derived in Supplemental Material Sec.~\ref{sec:sm_rte}.

Although Eqs.~(\ref{eq:dyson}) and (\ref{eq:lorentzian}) describe ensemble-averaged quantities, the on-shell concentration persists in individual realizations. The dispersion relation constrains propagating modes to $|\vec{k}^\prime|\approx k$, while scattering redistributes amplitude azimuthally around the shell rather than radially across it. The shell structure is thus a property of individual fields, not merely a statistical artifact of averaging. Secondly, Eq.~\eqref{eq:green_avg} is valid for systems with size $L$ much greater than $\ell_s$, making the shell isotropic on average. In Sec.~\ref{sec:3d} we demonstrate that this latter condition is not required for applicability of OSCAR.

\begin{figure*}[ht!]
\centering
\includegraphics[width=\textwidth]{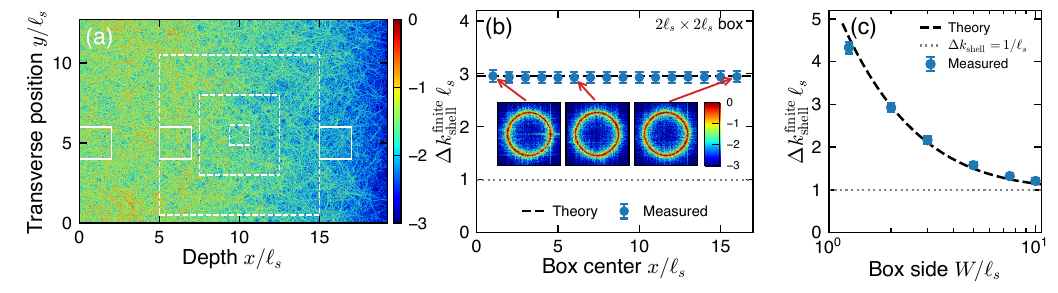}
\caption{On-shell structure of a scattered wave field.
(a)~Intensity $|E(x,y)|^2$ ($\log_{10}$, three decades) of one realization of the 2D medium under plane-wave illumination, with a series of boxes of fixed size $2\ls \times 2\ls$ taken at increasing depths (solid) and another series of boxes of increasing size (dashed).
(b)~The measured shell width $\dksf$, the full width at half maximum of the $k$-shell (ring) measured in $K(\vec k')$ of Eq.~\eqref{eq:discrete_dispersion}, vs box depth for boxes from the depth series, mean $\pm$ standard deviation over $10^3$ realizations and five transverse positions.
The width does not change with depth within the realization spread.
Dashed line: the parameter-free quadrature prediction combining the intrinsic width $1/\ls$ with broadening due to the finite size of the box. Dotted line: $1/\ls$.
Insets: select box spectra.
(c)~The measured $\dksf$ approaches $\dks = 1/\ls$ (dotted) as the finite-size broadening is reduced in the expanding box series in panel (a), where $W$ is the box side. Dashed line as in (b). Error bars are the sample-to-sample standard deviation.}
\label{fig:onshell_schematic}
\end{figure*}

This sample-independent property of fields in weakly scattering media forms the theoretical basis for the compression scheme. Indeed, retaining only spatial Fourier components within the shell of width $\sim 1/\ell_s$ reduces data by a factor $\propto k\ell_s \gg 1$ while preserving physically relevant information in each realization. 
Figure~\ref{fig:onshell_schematic} illustrates this structure directly in a solution of Eq.~\eqref{eq:helmholtz} obtained from one realization of the 2D medium of Sec.~\ref{sec:2d_setup} under plane-wave illumination.
Spatial spectra of boxes of side $2\ls$ show the ring at every depth, and its measured width does not change with depth within the realization spread, see Fig.~\ref{fig:onshell_schematic}(b), as predicted by Eq.~\eqref{eq:lorentzian}.
The measured width includes the contributions from the intrinsic width $1/\ls$ as well as the spectral broadening due to the finite area, cf.\ Supplemental Material Sec.~\ref{sec:sm1_estimator}.
The latter contribution diminishes with increasing box size, and the measured width approaches the intrinsic value $1/\ls$, cf.\ Fig.~\ref{fig:onshell_schematic}(c).

To obtain a more quantitative estimate for the expected compression ratio, we now introduce two parameters:
(1)~$\alpha\!=\!\dkm\,\ell_s$, where $\dkm$ is the full width of the mask, and
(2)~$n_{\Delta x}\!\equiv\!\lambda/\Delta x$, the number of grid points per wavelength.
Taking the ratio between the volume of $k$-space in $d$ dimensions $(2\pi/\Delta x)^d$ and the volume of the (isotropic) mask $2\pi k \dkm$ (2D) or $4\pi k^2 \dkm$ (3D), we arrive at
\begin{equation}
    C_{\rm OSCAR}\simeq\begin{cases} 
    k\ell_s\times\displaystyle\frac{n_{\Delta x}^2}{2\pi\alpha} & \quad\text{(2D)}  \\ 
    \,\\
    k\ell_s\times\displaystyle\frac{n_{\Delta x}^3}{4\pi\alpha} & \quad\text{(3D)} 
    \end{cases}
    \label{eq:compression}
\end{equation}
Here the compression ratio $C_{\rm OSCAR}$ is the number of complex amplitudes on the full computational grid divided by the number retained in the shell.
The two factors in Eq.~\eqref{eq:compression} have distinct physical origins.
The factor $k\ell_s$ reflects the spectral concentration imposed by weak scattering. The shell width $\sim\!1/\ell_s$ is much narrower than its radius $k$.  
The factor $n_{\Delta x}^d$ arises because the computational $k$-space extends to the Nyquist limit $k_{\rm max}\!=\!\pi/\Delta x\!=\!\pi n_{\Delta x}/\lambda$, which exceeds $k$ by a factor $n_{\Delta x}/2$.  
Since the shell occupies only a narrow annulus near $|\vec{k}'|=k\ll k_{\rm max}$, most of the
discretized $k$-space lies far off-shell and carries negligible spectral weight.  
Consequently, simulations that employ fine grids to resolve sub-wavelength structure of the scatterers (large $n_{\Delta x}$) benefit from proportionally larger compression ratios.
In particular, Yee-type discretization~\cite{1966_Yee_FDTD}, where $n_{\Delta x}\!=\!20$ is common, can yield substantial compression ratios due to the second factor in Eq.~\eqref{eq:compression} alone, the $C \simeq 54$ of the $k\ls \approx 300$ slab of Sec.~\ref{sec:3d} at $n_{\Delta x} = 3$ extrapolating to about $1.6 \times 10^4$ at $n_{\Delta x} = 20$.
We further note that Eq.~\eqref{eq:compression} assumes a uniform square (cubic) mesh. 
For non-uniform or unstructured grids, the compression ratio (the second factor) depends on the specific discretization scheme but the underlying on-shell structure (the first factor) remains unchanged.
Equation~\eqref{eq:compression} also presumes that the stored box resolves the shell, i.e., the mask width exceeds the spectral resolution of the box, $\dkm \gtrsim 2\pi/L$.
In a small system or box, this aperture limit, not the value of $k\ls$, limits the attainable ratio, cf.\ Supplemental Material Sec.~\ref{sec:sm1_estimator}.

\subsection{Corrections for finite discrete systems\label{sec:corrections}}
\noindent The analysis of Sec.~\ref{sec:onshell} assumes continuous fields in infinite media. 
Practical implementation of the compression technique requires corrections for 
(i) finite spectral weight captured by the shell, 
(ii) discrete dispersion relations, and 
(iii) boundary-induced anisotropy. 
Below we discuss approaches to mitigating each issue.

\emph{Power normalization.}
Retaining spatial Fourier components within a mask of full width $\dkm$ centered at $|\vec{k}^\prime|=k$ captures only a part of the total spectral weight in Eq.~\eqref{eq:lorentzian}:
\begin{equation}
\frac{P}{P_\infty} = \frac{2}{\pi}\tan^{-1}\!\left(\alpha\right) + \mathcal{O}\left((k\ell_s)^{-1}\right),
\label{eq:power_ratio}
\end{equation}
where $\alpha=\dkm\,\ell_s$, and we denote the retained fraction $P_{\rm in} \equiv P/P_\infty$.
For $\alpha\!=\!1$, $50\%$ of the spectral weight is retained, while $\alpha\!=\!3$ recovers $\approx\!80\%$. Reconstructed fields must be rescaled by $\sqrt{P_\infty/P}$ to preserve total energy, on average. 
For systems with $\ell_s$ much larger than $\lambda$, this normalization factor is essentially sample-independent, depending only on the choice of $\alpha$ and the value of $\ell_s$.
The normalization options and their numerical verification are presented in Supplemental Material Sec.~\ref{sec:sm_metricverify}.

\emph{Discrete dispersion relation.}
Finite-difference discretization modifies the dispersion relation. For a uniform grid with spacing $\Delta x$, a plane wave of grid wave vector $\vec k'$ is an eigenfunction of the discrete Laplacian with eigenvalue $-K^2(\vec k')$, where~\cite{2005_Taflove_FDTD}
\begin{equation}
K^2(\vec k') = \frac{4}{\Delta x^2}\sum_{i=1}^{d}\sin^2\!\left(\frac{k'_i\Delta x}{2}\right),
\label{eq:discrete_dispersion}
\end{equation}
which replaces the continuum relation $K^2(\vec k')= |\vec k'|^2$ and reduces to it as $\Delta x\to 0$. The on-shell ring is the iso-surface $K(\vec k') = k$. 
Expanding Eq.~\eqref{eq:discrete_dispersion} for small $k'_i\Delta x$ gives, in 2D, a relative deviation $(k\Delta x)^2/24$ along the axes and $(k\Delta x)^2/48$ along the diagonals.

For $\Delta x=\lambda/n_{\Delta x}$, the deviation scales as $\sim(\pi/n_{\Delta x})^2$. 
In the 2D simulations of Sec.~\ref{sec:2d}, with $k\ls \approx 90$ and seven grid points per wavelength in the medium, the relative shell width $1/(k\ls) = 1.1\%$ and the grid anisotropy $(k\Delta x)^2/48 = 1.7\%$ are comparable, and the measured fourfold modulation of the shell radius, $0.8\%$ of $k$, cf.\ Supplemental Material Sec.~\ref{sec:sm1_params}, is of the order of the shell width.
The shell mask therefore follows the discrete dispersion iso-surface of Eq.~\eqref{eq:discrete_dispersion} rather than a circle in every 2D calculation of this work, whereas the spectral solver of Sec.~\ref{sec:3d} is free of this anisotropy.
We revisit these considerations in Secs.~\ref{sec:2d} and~\ref{sec:3d}.
A high-precision test of this dispersion compensation, together with the removal of the finite-size broadening, confirms the Lorentzian shape of the ring, see Supplemental Material Sec.~\ref{sec:sm_rte_numerics}.

\emph{Boundary- and source-induced anisotropy.}
In finite systems with directional or spatially confined sources, the azimuthal distribution of spectral weight along the shell might become nonuniform. 
A Gaussian beam impinging on a semi-infinite scattering medium is an example of such a situation.
In this case, fields near the injection boundary or in the dominant direction $\vec{k}_{\rm in}$ carry larger amplitude than those propagating in the bulk of the sample, producing intensity buildup in $|\tilde{E}(\vec{k}^\prime)|^2$ close to $\vec{k}_{\rm in}$ in Fourier space. 
This effect reflects the exponential decay of ballistic intensity with depth~\cite{2007_Akkermans_Mesoscopic_Physics,1999_vanRossum_diffuse_waves}.
Such anisotropy affects only the magnitude distribution along the shell, whereas the characteristic radial spread $\dks \sim 1/\ell_s$ remains determined by bulk scattering properties and is independent of the boundary and source geometry. 
Each of these finite-size and source effects is quantified in this work.
The sharp spectral truncation affects the field only within a boundary layer of depth $1/\dkm$ suppressed by increasing the mask width, cf.\ Sec.~\ref{sec:3d_results} and Supplemental Material Secs.~\ref{sec:sm_metriczones}, \ref{sec:sm_crossover}.
The angle-resolved support of the shell isotropizes once the transport depth exceeds unity, cf.\ Supplemental Material Sec.~\ref{sec:sm_mask}, and the reconstruction error of the fields is set by the discarded power at every transport depth, cf.\ Supplemental Material Sec.~\ref{sec:sm_crossover}.
One isotropic mask therefore applies to every medium, source, and regime we tested.

\subsection{Extension to second-order quantities\label{sec:secondorder}}
\noindent Physical observables such as intensity $I(\vec{r})=|E(\vec{r})|^2$, spatial correlations of the kind $E(\vec{r})E^*(\vec{r}^\prime)$, and sensitivity functions involve products of two or more fields~\cite{1999_vanRossum_diffuse_waves,2023_Jara_Simulation_Coherent_Remission,2022_Bender_Coherent_Enhancement,2025_Jara_coherent_sensing}. 
Computing these quantities directly requires decompressing fields to real space, multiplying, and (potentially) averaging over disorder configurations. 
Although averaging can be accumulated incrementally, without the need to store all decompressed fields simultaneously, the very process of decompressing all fields could be resource-constrained. 
The convolution structure of Fourier transforms permits an alternative proposed below. Second-order (as well as higher-order) quantities can be computed entirely in compressed $k$-space, with only the single result decompressed.

\emph{Intensity as convolution.}\label{sec:intensity_convolution}
As an example, we consider intensity of a scalar field $E(\vec{r})$ found in a specific realization of disorder (no ensemble averaging)
\begin{equation}
I(\vec{r})=E(\vec{r})E^*(\vec{r}).
\label{eq:intensity_def}
\end{equation}
In Fourier space, this product becomes a convolution:
\begin{equation}
\tilde{I}(\vec{q})=\int\frac{d^dk^\prime}{(2\pi)^d}\,\tilde{E}(\vec{k}^\prime)\,\tilde{E}^*(\vec{k}^\prime-\vec{q}),
\label{eq:intensity_convolution}
\end{equation}
where $\vec{q}$ is the spatial frequency of the intensity pattern and $d = 2, 3$ is the dimensionality of the space. 
The support of this convolution extends slightly beyond $2k$, because two on-shell wave vectors separated by a large angle interfere at half-wavelength scale, and the shell coefficients carry this structure in full, cf.\ Supplemental Material Sec.~\ref{sec:sm_specsupport}.
Evaluating Eq.~\eqref{eq:intensity_convolution} on the retained components therefore differs from the exact product only through terms containing the discarded amplitudes, and their spectral weight is the small fraction $P_{\rm out}$ quantified below.
Averaging over cells of size $L_c \gtrsim 1/\dkm$ confines the product to $|\vec{q}| \lesssim \dkm$ and suppresses the cross terms, reducing its error well below the field-level value, cf.\ Sec.~\ref{sec:3d_secondorder} and Supplemental Material Sec.~\ref{sec:sm_cg}.
The runtime cost of the compressed convolution, at native and at envelope resolution, is measured in Sec.~\ref{sec:sm_benchcost}.

Computing products in compressed $k$-space yields two computational advantages.
First, the convolution Eq.~\eqref{eq:intensity_convolution} can be evaluated using only the compressed shell data, without reconstructing the full field.
Second, when the observable is needed only at the envelope resolution, the output occupies $(L/L_c)^d$ points rather than $(n_{\Delta x}L/\lambda)^d$, a reduction of the final slowly varying quantity of interest beyond the field compression of Eq.~\eqref{eq:compression}.

\emph{Optical sensitivity.}
$S(\vec{r}_0)$ quantifies the response of detected flux to a localized absorptive perturbation at position $\vec{r}_0$ inside the medium~\cite{1999_Arridge_optical_tomography,2009_Arridge_Schotland_review,2010_Durduran_DOT}. 
For a point perturbation $\delta\varepsilon(\vec{r})=i\,\delta\epsilon\,\delta V\,\delta(\vec{r}-\vec{r}_0)$, it takes the form
\begin{equation}
S(\vec{r}_0)=-k_0^2\,\delta\epsilon\,\delta V\,\mathrm{Re}\left[E_{\rm adj}(\vec{r}_0)E_{\rm fwd}(\vec{r}_0)\right],
\label{eq:sensitivity}
\end{equation}
where $E_{\rm fwd}(\vec{r}_0)$ is the field at $\vec{r}_0$ due to illumination from the source and $E_{\rm adj}(\vec{r}_0)=\sum_b u_b^*\phi_b(\vec{r}_0)$ is a back-propagating field constructed from the output amplitudes $u_b$ and the fields $\phi_b(\vec{r}_0)$ excited at $\vec{r}_0$ by illumination from detector channel $b$.
This expression involves a product of two fields at the same point, analogous to Eq.~\eqref{eq:intensity_convolution}, and holds for arbitrary illumination. 
When excitation is unrelated to the specific disorder realization, disorder averaging reduces Eq.~\eqref{eq:sensitivity} to a product of two mean intensities, each obtainable from the diffusion equation or the radiative transfer equation~\cite{1999_Arridge_optical_tomography,2009_Arridge_Schotland_review}.
This equivalence has been established analytically~\cite{2025_Jara_coherent_sensing}, supporting the use of diffusion-based sensitivity under random illumination.

The situation changes fundamentally when the incident wavefront is tailored to the specific disorder realization.
In this case the configuration-dependent phase between $E_{\rm fwd}$ and $E_{\rm adj}$ contributes to the sensitivity and cannot be recovered from mean intensities.
Consequently, Eq.~\eqref{eq:sensitivity} must be evaluated at the wave-field level~\cite{2025_Jara_coherent_sensing}.
We refer to this setting as wavefront-shaped diffuse optical tomography (WFS-DOT), which is not a separate modality but DOT operated under medium-specific excitation, for which a wave-field computation is unavoidable.
Biological tissue at near-infrared wavelengths has $k\ell_s\sim500$--$1000\gg1$~\cite{2013_Jacques_optical}, so the wave fields in WFS-DOT are concentrated sharply on-shell in $k$-space, and OSCAR compression applies.

In Fourier space, Eq.~\eqref{eq:sensitivity} also becomes a convolution of the transforms $\tilde{E}_0$ and $\tilde{E}_{\mathrm{out}}$, each confined to a shell of width $\sim 1/\ell_s$. 
As for the intensity, the cell-averaged sensitivity map is reproduced faithfully at cell sizes $L_c \gtrsim 1/\dkm$, permitting coarser spatial sampling of the output.
The convolution can be evaluated directly from compressed data, and the output requires only $(L/L_c)^d$ points, where $L$ is the system size. 
This enables efficient computation of microscopic sensitivity maps, essential to tomographic reconstruction, without decompressing individual field realizations.

\emph{Ensemble averaging without decompression.}
For disorder-averaged quantities $\overline{I}(\vec{r})$ or correlation functions $\overline{E(\vec{r})E^*(\vec{r}^\prime)}$, the sample-independence of shell geometry becomes essential. 
Because all compressed quantities in all disorder realizations share the same shell structure, ensemble operations can be performed on compressed data by averaging the convolution results over all realizations without ever reconstructing individual fields, since the projection is linear and the mask is common to the ensemble, cf.\ Supplemental Material Sec.~\ref{sec:sm1_params}. 
The tested compression schemes lack this property because each realization compresses differently, precluding direct operations involving different members of the ensemble.

The combination of shell-based compression, $k$-space convolution, and coarse output discretization transforms the computational scaling from $\propto(n_{\Delta x}L/\lambda)^d\times N_{\rm rlz}$ to $\propto(L/\ell_s)^d\times N_{\rm rlz}$ for ensemble-averaged second-order quantities, where $N_{\rm rlz}$ is the number of disorder realizations. 
At the coarsest cells of Sec.~\ref{sec:3d_secondorder}, of side $\approx 0.9\,\ell_s$, this amounts to a factor of $2\times10^6$ in 3D.

\begin{figure*}[ht!]
\centering
\includegraphics[width=\textwidth]{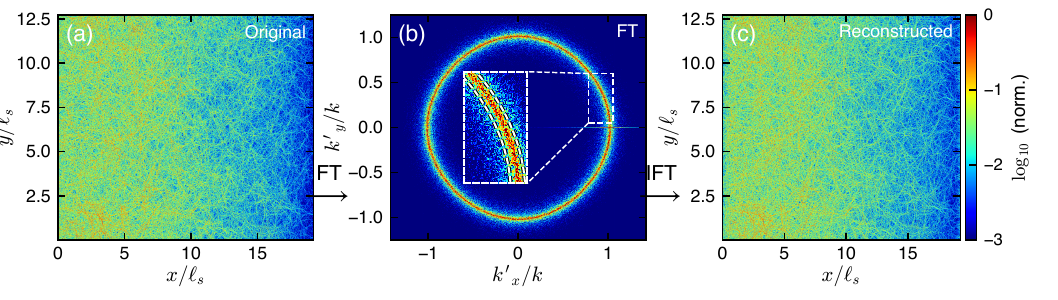}
\caption{Schematic of the OSCAR method.
(a) Real-space intensity $|E(\vec{r})|^2$ ($\log_{10}$, three decades) from a 2D simulation under plane-wave illumination.
(b) Spatial Fourier transform $|\tilde{E}(\vec{k}^\prime)|^2$ ($\log_{10}$, three-decade range) showing field amplitude concentrated on the characteristic shell $|\vec{k}^\prime| = k$.
The magnified segment (inset) shows the dispersion-shaped mask edges (white dashed, $\alpha = 5$, $C = 132$).
(c) Reconstructed field after inverse Fourier transform of the compressed data, preserving speckle structure with $\epsilon_\varphi = 0.063$ ($\epsilon_{L2} = 0.35$). Parameters are defined in Sec.~\ref{sec:2d_setup}.
\label{fig:OSCAR_schematic}}
\end{figure*}

\section{Two-Dimensional Validation\label{sec:2d}}
\noindent In this section we validate the OSCAR algorithm using numerical simulations of scalar waves in two-dimensional (2D) disordered media. This geometry permits direct comparison between compressed and uncompressed fields across large ensembles of disorder realizations, while capturing the essential physics of multiple scattering and interference. The 2D geometry also corresponds to experimental realizations in planar photonic structures, where in-plane confinement reduces the electromagnetic problem to a scalar wave equation~\cite{2016_Sarma_Open_Channels,2020_Bender_Eigenchannels,2022_Bender_Coherent_Enhancement,2023_Jara_Simulation_Coherent_Remission,2025_Jara_coherent_sensing}.

\subsection{Simulation configuration\label{sec:2d_setup}}
\noindent We consider transverse magnetic (TM) polarization, for which the out-of-plane electric field component satisfies the 2D scalar Helmholtz equation~\eqref{eq:helmholtz}. Numerical solution is obtained using the MESTI software package~\cite{2022_Lin_MESTI_Augmented}, which employs an augmented partial factorization method enabling efficient multi-source simulations on a computer cluster~\cite{2024_Cluster}.

All 2D simulations share the solver settings: background refractive index $n_{\mathrm{eff}} = 2.85$, vacuum wavelength $\lambda_0 = 1.55\um$ (in-medium wavelength $\lambda \approx \lambda_0/n_{\mathrm{eff}} = 0.54\um$), and grid spacing $\Delta x = \lambda_0/20$, which gives $n_{\Delta x} \simeq 7$ points per in-medium wavelength.
Three geometries are used.
Every transport parameter quoted below is measured as described in Supplemental Material Sec.~\ref{sec:sm_media}.
Ensemble sizes (number of disorder realizations) are stated per figure.

(i)~The periodic plane-wave box behind Figs.~\ref{fig:onshell_schematic} and \ref{fig:OSCAR_schematic} is a $99.2 \times 150\um$ domain with periodic transverse boundaries, illuminated by a plane wave through a line source spanning the full width.
Disorder consists of randomly positioned sub-wavelength air holes (radii $40$--$65$\,nm) at a filling fraction of $0.1$, giving $\ls = 7.8\um$ from the ballistic decay of the coherent field and $k\ls \approx 90$.
The measured on-shell anisotropy, the mean cosine of the scattering angle, is $g \approx 0.03$, so the individual scatterers are nearly isotropic.

\begin{figure}[ht!]
\centering
\includegraphics[width=\columnwidth]{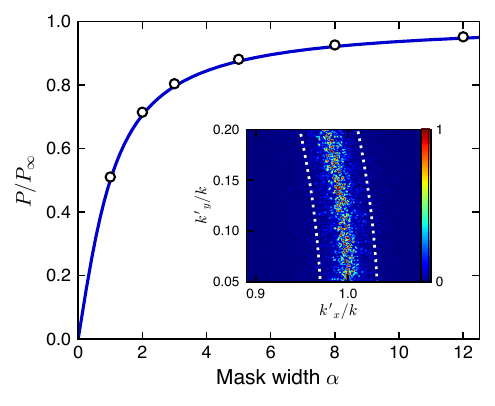}
\caption{Power fraction $P/P_\infty$ retained within the mask as a function of the normalized mask width $\alpha = \dkm\,\ls$.
Solid line: analytical prediction, Eq.~\eqref{eq:power_ratio}.
Circles: the square plane-wave ensemble mean over $40$ realizations, with error bars, smaller than the markers, the ensemble standard deviation.
Inset: spectral density of one realization near the shell, in a window off the ballistic line, with the dispersion-shaped mask edges at $\alpha = 5$ (dotted).
\label{fig:intensity_correction}}
\end{figure}

(ii)~The square plane-wave ensemble behind Figs.~\ref{fig:intensity_correction} and \ref{fig:boundary_correction_2d} is the same medium in a $300 \times 300\um$ box, $40$ realizations.
The square shape removes the box-anisotropy distortions of the rectangular geometry, and its size keeps the finite-size window $2\pi\ls/L = 0.16$ small in units of $1/\ls$, cf.\ the aperture limit of Sec.~\ref{sec:onshell}.

(iii)~The remission geometry behind Fig.~\ref{fig:second_order_quantities} is a $250 \times 300\um$ slab surrounded by perfectly matched layers~\cite{2005_Taflove_FDTD}, with light injected through a finite-size input region of width $W_1 = 10\um$ on the front interface, supporting $N_1 = kW_1/\pi \approx 37$ transverse modes.
Its medium is identical to that of geometries (i) and (ii), with the same measured $\ls = 7.8\um$ and $k\ls \approx 90$, see the remission row of Supplemental Material Table~\ref{tab:sm3_2d}.

The OSCAR procedure, illustrated schematically in
Fig.~\ref{fig:OSCAR_schematic}, is applied as follows:
the computed field is Fourier-transformed, masked to the
characteristic shell, renormalized, and reconstructed via
inverse transform.

\subsection{Field reconstruction fidelity\label{sec:2d_field}}
\noindent We quantify reconstruction accuracy with three complementary metrics, ordered from the strictest to the most forgiving.
Here $E_0$ denotes the original field and $E_c$ the compressed-reconstructed one.
The relative $L_2$ error
\begin{equation}
  \epsilon_{L2} = \left[\frac{\int\left|E_c(\vec{r}) - E_0(\vec{r})\right|^2 d\vec{r}}{\int|E_0(\vec{r})|^2\,d\vec{r}}\right]^{1/2}
  \label{eq:eps_L2}
\end{equation}
penalizes every difference between the two fields.
The overlap error
\begin{equation}
  \epsilon_\varphi = 1 -
  \frac{\left|\int E_0^*(\vec{r})\,E_c(\vec{r})\,d\vec{r}\right|}
  {\sqrt{\int|E_0(\vec{r})|^2\,d\vec{r}\;\int|E_c(\vec{r})|^2\,d\vec{r}}}
  \label{eq:eps_varphi}
\end{equation}
is insensitive to an overall complex pre-factor but penalizes local phase errors through the complex inner product.
The amplitude metric
\begin{equation}
  \epsilon_A = 1 - 
  \frac{\int\left|E_0^*(\vec{r})E_c(\vec{r})\right|d\vec{r}}
  {\sqrt{\int|E_0(\vec{r})|^2\,d\vec{r}\;\int|E_c(\vec{r})|^2\,d\vec{r}}}
  \label{eq:eps_A}
\end{equation}
compares the local magnitude profiles. With the absolute value inside the integral, it is insensitive to local phase errors and is the most forgiving of the three.
The last two are non-negative by the Cauchy--Schwarz inequality, applied to the fields and to their magnitudes respectively, while $\epsilon_{L2}$ is non-negative by construction.
All three vanish for $E_c = {\rm const}\times E_0$, with ${\rm const} = 1$ required for $\epsilon_{L2}$.
For the on-shell projection with the power restored, the two strict metrics follow exactly from the discarded spectral weight $P_{\rm out} = 1 - P/P_\infty$:
\begin{equation}
  \epsilon_\varphi = 1 - \sqrt{1 - P_{\rm out}}, \qquad
  \epsilon_{L2} = \sqrt{2\,\epsilon_\varphi},
  \label{eq:metric_identities}
\end{equation}
while the $\epsilon_A$ prediction involves complete elliptic integrals of the retained fraction, cf.\ Appendix~\ref{app:metrics}, which derives it under an assumption of locally Gaussian speckle statistics and derives Eq.~\eqref{eq:metric_identities} exactly.
All analytic expressions are tested numerically and verified, cf.\ Supplemental Material Sec.~\ref{sec:sm_metricverify}.

Figure~\ref{fig:intensity_correction} shows the fractional power $P/P_\infty$ retained within a mask of full width $\dkm$ as a function of the normalized mask width $\alpha = \dkm\,\ls$, comparing the analytical prediction Eq.~\eqref{eq:power_ratio} with the square plane-wave ensemble.
The dispersion-shaped mask follows the discrete dispersion iso-surface of the finite-difference grid, with a single ensemble-fitted mask applied to every realization, cf.\ Supplemental Material Secs.~\ref{sec:sm_media} and \ref{sec:sm_rte_numerics}.
Explicitly, a coefficient at grid wave vector $\vec k'$ is retained when $|K(\vec k') - k_p| < \dkm/2$, with $K(\vec k')$ of Eq.~\eqref{eq:discrete_dispersion} and $k_p$ the ring radius in $K(\vec k')$, fitted once on the ensemble, cf.\ Supplemental Material Sec.~\ref{sec:sm1_params}. The shell width $1/\ls$ of Eq.~\eqref{eq:lorentzian} is a width in $K(\vec k')$, which makes it the proper coordinate to define the mask.
The agreement confirms that the Lorentzian spectral density of Eq.~\eqref{eq:lorentzian} accurately describes the radial distribution of the spectral weight, within $2\%$ at $\alpha = 1$ and closer at every wider mask.
The per-realization spread of $P_{\rm out}$ is at or below the percent level across the ensembles, rising above it only in the smallest, size-limited 2D box, cf.\ Supplemental Material Sec.~\ref{sec:sm_metricverify}.
This sample independence ensures robustness of the disorder-averaged power normalization.

\begin{figure}[t]
\centering
\includegraphics[width=\columnwidth]{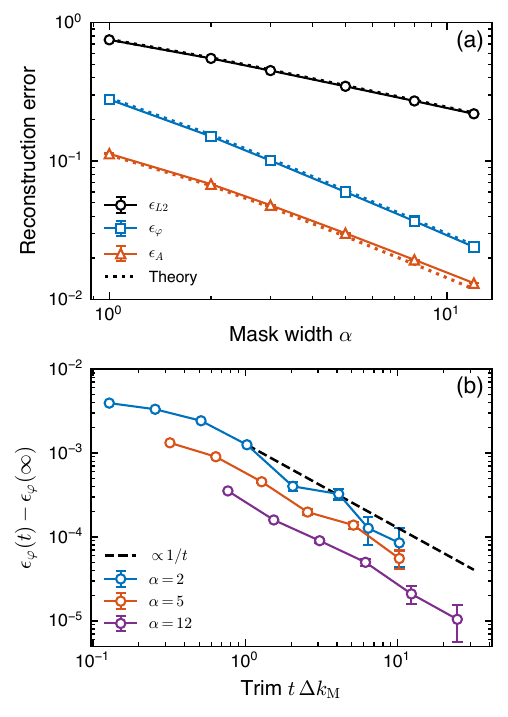}
\caption{Reconstruction accuracy in 2D, square plane-wave ensemble of $40$ realizations, with error bars the ensemble standard deviation.
(a)~The three field-error metrics in Eqs.~\eqref{eq:eps_L2}--\eqref{eq:eps_A} vs the normalized mask width $\alpha = \dkm\,\ls$, strictest to most forgiving, namely $\epsilon_{L2}$ (with the disorder-averaged power-restoring constant $c_2 = (1-\overline{P}_{\rm out})^{-1/2}$), $\epsilon_\varphi$, and $\epsilon_A$.
Dotted lines: the parameter-free predictions evaluated at the measured $\overline{P}_{\rm out}$, Eqs.~\eqref{eq:metric_identities} and the $\epsilon_A$ expression of Appendix~\ref{app:metrics}.
(b)~The excess overlap error $\epsilon_\varphi(t) - \epsilon_\varphi(\infty)$ vs boundary trim $t\,\dkm$ at $\alpha = 2, 5, 12$, showing the excess saturating toward its finite $t \to 0$ value and decaying as $1/t$ beyond $t\,\dkm \sim 1$ (dashed guide), which confirms that Gibbs artifacts are confined to a boundary layer of depth $1/\dkm$.
\label{fig:boundary_correction_2d}}
\end{figure}

The trade-off between compression and reconstruction fidelity is controlled by the mask width $\alpha$, and its choice relative to the measured shell support is discussed in Supplemental Material Sec.~\ref{sec:sm_mask}.
Figure~\ref{fig:boundary_correction_2d}(a) shows all three metrics computed in the square plane-wave ensemble, each falling on its parameter-free prediction evaluated at the measured $\overline{P}_{\rm out}$, with no fitted constants.
At $\alpha = 5$, the mask width of Fig.~\ref{fig:OSCAR_schematic}, the square ensemble gives $\overline{P}_{\rm out} = 0.12$ and the ensemble means $\epsilon_{L2} = 0.35$, $\epsilon_\varphi = 0.060$, and $\epsilon_A = 0.030$, cf.\ Appendix~\ref{app:metrics}.
Despite these relatively high numerical values, visual inspection of Fig.~\ref{fig:OSCAR_schematic}(a,c) shows that the speckle pattern is preserved down to individual grains.
At $\alpha = 12$ they reach $0.22$, $0.024$, and $0.013$.
The asymptotic scalings follow from $P_{\rm out} \simeq 2/(\pi\alpha)$, with $\epsilon_\varphi \to 1/(\pi\alpha)$ while the strict $\epsilon_{L2} \to \sqrt{2/(\pi\alpha)}$ decays only as $\alpha^{-1/2}$.
The same parameter-free relationships hold across the quasi-ballistic to diffusive crossover in both 2D and 3D, cf.\ Supplemental Material Sec.~\ref{sec:sm_crossover}.

Sharp $k$-space truncation produces Gibbs-like artifacts near the domain boundaries.
Figure~\ref{fig:boundary_correction_2d}(b) isolates them through the excess overlap error $\epsilon_\varphi(t) - \epsilon_\varphi(\infty)$ as a function of the boundary trim $t$, normalized by the mask width $\dkm$, cf.\ Supplemental Material Sec.~\ref{sec:sm_metriczones}.
The excess saturates toward its finite $t \to 0$ value for $t\,\dkm \ll 1$ and decays as $1/t$ beyond, confirming that the affected layer has depth $1/\dkm = \ls/\alpha$, independent of wavelength and grid spacing.
Quantities identified as interior in this paper use the trim $t = 3/\dkm$, except at the longitudinal faces of the 3D slabs, cf.\ Supplemental Material Sec.~\ref{sec:sm_metriczones}.

These results show that OSCAR achieves $\epsilon_\varphi = 0.03$ at $\alpha \approx 10$, corresponding to the compression ratio $C_{\rm OSCAR} \simeq 68$ for this medium, and $\epsilon_\varphi = 0.063$ at $\alpha = 5$, $C_{\rm OSCAR} \simeq 132$, both for $k\ls = 90$ in the plane-wave box system of Fig.~\ref{fig:OSCAR_schematic}. Both ratios fall about $6\%$ below Eq.~\eqref{eq:compression}, with the deviation attributed to the difference between $K(\vec k')$ and $|\vec k'|$ on the finite-difference grid.
\subsection{Second-order quantities\label{sec:2d_intensity}}
\noindent To validate the $k$-space convolution approach of Sec.~\ref{sec:secondorder} for computing second-order quantities directly from compressed data, without reconstructing full real-space fields, we consider the optical sensitivity $S(\vec{r}_0)$, Eq.~\eqref{eq:sensitivity}.
It involves the product of two distinct fields: the forward-propagating field $E_{\rm fwd}(\vec{r}_0)$ due to the source and the back-propagating field $E_{\rm adj}(\vec{r}_0)$ constructed from detector channels.
This quantity is directly relevant to WFS-DOT~\cite{2025_Jara_coherent_sensing} and provides a stringent test of OSCAR, as it requires accurate preservation of the relative phase between independently computed fields.
We compute sensitivity similarly to the convolution Eq.~\eqref{eq:intensity_convolution} using only the shell-compressed Fourier components.

\begin{figure*}[ht!]
\centering
\includegraphics[width=\textwidth]{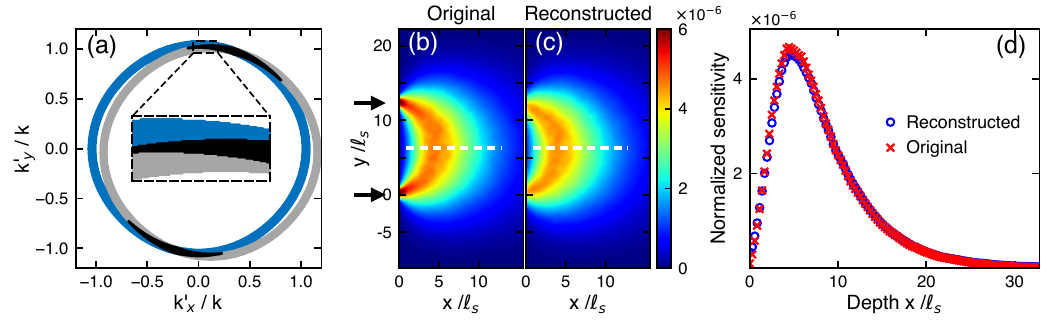}
\caption{Second-order quantities from compressed 2D fields.
(a)~Schematic of the convolution geometry, the overlap of two shells of width $\sim 1/\ell_s$ (blue and gray) and their common sites (black), with the overlap region magnified in the inset.
(b,c)~The sensitivity $S(\vec{r}_0)$ computed from original~(b) and reconstructed~(c) fields via $k$-space convolution, with mask width $\alpha=2.4$, averaged over $1000$ disorder realizations. Arrows in (b) mark the input and output ports of the remission geometry~\cite{2025_Jara_coherent_sensing}.
Discrepancies are confined to the boundaries due to Gibbs artifacts.
Dashed line indicates the cross-section shown in~(d).
(d)~Depth profile of $S(\vec{r}_0)$ along the dashed line in~(b,c), extended over the full depth of the sample, comparing original (crosses) and reconstructed (circles) results. The two curves overlap throughout the interior, confirming that coherent interference between independently compressed fields is faithfully retained by the $k$-space convolution.
The two profiles differ by $3\%$ in relative rms.
The spatial axes of (b--d) are in units of $\ell_s$, parameters as in Sec.~\ref{sec:2d_setup}.
\label{fig:second_order_quantities}}
\end{figure*}

Figure~\ref{fig:second_order_quantities}(b,c) compares the sensitivity map computed from uncompressed (raw) fields with that obtained entirely from compressed data via $k$-space convolution, for a mask width as small as $\alpha=2.4$. 
At this narrow mask, the compressed sensitivity map agrees with the original throughout the sample interior, see Fig.~\ref{fig:second_order_quantities}(d), with discrepancies confined to a narrow boundary layer, which we attribute to Gibbs artifacts, cf.\ Fig.~\ref{fig:boundary_correction_2d}(b).
The sensitivity computed via the OSCAR route in the individual disorder realizations reproduces the wavelength-scale features (in both 2D and 3D), as detailed in Supplemental Material Sec.~\ref{sec:sm_cg}.
This is possible because the shell coefficients carry the full $|\vec{q}| > 2k$ support of the kernel, as measured in Supplemental Material Sec.~\ref{sec:sm_specsupport}.

The agreement in Fig.~\ref{fig:second_order_quantities}(b--d) confirms that OSCAR indeed preserves the coherent interference between independently computed and compressed fields, i.e., the phase relationship between the forward field $E_{\rm fwd}(\vec{r}_0)$ and the back-propagating field $E_{\rm adj}(\vec{r}_0)$, each compressed separately, is faithfully retained in their product.
This is essential for applications such as WFS-DOT, which requires products of fields from different input channels.
The same test on a twenty-realization plane-wave ensemble of the same medium, at every mask width, is Supplemental Material Fig.~\ref{fig:sm5_sens2d}(a).

The sensitivity maps computed here via convolution of compressed data, and intensity maps by the identical construction, admit discretization on a coarse output grid of cell size $L_c \gtrsim 1/\dkm$ rather than the sub-wavelength grid of the original fields.
Combined with the shell compression of the input fields, this yields the overall scaling reduction from $(n_{\Delta x}L/\lambda)^d$ to $(L/L_c)^d$ for second-order observables, as predicted in Sec.~\ref{sec:secondorder}.

\section{Three-Dimensional Validation\label{sec:3d}}
\noindent We now apply the OSCAR method to the three-dimensional (3D) fields obtained in the full-wave vector electromagnetic simulations for a system representative of biological tissue. 
The extension from 2D to 3D is essential for practical applications where tissue volumes probed by near-infrared light in biomedical optics are inherently three-dimensional, and vector polarization effects absent in scalar 2D models can influence scattering statistics~\cite{2013_Jacques_optical}.

The compression efficiency of OSCAR scales as $k\ell_s$ regardless of dimensionality, predicting ratios of $\sim$100--1000 for typical tissue parameters where $\ls \sim 100\um$ and $\lambda \sim 1\um$. 
Below we validate the method on plane-wave ensembles spanning the transport crossover from quasi-ballistic to diffusive ($\gamma \equiv L(1-g)/\ls = 0.011$--$10.4$, with $g$ the single-scattering anisotropy), with two disorder generators of closed-form spectra and sixteen systems in total, see Supplemental Material Sec.~\ref{sec:sm_media}.
Section~\ref{sec:3d_secondorder} also demonstrates second-order observables computed from the compressed fields.
Importantly, the mask geometry remains universal across all samples (i.e., disorder realizations) within a given simulation configuration. 
The sample-independent structure required for ensemble post-processing, cf.\ Sec.~\ref{sec:secondorder}, is therefore preserved, enabling calculations of the second-order quantities and disorder-averaging to be computed entirely in compressed $k$-space without the need for decompression of individual fields.

\subsection{Simulation configuration\label{sec:3d_setup}}
\noindent 3D simulations are performed using WaveSim~\cite{2016_Osnabrugge_wavesim}, a modified Born series solver implemented on graphics processing units (GPUs) via the CuPy GPU array library. 
The method enables efficient computation of vector electromagnetic fields in large scattering volumes aided, in part, by domain decomposition~\cite{2026_Mache_domain_decomposition}.

\begin{figure}[t]
\centering
\includegraphics[width=\columnwidth]{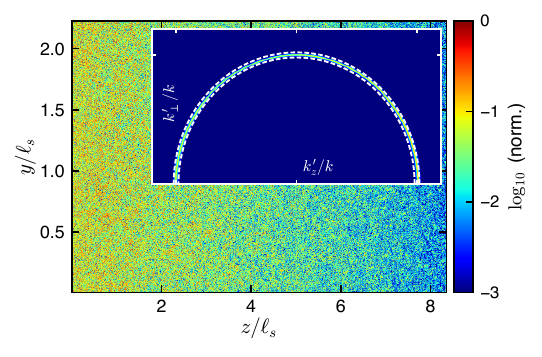}
\caption{A raw (uncompressed) plane-wave 3D field and its on-shell spectrum.
Main panel: intensity cut $\log_{10} I(x_0, y, z)$ at the mid-$x$ plane of a diffusive slab ($k\ls \approx 300$, $\gamma = 4.11$), axes in units of $\ls$.
Inset: azimuthally averaged spectral density in the $(k_z^\prime, k_\perp^\prime)$ plane ($k_\perp^\prime$ from $0$ to $1.2k$, $k_z^\prime$ from $-1.2k$ to $1.2k$), axes normalized by the measured shell radius, with the isotropic mask edges at $\alpha = 12$ (dashed, shown for reference).
Both the main panel and the inset use a $\log_{10}$ color scale over three decades.
\label{fig:3D_beam}\label{fig:3D_shell}}
\end{figure}

The simulation models near-infrared light propagation of wavelength $\lambda_0 = 0.85\um$ in tissue-like media. 
The background refractive index $n_{\rm bkg} = 1.35$ gives in-medium wavelength $\lambda \approx \lambda_0/n_{\rm bkg} = 0.63\um$ and wavenumber $k = 2\pi n_{\rm bkg}/\lambda_0 \approx 10\um^{-1}$. 
Voxel resolution $\Delta x = \lambda_0/(3n_{\rm bkg}) \approx 0.21\um$ ensures rapid convergence of the numerical algorithm~\cite{2016_Osnabrugge_wavesim}.
The solver applies the Laplacian spectrally, so the discrete dispersion of Eq.~\eqref{eq:discrete_dispersion} does not arise and the isotropic mask needs no dispersion correction.
The main considered geometry is a slab of $320^2 \times 1200$ voxels $= 67 \times 67 \times 252\um$, periodic transversely and absorbing along $z$~\cite{2026_Mache_domain_decomposition}, illuminated by an $x$-polarized plane wave incident along $+z$.
A deep slab of $128^2 \times 3000$ voxels $= 27 \times 27 \times 630\um$ extends the thickness to $21\ls$ for its medium.
For the diffusive system featured in Fig.~\ref{fig:3D_beam}, the transverse extent is $2.2\ls$, with the periodic boundaries acting as a supercell period rather than an aperture, cf.\ Supplemental Material Sec.~\ref{sec:sm_media}.
The same medium in the $27\um$ deep slab, $0.9\ls$ wide, gives an interior error ratio $\epsilon_{L2}/\sqrt{P_{\rm out}}$ within the realization spread of the $67\um$ boxes, cf.\ Supplemental Material Sec.~\ref{sec:sm_crossover}.

Disorder is a spatially correlated refractive-index field $n(\vec{r}) = n_{\rm bkg} + \delta n(\vec{r})$ built from unit-variance filtered Gaussian noise, using two generators with closed-form spectra.
Gaussian-correlated disorder of correlation length $\ell_c$ has anisotropy $g = \coth a - 1/a$ with $a = 2(k\ell_c)^2$.
Spectral top-hat disorder, constant spatial spectrum up to a cutoff $q_c$, has $g = 1 - q_c^2/4k^2$ exactly~\cite{1978_Ishimaru_Wave_Propagation}.
Varying the generator, its scale, and the amplitude $\delta n$ produces sixteen systems spanning $\ls = 5.8$--$299\um$ ($k\ls \approx 60$--$3000$), $g = 0.51$--$0.999$, and transport depths $\gamma = L(1-g)/\ls$ from $0.011$ to $10.4$.
The detailed description of definitions and measurement procedure for obtaining each parameter is given in Supplemental Material Sec.~\ref{sec:sm_media}, cf.\ Table~\ref{tab:sm3_3d}.
The value of $\ell_s$ is determined from the exponential decay of the transversely averaged (denoted as $\langle ...\rangle_{x,y}$) co-polarized coherent field, $\langle E_x(\mathbf{r})\rangle_{x,y} \propto e^{-z/2\ell_s}$, under plane-wave illumination~\cite{1999_vanRossum_diffuse_waves}.

\subsection{Mask selection considerations\label{sec:3d_shell}}

\noindent Figure~\ref{fig:3D_beam} shows a representative (raw) diffusive field ($k\ls \approx 300$, $\gamma = 4.11$).
Its spatial spectrum (inset) is indeed concentrated on the shell $|\vec{k}^\prime| = k$ of width $\sim 1/\ls$, azimuthally averaged in the $(k_z^\prime, k_\perp^\prime)$ plane, with typical isotropic mask edges at $\alpha = 12$ shown for reference, also cf.\ Figs.~\ref{fig:onshell_schematic}--\ref{fig:intensity_correction} in 2D.

In the regime of underdeveloped speckle~\cite{2007_Goodman_speckle}, an anisotropic $k$-shell persists over the first few transport depths.
In Supplemental Material Sec.~\ref{sec:sm_mask}, we quantitatively investigate this effect by measuring the angle-resolved support of the shell in samples across the transition from quasi-ballistic to diffusive transport, cf.\ Supplemental Material Sec.~\ref{sec:sm_media}.
We confirm the shell isotropization when the transport depth $\gamma = L(1-g)/\ls$ exceeds unity.
Importantly, even at $\gamma \lesssim 1$, OSCAR reaches the target error with an isotropic mask, cf.\ Sec.~\ref{sec:benchmarks}, and the mask itself does not depend on the disorder realization.
There, an anisotropic mask can make the compression even more efficient (measured up to a factor of $22$ at $\gamma \approx 0.01$), an option we benchmark in Sec.~\ref{sec:benchmarks}.
However, a judiciously chosen isotropic mask is shown to be sufficient to cover the measured support in every regime we tested, the property of OSCAR that we refer to as `universal compression'.
Therefore, as in 2D, we default throughout to the \emph{isotropic mask} $\bigl||\vec{k}^\prime| - k\bigr| < \dkm/2$. The geometry and parameters of such a mask depend only on the measured $k$ and the chosen $\alpha$, and not on the source, polarization, or boundary geometry.

\begin{figure*}[t]
\centering
\includegraphics[width=\textwidth]{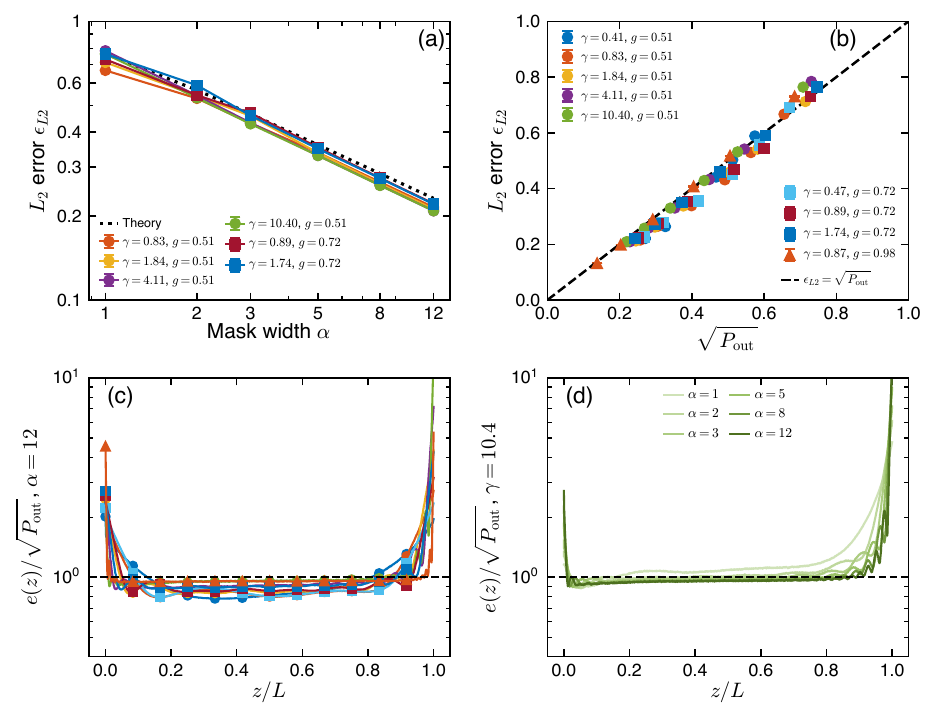}
\caption{Field reconstruction in 3D across the transport crossover.
(a)~Interior $\epsilon_{L2}$ vs the mask width for the six systems with $\gamma \ge 0.83$ and $g \le 0.723$, listed in the legend.
The dotted line is the theory prediction of Eqs.~\eqref{eq:power_ratio} and \eqref{eq:metric_identities}.
(b)~A range of quasi-ballistic to diffusive systems, listed in the legend, plotted vs $\sqrt{P_{\rm out}}$, collapsing onto the theoretical relationship (dashed).
Panel (c) shows the depth-resolved error at fixed $\alpha = 12$ for all systems of (b). Panel (d) shows the same in one system ($\gamma = 10.4$, $g = 0.51$, the deep slab of depth $21\ls$) for different values of $\alpha$.
The depth-resolved error $e(z)$, Eq.~(\ref{eq:sm2_depthres}) of the Supplemental Material, is normalized by the theoretical prediction $\sqrt{P_{\rm out}}$.
The error is depth-independent in the interior of the system, with deviations due to the Gibbs effect, cf.\ Supplemental Material Sec.~\ref{sec:sm_metriczones}, confined to the boundary layers.
\label{fig:3d_uncompression}\label{fig:boundary_correction_3D}}
\end{figure*}

\subsection{Field reconstruction fidelity\label{sec:3d_results}}
To quantify the reconstruction fidelity we consider sixteen different systems spanning the transition from quasi-ballistic to diffusive transport, cf.\ Supplemental Material Sec.~\ref{sec:sm_media}, nine of which appear in Fig.~\ref{fig:3d_uncompression}, with $10$ realizations per system used to compute the standard deviation shown as error bars.
Figure~\ref{fig:3d_uncompression}(a) shows the interior $\epsilon_{L2}$ vs the mask width in six different systems with $\gamma = 0.83$--$10.4$ and $g = 0.51$--$0.72$.
Every system follows its parameter-free prediction $\sqrt{P_{\rm out}}$, and plotting these data together with the three remaining systems vs $\sqrt{P_{\rm out}}$ collapses all systems, generators, and mask widths onto the relationship, see Fig.~\ref{fig:3d_uncompression}(b).
This result confirms that the compression error is set by the discarded spectral power across the tested regimes of wave transport.
The depolarized power fraction, the share of the power in the two components orthogonal to the incident polarization, is substantial in these diffusive media, rising to $0.60$ at $\gamma = 10.4$.
For the $\gamma = 4.11$ system at $\alpha = 12$, $P_{\rm out} = 0.052$ and $\epsilon_{L2} = 0.21$ ($\epsilon_\varphi = 0.021$), at the compression ratio $C_{\rm OSCAR} \simeq 54$ ($52$ measured) for $k\ls \approx 300$ and $n_{\Delta x} = 3$.
The same mask in the $k\ls \approx 3000$ system yields $C_{\rm OSCAR}$ in the hundreds, see Sec.~\ref{sec:benchmarks}.

Reconstruction uses the stored bare projection with the disorder-averaged constant $c_2 = (1 - \overline{P}_{\rm out})^{-1/2}$ of Appendix~\ref{app:metrics} where an observable is power-sensitive.
Nothing is stored per realization, and the per-realization retained power varies by well under $1\%$ across every 3D ensemble, see Supplemental Material Sec.~\ref{sec:sm_metricverify}.

Depth resolution of the error confirms that the interior error is depth-independent.
Normalized by $\sqrt{P_{\rm out}}$, all systems collapse onto unity across the slab, with Gibbs layers confined to the faces, see Fig.~\ref{fig:3d_uncompression}(c).
The boundary layers have the same $1/\dkm$ depth as in the 2D analysis of Fig.~\ref{fig:boundary_correction_2d}(b), with the trim law quantified in Supplemental Material Sec.~\ref{sec:sm_metriczones}.
The mask sequence of the deep slab shows the same structure at every $\alpha$, with no accumulation of the error over its $21\ls$ thickness, see Fig.~\ref{fig:3d_uncompression}(d).

\subsection{Second-order quantities\label{sec:3d_secondorder}}
\noindent Here we compute the \emph{transmission-sensitivity} kernel $S(\vec{r}_0)$, the analog of Eq.~\eqref{eq:sensitivity},
\begin{equation}
S(\vec r) = -\mathrm{Re}\bigl[E_{\rm adj}(\vec r)\, E_{\rm fwd}(\vec r)\bigr],
\label{eq:sm4_kernel}
\end{equation}
where constant prefactors such as $k_0^2\,\delta\epsilon\,\delta V$ are dropped for convenience and the minus sign is retained to keep $S$ positive.
$E_{\rm fwd}$ is the field injected from the left via a plane wave.
The transmitted wavefront recorded at the exit plane of the forward solve is conjugated and injected as the adjoint source to obtain $E_{\rm adj}$.
In this arrangement we consider a 3D slab with open boundary conditions longitudinally and periodic boundary conditions in the transverse directions. 
The kernel is constructed from a forward (excitation at the front face of the slab) and adjoint (excitation at the back face of the slab) pair of fields of the $\gamma = 4.11$ system, cf.\ Supplemental Material Sec.~\ref{sec:sm_cg_kernel}.
Both fields are compressed independently with the isotropic mask, the kernel is formed from the compressed data via the convolution of Sec.~\ref{sec:secondorder}, and the result is averaged over coarse-graining cells of side $L_c$.
The corresponding error is computed on the coarse-cell-averaged grid,
\begin{equation}
\epsilon_S(L_c) = \left[\frac{\sum_{\vec r \in V}\bigl|\langle S_c\rangle_{L_c} - \langle S\rangle_{L_c}\bigr|^2}{\sum_{\vec r \in V}\bigl|\langle S\rangle_{L_c}\bigr|^2}\right]^{1/2},
\label{eq:sm5_epss}
\end{equation}
where $\langle\,\cdot\,\rangle_{L_c}$ is the block mean over cells of side $L_c$, $S_c$ is the compressed kernel, and $V$ is the interior of the system.

\begin{figure}[t]
\centering
\includegraphics[width=\columnwidth]{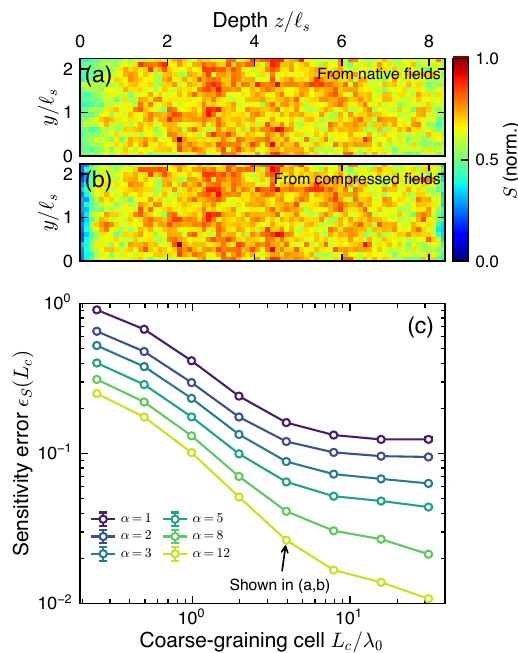}
\caption{Second-order observables from compressed 3D fields.
(a)~The transmission-sensitivity kernel of the $\gamma = 4.11$ slab (transverse extent $67\um = 2.2\ls$), constructed from a forward/adjoint pair of fields, cf.\ Supplemental Material Sec.~\ref{sec:sm_cg_kernel}, coarse-grained onto cells of side $L_c = 3.4\um \approx 4\lambda_0$, and computed from the native fields.
Axes $y, z$ are in units of $\ls$, and the color scale is normalized to the native-map maximum.
(b)~The same kernel from the fields compressed at $\alpha = 12$ (isotropic mask), on the same color scale.
Visible discrepancies are confined to the Gibbs boundary layers at the slab faces.
(c)~The sensitivity error vs coarse-graining cell size for the isotropic mask sequence, same system and geometry.
Symbols are means over the ten forward/adjoint pairs of the ensemble, with error bars, the realization spread of at most $3\%$ of the mean, covered by the symbols.
The arrow marks the cell size of the maps in (a,b).
\label{fig:second_order_3d}}
\end{figure}

Figure~\ref{fig:second_order_3d}(a,b) compares the cell-averaged kernels from the native and the compressed fields at $\alpha = 12$ and $L_c \approx 4\lambda_0$.
The two maps are visually indistinguishable, with the exception of the boundary layers affected by the Gibbs effect, and the compressed-map maximum is within $1\%$ of the native one.
The wavelength-scale features of the kernel are reproduced faithfully as well, see the wavelength-scale ($L_c = \lambda_0$) maps of the quasi-ballistic and the deep-slab systems in Supplemental Material Fig.~\ref{fig:sm4_maps3d} and Sec.~\ref{sec:sm_cg_kernel}.
Panel (c) quantifies the convergence with the cell size.
At wavelength-scale cells the sensitivity error $\epsilon_S$ reflects the field-level error, and cell averaging suppresses the cross terms as derived in Appendix~\ref{sec:sm_metricsecond}, so that by $L_c \approx 4\lambda_0$ the interior error falls to $\epsilon_S = 0.026$ at $\alpha = 12$, and to $\epsilon_S \approx 0.01$ at the coarsest cells.
Across the crossover from the quasi-ballistic to the diffusive regime of wave transport, the gain of coarse graining grows with the sample thickness at fixed cell size, exceeding fiftyfold in the most diffusive systems of both dimensionalities, see Supplemental Material Fig.~\ref{fig:sm5_sens2d}(e,f).
Coarse graining is effective when the mask covers the shell, the box resolves the mask, and the cell exceeds the error correlation length $1/\dkm$, cf.\ Supplemental Material Sec.~\ref{sec:sm_cg}.
Source-type independence of this result is verified in Supplemental Material Sec.~\ref{sec:sm_metricverify}, where all $25$ input--output kernel pairs agree within realization spreads. The equivalence of the compressed-space convolution with decompression followed by multiplication is verified in Sec.~\ref{sec:sm_specsupport}.
Additionally, Supplemental Material Sec.~\ref{sec:sm_cg} compares the transmission sensitivity in 2D and 3D systems, confirming the generality of our conclusions.

\begin{figure*}[t]
\centering
\includegraphics{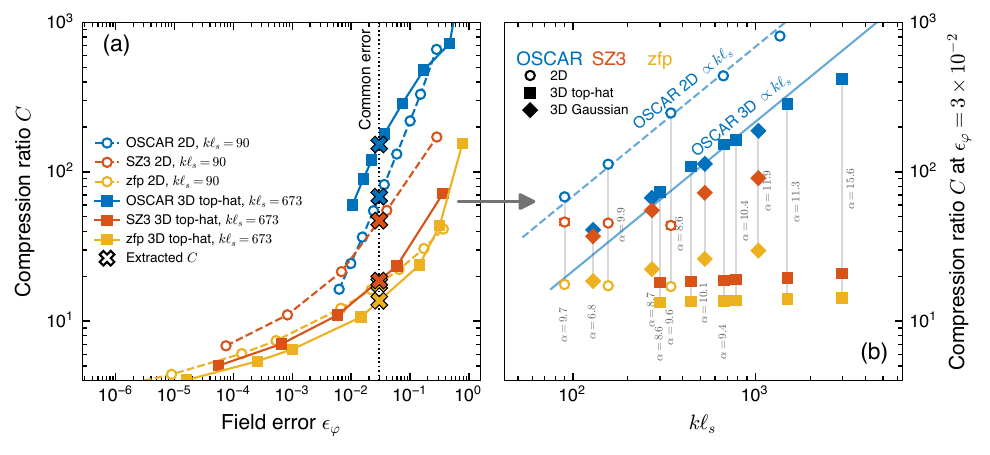}
\caption{The field-reconstruction benchmark.
In this figure, 2D data are shown with open symbols and dashed lines and 3D data with filled symbols and solid lines, while methods are identified by color.
Panel (a) shows examples (one 2D and one 3D medium) of the procedure used to obtain an equivalent compression.
$C$ is measured vs its own field error $\epsilon_\varphi$ using each compression scheme, one point per setting of its parameter sequence.
The dotted vertical line is the common error $\epsilon_\varphi = 3\times10^{-2}$, and the crosses mark each method's compression at that error.
This specific value of the error is chosen for the construction of (b).
Panel (b) depicts the extracted matched-error compression ratios, cf.\ Supplemental Material Tables~\ref{tab:sm3_2d} and \ref{tab:sm3_3d}. Each point is the mean over the ten realizations of its system, every scheme applied to the co-polarized component of the forward field. Error bars are the realization spreads, below $1\%$ of the mean for OSCAR and $3\%$ for the alternative schemes in 3D and at most $2\%$ and $6\%$ in 2D, and are covered by the symbols.
The OSCAR mask width $\alpha$ at the matched error is printed at each connector, the two highest 2D levels having no alternative-scheme point and no connector.
For a given medium with its specific value of $k\ls$, the methods are stacked on a vertical connector, so methods are compared symbol by symbol in a fixed system.
The scaling $C\propto k\ls$ predicted by Eq.~(\ref{eq:compression}) is shown by the guides for 2D (dashed line, cf.\ blue open circles) and 3D (solid line, cf.\ filled symbols).
The on-shell data points follow the scaling within each family without lying exactly on the fitted guides, because the extraction uses the measured error-vs-shell-width relations, with the matched-error $\alpha$ rising with $k\ls$ as the box approaches the aperture limit of Sec.~\ref{sec:onshell}.
For the alternative schemes, compression ratios are nearly flat in $k\ls$, so OSCAR becomes advantageous for $k\ls \gtrsim 100$ even in the Gaussian-correlated family.}
\label{fig:sm6_headline}
\end{figure*}

\section{Benchmarks against generic compression and measured costs}
\label{sec:benchmarks}\label{sec:sm_bench}
{

In this section we benchmark OSCAR against generic lossy compression schemes, including a low-rank baseline, and report the measured storage and runtime costs of the complete store-and-read procedure.
To quantify the comparison, we use simulations of the media characterized in Supplemental Material Sec.~\ref{sec:sm_media} and the error metrics of Sec.~\ref{sec:2d_field}. The exact relationships for the latter are derived in Appendix~\ref{app:metrics}.

\subsection{Benchmark procedure}\label{sec:sm_benchproto}

\begin{figure}[t]
\centering
\includegraphics[width=\columnwidth]{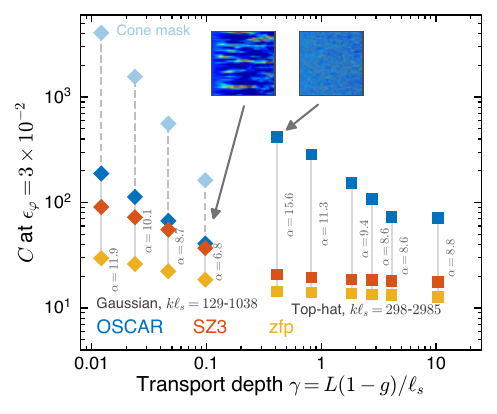}
\caption{Matched-error compression vs the transport depth $\gamma = L(1-g)/\ls$, for the 3D systems of Supplemental Material Table~\ref{tab:sm3_3d} at the common field error $\epsilon_\varphi = 3\times10^{-2}$.
Unlike $k\ls$ in Fig.~\ref{fig:sm6_headline}(b), $\gamma$ discriminates the transport regime, cf.\ the intensity snippets of the $\gamma = 0.098$ and $\gamma = 0.41$ fields, with arrows to the corresponding symbols.
The alternative schemes follow a single $\gamma$ trend, efficient mainly for underdeveloped speckle, while the OSCAR compression, diamonds for the Gaussian-correlated sweep and squares for the top-hat systems, is determined by each medium's $k\ls$ through Eq.~(\ref{eq:compression}) and does not depend on $\gamma$.
The mask width $\alpha$ at matched error is shown at each connector.
Pale diamonds are OSCAR with the anisotropic cone mask, cf.\ Sec.~\ref{sec:sm_benchres}, connected to the isotropic points by broken lines. Every symbol is the mean over the ten realizations of its system, with error bars, below $3\%$ of the mean, covered by the symbols.}
\label{fig:sm6_gamma}
\end{figure}

SZ3 and zfp are the two standard error-bounded compressors for scientific floating-point arrays, developed specifically for simulation data in high-performance computing, and truncated SVD is the simplest low-rank baseline.
These schemes guarantee a pointwise error bound, which is required by the matched-error protocol described below.

SZ3~\cite{sm_sz3, sm_sz3b} exploits the local predictability of the array, storing rounded prediction residuals under a user-set pointwise tolerance.
zfp~\cite{sm_zfp} exploits block smoothness, transforming $4^d$ blocks and truncating the trailing bits of the coefficients.
The truncated SVD~\cite{sm_svd} exploits rank decay, storing the leading singular vectors of the field treated as a matrix, and tests whether multiply scattered fields retain the low-rank structure that seismic applications commonly exploit.
In 3D, each field component is reshaped to a matrix with the transverse plane as rows and the depth as columns before the decomposition.

SZ3 (v3.3.2) and zfp (v1.0.1) are compiled from source and applied to the real and imaginary parts of each field as two separate float32 arrays at a common absolute tolerance, with the two stream sizes added.
The SVD is computed in double precision with the standard LAPACK routines~\cite{sm_lapack}.

The OSCAR compression ratio has the closed-form expression of Eq.~\eqref{eq:compression}.
The stored set is a shell of radius $k$ and full width $\dkm = \alpha/\ls$, and the ratio is independent of the numerical precision and of the system size, for boxes large enough to resolve the shell, cf.\ Supplemental Material Sec.~\ref{sec:sm_cg}.
We confirm it directly in Sec.~\ref{sec:sm_benchcost} below.
The other compression algorithms do not have explicit expressions for their compression ratios, so we obtain them empirically on the same data arrays, as a ratio between the compressed and the raw complex64 data sets.

The accuracy parameter is the mask width $\alpha$ for OSCAR and the error tolerance for SZ3 and zfp.
To compare the schemes under the same conditions, we read their compression ratios off the curves of Fig.~\ref{fig:sm6_headline}(a) either at the same error or at the same storage, interpolating between the computed points when necessary.
The cost table of Sec.~\ref{sec:sm_benchcost} instead lists each scheme at its own native settings.

Second-order (sensitivity) tests use forward/adjoint pairs of fields, with both fields compressed by the same method at the same setting.
All central processing unit (CPU) benchmarks are obtained on two-socket Intel Xeon Skylake nodes of the Mill cluster~\cite{2024_Cluster}, with eight threads, and the 3D solves use one NVIDIA V100 GPU, while the 2D direct solves run on CPU nodes of the same cluster.

\subsection{Results at matched error: wave fields}\label{sec:sm_benchres}

\begin{figure*}[t]
\centering
\includegraphics{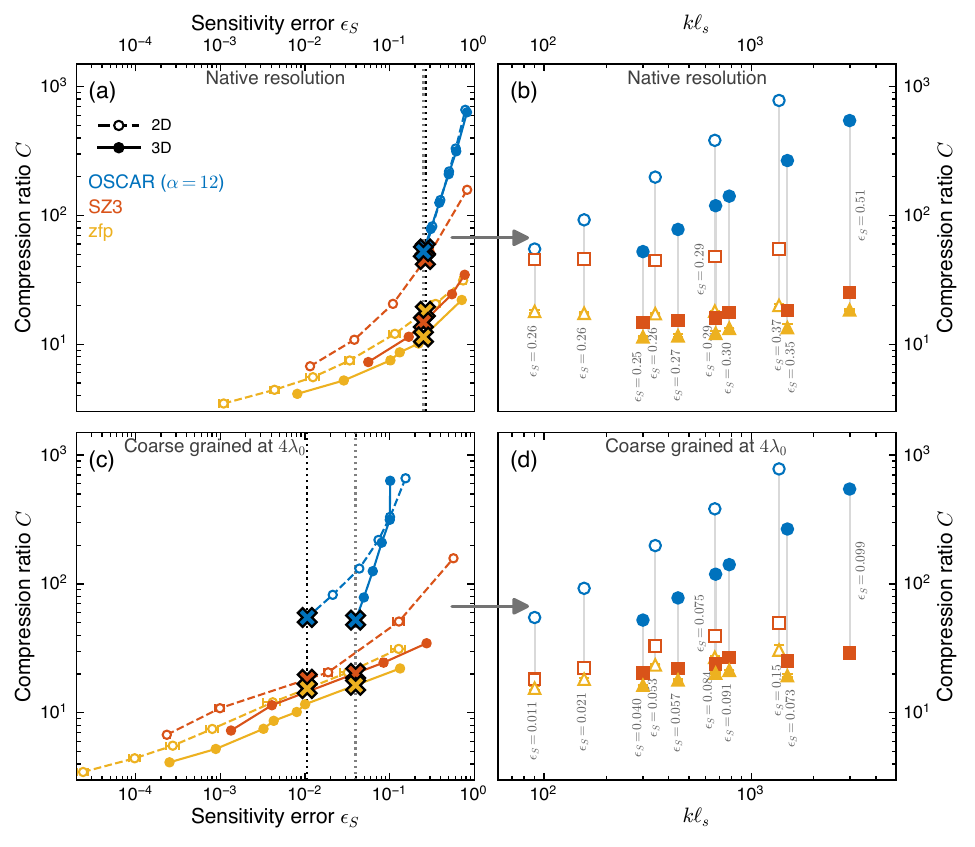}
\caption{Matched-error benchmark of the sensitivity kernel.
The top row compares kernel calculations at the native sampling and the bottom row after coarse graining at $L_c = 4\lambda_0$, as indicated by the panel headings.
(a, c) The construction, shown for the 2D Level~1 medium of Supplemental Material Table~\ref{tab:sm3_2d}, open symbols and dashed lines, $k\ls \approx 90$, ten pairs of fields, one per realization. The 3D data are the $\gamma = 4.11$ system of Supplemental Material Table~\ref{tab:sm3_3d}, filled symbols and solid lines, $k\ls = 300$, ten pairs.
Compression ratio $C$ is computed in each scheme vs its kernel error $\epsilon_S$, one point per setting, with saturated cases ($\epsilon_S \ge 0.9$) excluded.
The dotted lines mark the matched error of each system, the $\epsilon_S$ of the $\alpha = 12$ mask in OSCAR, and the crosses mark each scheme's compression at that error.
Error bars are sample-to-sample standard deviations over the ten pairs of fields of each system, below $7\%$ of the mean and covered by the symbols, except at $k\ls = 1362$ after coarse graining, where they reach $7.5\%$ for SZ3 and $9\%$ for zfp.
(b, d) The matched-error compressions for all systems, the five 2D dilution levels of Supplemental Material Table~\ref{tab:sm3_2d} and the six 3D systems of Supplemental Material Table~\ref{tab:sm3_3d}, with the matched $\epsilon_S$ printed at each gray connector.
The arrows indicate that the crossings of (a,c) supply the points for (b,d), where circles are OSCAR, squares SZ3, and triangles zfp.
A missing symbol means that the scheme does not reach the matched error at any setting.}
\label{fig:sm6_sens}
\end{figure*}

Figure~\ref{fig:sm6_headline}(a) describes our procedure for obtaining equivalent compression ratios by matched error in recovering individual wave fields.
For each simulated system we measure each scheme's compression ratio $C$ and plot it vs its field error $\epsilon_\varphi$, one point per setting.
The crossings with the common error $\epsilon_\varphi = 3\times10^{-2}$ (vertical dotted line) give the matched-error ratios for each scheme.
This specific error is a representative few-percent target, fixed once for the analysis in this section.
From the relationships of Appendix~\ref{app:metrics}, this error corresponds to $\epsilon_{L2} \approx 0.24$.

Figure~\ref{fig:sm6_headline}(b) compiles the matched-error compression ratios.
The 2D data are the five dilution levels of Supplemental Material Table~\ref{tab:sm3_2d}, $k\ls = 90$ to $1362$, with ten realizations each. The alternative schemes are benchmarked at the three lowest levels only. OSCAR reaches the matched error at $\alpha = 10.4$ and $11.6$ for $k\ls = 667$ and $1362$, giving $C = 440$ and $812$.
The 3D data are the ten systems of Supplemental Material Table~\ref{tab:sm3_3d} marked for this figure, with ten realizations each, and every scheme is applied to the co-polarized component of the forward field.
The OSCAR ratio grows in proportion to $k\ls$, cf.\ Eq.~(\ref{eq:compression}), while the ratios for the alternative compression schemes are essentially flat.
The rising trend (OSCAR) and near-flat trend (other schemes) cross below $k\ls \approx 100$ for the Gaussian-correlated family.
Below the crossing, SZ3 is competitive on the field metric.
Above it, OSCAR leads, by up to a factor of $20$ over SZ3 on the 3D top-hat media.
The prefactor of the OSCAR trend differs slightly between disorder families.
The SZ3 ratio drops by a factor of three to five from Gaussian-correlated to top-hat media at fixed $k\ls$.
At fixed sample thickness the two parameters are not independent, since $\gamma \propto L/\ls$, so the largest $k\ls$ of this study belongs to the least developed speckle.
zfp does not reach competitive ratios in either dimensionality in the studied regimes.
In the same range, truncated SVD is also not competitive, because fully developed speckle has a nearly flat singular-value spectrum.
One realization of the 2D remission system ($k\ls \approx 90$) requires rank $611$ of $3226$ for $99\%$ of the energy, and at the error of the $\alpha = 12$ mask the truncated SVD reaches only $C = 4.4$.

Figure~\ref{fig:sm6_headline}(b) also shows that the 3D Gaussian media, depicted as diamonds, are amenable to efficient compression by SZ3.
SZ3 reaches OSCAR's compression ratio, particularly at low values of $k\ls$.
The reason is that $k\ls$ on the horizontal axis of Fig.~\ref{fig:sm6_headline}(b) does not discriminate the transport regime.
The Gaussian-correlated model scatters very anisotropically, $g \ge 0.98$, so its fields remain quasi-ballistic even at large $k\ls$.
The transport depth $\gamma = L(1-g)/\ls$ discriminates better.
Speckle at $\gamma < 1$ can be interpreted as underdeveloped and becomes fully developed once $\gamma$ exceeds a few.
Figure~\ref{fig:sm6_gamma} replots the matched-error compressions of the 3D systems vs $\gamma$.
The alternative schemes now follow a single trend that does not distinguish the two disorder families.
They are efficient in the regime of underdeveloped speckle, up to $C = 91$ at $\gamma = 0.012$, and become less efficient as speckle develops, down to $C = 18$ in the diffusive regime.
The two intensity snippets illustrate the transition, from the forward-streaked pattern at $\gamma = 0.098$ to the speckle pattern at $\gamma = 0.41$.
OSCAR itself can be made more efficient in the quasi-ballistic regime.
We take advantage of the predominantly forward propagation (visible in the left snippet) by making the mask anisotropic, i.e., the intersection of the shell with the forward cone. For such a construction, the mask retains the radial width $\dkm = \alpha/\ls$ but keeps only the directions with $\mu \ge \mu_{\rm min}$, where $\mu = k_z'/|\vec k'|$ and $\mu_{\rm min}$ is chosen for each medium so that the mask contains $99.5\%$ of the scattered power, the ballistic component excluded.
Implementing this cone mask greatly improves the compression, shown by the pale diamonds.
The ratio exceeds $4000$ at $\gamma = 0.012$, a factor of $22$ above the isotropic shell, with the gain shrinking as the support isotropizes, cf.\ Supplemental Material Sec.~\ref{sec:sm_mask}.
An underdeveloped field is thus compressible by generic means as well, but OSCAR with the matched cone mask retains a $45$-fold advantage there, while fully developed speckle is compressed far more efficiently through its on-shell structure.

\subsection{Results at matched error: second-order quantities}\label{sec:sm_benchres2}

\begin{figure*}[t]
\centering
\includegraphics{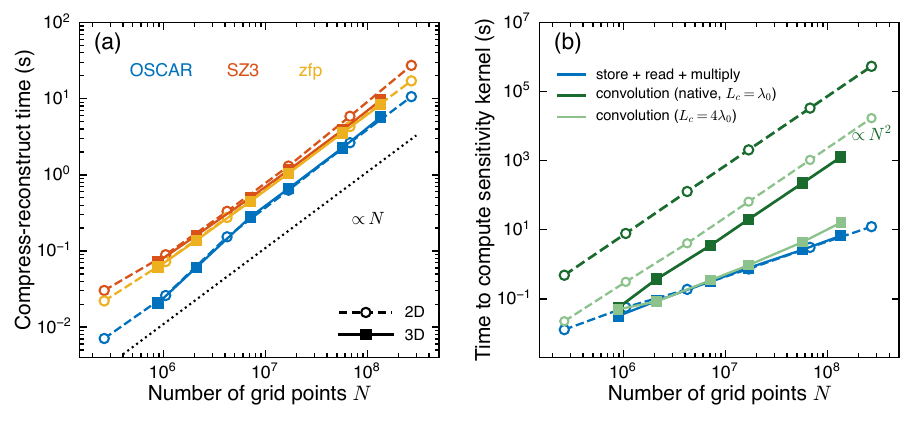}
\caption{Computational cost of the store-and-read procedure.
(a) The measured wall time to compress and to recover the two fields, vs the number of grid points $N$, for synthetic on-shell fields at $k\ls = 100$ in 2D and $300$ in 3D, $\alpha = 12$, the compress time including the construction of the mask from the header. SZ3 and zfp run at a fixed absolute error bound of $0.1$ of the peak field amplitude, on the nodes and with the thread count of Table~\ref{tab:sm4_audit}.
In this figure, 2D data are depicted with open symbols and dashed lines and 3D data with filled symbols and solid lines.
All schemes follow the linear scaling shown by the dotted line.
The final multiplication into $S(\vec r)$, common to all schemes, is not included.
(b) The compressed-space convolution route to the sensitivity kernel, unique to OSCAR, for the same fields.
The OSCAR curves of (a), with the final multiplication into the kernel added, are repeated for reference, and the label $\propto N^2$ marks the quadratic scaling.
Two evaluation modes are measured and the one with the lower cost is shown, dark green for the pair enumeration, independent of the output resolution and therefore serving both the native output and $L_c = \lambda_0$, and pale green for the per-output summation at $L_c = 4\lambda_0$, see text.}
\label{fig:sm6_cost}
\end{figure*}

We repeat the matched-error comparison using the transmission sensitivity kernel, see Fig.~\ref{fig:sm6_sens}.
The kernel is computed via Eq.~(\ref{eq:sm4_kernel}), cf.\ Sec.~\ref{sec:3d_secondorder} and Supplemental Material Sec.~\ref{sec:sm_cg_kernel}, from a forward/adjoint pair of fields, with both compressed and decompressed by the same scheme at the same setting, and its error $\epsilon_S$ is computed via Eq.~(\ref{eq:sm5_epss}).
The top row of Fig.~\ref{fig:sm6_sens} compares the kernel at the native sampling and the bottom row after coarse graining at $L_c = 4\lambda_0$.
The 2D systems are the five dilution levels of Supplemental Material Table~\ref{tab:sm3_2d}, with ten pairs of fields each.
The 3D systems are six members of Supplemental Material Table~\ref{tab:sm3_3d} spanning $k\ls = 300$--$2985$, with ten pairs of fields each.

Figure~\ref{fig:sm6_sens}(a,c) shows the construction for one 2D and one 3D system, similar to that in Fig.~\ref{fig:sm6_headline}(a).
Each scheme traces its compression ratio vs $\epsilon_S$, one point per setting.
Cases where error saturates, $\epsilon_S \ge 0.9$, produce a reconstruction indistinguishable from noise and are therefore excluded.
The matched error of each system is the $\epsilon_S$ of its $\alpha = 12$ mask, marked by the dotted lines in (a,c) and printed at each connector in (b,d).
The crossings of the curves with this error give the matched-error compressions collected in Figs.~\ref{fig:sm6_sens}(b,d).
A missing symbol in (b,d) means that the scheme does not reach the matched error at any meaningful setting.

By construction, the OSCAR points in (b,d) are the $\alpha = 12$ compressions and follow the $C \propto k\ls$ scaling of Eq.~(\ref{eq:compression}).
At the native sampling, SZ3 stays within a factor of two of OSCAR at the lowest 2D $k\ls$ and falls behind with increasing $k\ls$ in both dimensionalities, reaching $C = 15$--$25$ in 3D as OSCAR values grow.
After coarse graining, the OSCAR points do not change (both $C$ and $k\ls$ are independent of the error), again by construction shown in Fig.~\ref{fig:sm6_sens}(c), whereas the corresponding $\epsilon_S$ does decrease, cf.\ Fig.~\ref{fig:sm6_sens}(b,d).
Meanwhile, the SZ3 compression in 2D drops from $46$ to $18$ at $k\ls = 90$, while in 3D it stays below $C \approx 30$ for both SZ3 and zfp.
Coarse graining therefore reduces the OSCAR error by a factor of $2.4$ to $25$ at unchanged compression, because its error is off-shell and averages away, cf.\ Supplemental Material Sec.~\ref{sec:sm_cg}, while the residuals of the other schemes retain power at low spatial frequencies and benefit far less.
zfp does not reach the matched error after coarse graining at $k\ls = 2985$ in 3D.

\subsection{Measured storage and runtime costs}\label{sec:sm_benchcost}

\begin{table*}[t]
{
\footnotesize
\centering
\setlength{\tabcolsep}{3pt}
\caption{Measured storage, error, and wall time of the complete store-and-read procedure, for systems spanning quasi-ballistic to diffusive transport and for both reconstructed quantities (field and sensitivity), with the system parameters and the matched error printed in each block heading, cf.\ Supplemental Material Tables~\ref{tab:sm3_2d} and \ref{tab:sm3_3d}.
In the matched-error columns, every scheme is read at the common error of the block, $\epsilon_\varphi$ for the fields, cf.\ Sec.~\ref{sec:sm_benchres}, and the $\alpha = 12$ kernel error at the native sampling for the sensitivities, cf.\ Sec.~\ref{sec:sm_benchres2}, and the `Storage' column lists the space each scheme needs at that error.
The 3D sensitivity blocks are read on the pairs of the Time column, so their matched errors can differ from the ten-pair means of Fig.~\ref{fig:sm6_sens} in the second decimal.
All errors are whole-domain values, computed without the boundary trim.
The OSCAR mask width $\alpha$ of each block is listed in its heading.
In the matched-storage columns, every scheme is held at the OSCAR storage of the block and the $\epsilon$ column lists the error it yields.
The Time column lists the round-trip time, store plus read of the pair of fields, measured at the setting of its matching on the block's own fields. Times are measured with eight threads on the Skylake nodes of Sec.~\ref{sec:sm_benchproto}, as the mean over the timed pairs, ten in 2D, two for the deep 3D slab, and the one complete pair of the quasi-ballistic 3D system.
Raw is the uncompressed reference, disk read only.
The two 2D fields occupy $16N$ bytes and the two 3D fields, with three vector components each, $48N$ bytes.
Truncated SVD stores each field as its leading rank-one factors, cf.\ Sec.~\ref{sec:sm_benchproto}, with the rank as its setting, and its time is the factorization plus the reconstruction at the rank of the matching.
A dash marks a matched value the scheme cannot reach at any setting, together with the time of that row, or, in the matched-storage columns, an output that retains no overlap with the original.\label{tab:sm4_audit}}
\begin{tabular}{l|lll|lll||lll|lll}
\hline\hline
 & \multicolumn{6}{c||}{2D} & \multicolumn{6}{c}{3D} \\
\cline{2-13}
 & \multicolumn{3}{c|}{Matched error} & \multicolumn{3}{c||}{Matched storage} & \multicolumn{3}{c|}{Matched error} & \multicolumn{3}{c}{Matched storage} \\
Method & Storage & $\epsilon$ & Time & Storage & $\epsilon$ & Time & Storage & $\epsilon$ & Time & Storage & $\epsilon$ & Time \\
\hline\hline
\multicolumn{7}{c||}{field, transition to diffusion} & \multicolumn{6}{c}{field, quasi-ballistic} \\
\multicolumn{7}{c||}{($\gamma = 2.3$, $k\ls = 90$, $\epsilon_\varphi = 0.03$, $\alpha = 10.4$)} & \multicolumn{6}{c}{($\gamma = 0.41$, $k\ls = 2985$, $\epsilon_\varphi = 0.03$, $\alpha = 15.8$)} \\
\hline
Raw & 5.0~MB & $0.00$ & 2~ms & 5.0~MB & $0.00$ & 2~ms & 5.9~GB & $0.00$ & 1.8~s & 5.9~GB & $0.00$ & 1.8~s \\
OSCAR & 79~kB & $0.03$ & 20~ms & 79~kB & $0.03$ & 20~ms & 14~MB & $0.03$ & 16.0~s & 14~MB & $0.03$ & 16.0~s \\
SZ3 & 133~kB & $0.03$ & 85~ms & 79~kB & $0.08$ & 83~ms & 262~MB & $0.03$ & 58.9~s & 14~MB & $0.55$ & 45.7~s \\
zfp & 301~kB & $0.03$ & 52~ms & 79~kB & $0.62$ & 36~ms & 353~MB & $0.03$ & 35.5~s & --- & --- & --- \\
SVD & 1.1~MB & $0.03$ & 136~ms & 79~kB & $0.45$ & 135~ms & 2.0~GB & $0.03$ & 153~s & 14~MB & $0.24$ & 151~s \\
\hline\hline
\multicolumn{7}{c||}{field, diffusive} & \multicolumn{6}{c}{field, diffusive} \\
\multicolumn{7}{c||}{($\gamma = 18.6$, $k\ls = 90$, $\epsilon_\varphi = 0.03$, $\alpha = 9.7$)} & \multicolumn{6}{c}{($\gamma = 10.4$, $k\ls = 298$, $\epsilon_\varphi = 0.03$, $\alpha = 8.8$)} \\
\hline
Raw & 40~MB & $0.00$ & 10~ms & 40~MB & $0.00$ & 10~ms & 2.4~GB & $0.00$ & 0.7~s & 2.4~GB & $0.00$ & 0.7~s \\
OSCAR & 583~kB & $0.03$ & 170~ms & 583~kB & $0.03$ & 170~ms & 33~MB & $0.03$ & 6.1~s & 33~MB & $0.03$ & 6.1~s \\
SZ3 & 860~kB & $0.03$ & 436~ms & 583~kB & $0.06$ & 417~ms & 148~MB & $0.03$ & 24.7~s & 33~MB & $0.41$ & 19.9~s \\
zfp & 2.3~MB & $0.03$ & 252~ms & 583~kB & $0.73$ & 141~ms & 197~MB & $0.03$ & 16.0~s & 33~MB & $0.52$ & 9.9~s \\
SVD & 9.6~MB & $0.03$ & 2.5~s & 583~kB & $0.63$ & 2.5~s & --- & --- & --- & 33~MB & $0.80$ & 128~s \\
\hline\hline
\multicolumn{7}{c||}{sensitivity, transition to diffusion} & \multicolumn{6}{c}{sensitivity, quasi-ballistic} \\
\multicolumn{7}{c||}{($\gamma = 2.42$, $k\ls = 667$, $\epsilon_S = 0.29$, $\alpha = 12$)} & \multicolumn{6}{c}{($\gamma = 0.41$, $k\ls = 2985$, $\epsilon_S = 0.52$, $\alpha = 12$)} \\
\hline
Raw & 40~MB & $0.00$ & 10~ms & 40~MB & $0.00$ & 10~ms & 5.9~GB & $0.00$ & 1.8~s & 5.9~GB & $0.00$ & 1.8~s \\
OSCAR & 104~kB & $0.29$ & 173~ms & 104~kB & $0.29$ & 173~ms & 11~MB & $0.52$ & 15.9~s & 11~MB & $0.52$ & 15.9~s \\
SZ3 & 825~kB & $0.29$ & 436~ms & --- & --- & --- & 234~MB & $0.52$ & 57.9~s & --- & --- & --- \\
zfp & 2.2~MB & $0.29$ & 240~ms & --- & --- & --- & 322~MB & $0.52$ & 34.6~s & --- & --- & --- \\
SVD & 7.5~MB & $0.29$ & 2.5~s & 104~kB & $0.9$ & 2.5~s & 1.2~GB & $0.52$ & 152~s & 11~MB & $0.87$ & 151~s \\
\hline\hline
\multicolumn{7}{c||}{sensitivity, diffusive} & \multicolumn{6}{c}{sensitivity, diffusive} \\
\multicolumn{7}{c||}{($\gamma = 18.6$, $k\ls = 90$, $\epsilon_S = 0.26$, $\alpha = 12$)} & \multicolumn{6}{c}{($\gamma = 10.4$, $k\ls = 298$, $\epsilon_S = 0.22$, $\alpha = 12$)} \\
\hline
Raw & 40~MB & $0.00$ & 10~ms & 40~MB & $0.00$ & 10~ms & 2.4~GB & $0.00$ & 0.7~s & 2.4~GB & $0.00$ & 0.7~s \\
OSCAR & 723~kB & $0.26$ & 169~ms & 723~kB & $0.26$ & 169~ms & 45~MB & $0.22$ & 6.1~s & 45~MB & $0.22$ & 6.1~s \\
SZ3 & 872~kB & $0.26$ & 437~ms & 723~kB & $0.32$ & 425~ms & 166~MB & $0.22$ & 25.2~s & 45~MB & $0.77$ & 20.6~s \\
zfp & 2.2~MB & $0.26$ & 252~ms & --- & --- & --- & 213~MB & $0.22$ & 16.6~s & 45~MB & $0.86$ & 10.8~s \\
SVD & 9.9~MB & $0.26$ & 2.5~s & 723~kB & $0.95$ & 2.5~s & --- & --- & --- & 45~MB & $0.98$ & 128~s \\
\hline\hline
\end{tabular}
}
\end{table*}

All numbers are measured, on the hardware of Sec.~\ref{sec:sm_benchproto}, and collected in Table~\ref{tab:sm4_audit} and Fig.~\ref{fig:sm6_cost}.

The storage for OSCAR and any other method is set by the compression ratio alone.
Each grid point of each field carries $8$ bytes as complex64, so the two fields, forward and adjoint, occupy $16N$ bytes raw for $N$ grid points in 2D and, with three vector components each, $48N$ bytes in 3D, and a factor of $C$ less when compressed.
At matched sensitivity error, $C$ is read from the benchmark of Sec.~\ref{sec:sm_benchres2}, and Eq.~(\ref{eq:compression}) predicts it analytically.
The measured file sizes confirm Eq.~(\ref{eq:compression}) at every grid size, $85$ vs a predicted $84.9$ from $512^2$ to $16384^2$ and $338$--$344$ vs $340$ over $96^3$--$384^3$, both for synthetic fields at $k\ls = 100$ and $\alpha = 12$.
The stored object carries a ${\sim}170$-byte header of mask parameters, from which the mask is regenerated, and no indices.
Explicit int32 indices would lower $C \approx 55$ to $44$ in 2D and $52$ to $48$ in 3D, and a full bitmask would lower them to $38$ and $46$.
Neither overhead changes with the system size, since the indices cost four bytes per retained coefficient and the bitmask one bit per grid point, either stored once per pair of fields and shared by the two fields, and in 3D by their three components.

Figure~\ref{fig:sm6_cost}(a) shows the computational overhead, the wall time to compress and to recover the two fields, vs the number of grid points.
The timing runs use synthetic fields, so that scaling for large system sizes can be obtained.
Each synthetic field is generated directly in Fourier space, as a shell of radial width $1/\ls$ at the prescribed $k\ls$ with an independent random phase at every point, and inverse transformed to the grid.
Random on-shell phases produce fully developed speckle by construction.
The wall times depend on the grid size and on the retained fraction $1/C$ of Eq.~(\ref{eq:compression}), and only weakly on the realization, consistent with calculations based on the actual fields of Table~\ref{tab:sm4_audit}, when available.
The final multiplication into $S(\vec r)$ is common to all schemes and is omitted.
OSCAR costs $1.7$--$4.3\times$ less than SZ3 and $1.5$--$3.1\times$ less than zfp on these synthetic fields, with the margin narrowing as $N$ grows, since the transforms scale as $N\log N$ and the coders as $N$. On the stored fields of Table~\ref{tab:sm4_audit} the margins are $2.5$--$4.2\times$ and $1.4$--$2.7\times$ at matched error, because the coders run slower on the simulated speckle than on the synthetic fields at the same bound, while the OSCAR cost does not depend on the field.
All curves grow linearly in $N$ up to logarithmic corrections.
The natural reference is the solve that produced the fields.
Compressing one field with OSCAR costs $0.15$--$0.19\%$ of the 3D solve and $0.05\%$ of the 2D one, and even the slowest scheme in Fig.~\ref{fig:sm6_cost} stays below $0.7\%$ per field.

OSCAR additionally offers a route to the sensitivity that skips the decoding of the individual fields altogether.
The kernel of Eq.~(\ref{eq:sm4_kernel}) is a product of two fields in real space, so its Fourier coefficients are the convolution of the two stored shell sets, cf.\ Sec.~\ref{sec:intensity_convolution}.
When only the coarse-grained kernel is required, the convolution needs to be evaluated at the few low-$|\vec q|$ cells that survive the block average, and this requires even less computation than decoding $S$ at the native resolution and then coarse graining.
In the 2D $k\ls = 90$ system with $N = 1937\times1280 \approx 2.5\times10^6$ grid points, it costs $34$~ms for $78$ output cells vs the $66$~ms decode-and-multiply read at $C \approx 55$, and the two routes agree on the computed Fourier coefficients to better than $10^{-6}$.
Figure~\ref{fig:sm6_cost}(b) plots the measured compute cost of this route vs the grid size.
Two evaluation modes exist, and the figure shows the one with the lower cost for each output resolution.
Direct enumeration of the retained coefficient pairs costs $(N/C)^2$ operations whatever the output resolution, and is the best route at and near the native resolution.
The per-output-cell summation costs $N_{\rm out} \times N/C$ operations, where $N_{\rm out}$ is the number of output cells, and is the faster route once the output is coarse enough that $N_{\rm out} \ll N/C$.
Either way the cost is quadratic in $N$, with the compression ratio and the output resolution entering the prefactor.
In 3D at $L_c = 4\lambda_0$, convolution stays within a factor of $2.4$ of the compress-reconstruct approach over the entire studied range, while for output at or near the native resolution decoding is the faster route by a growing margin.

Table~\ref{tab:sm4_audit} compares OSCAR to the alternative compression schemes in eight blocks, at matched error and at matched storage.
At matched error, OSCAR stores the pair of fields in $1.2$--$1.7\times$ less space than SZ3 on the three 2D blocks at $k\ls = 90$, $7.9\times$ less on the 2D sensitivity block at $k\ls = 667$, and $4.5$--$19\times$ less on the 3D field blocks, e.g.\ $14$~MB vs $262$~MB for the $5.9$~GB quasi-ballistic pair of fields.
At matched storage, the SZ3 field error is $0.06$--$0.08$ in 2D vs OSCAR's $0.03$, and $0.41$--$0.55$ in 3D, while on the sensitivity blocks the alternatives at the OSCAR storage yield errors of $0.32$ or larger, or retain no overlap with the original at all (dashes).
Truncated SVD cannot reach the $3\%$ field error at any rank in the diffusive 3D system (fully developed speckle retains no low-rank structure).
The round-trip times of Table~\ref{tab:sm4_audit} favor OSCAR by $2.5$--$4.2\times$ over SZ3 and by $1.4$--$2.7\times$ over zfp at matched error, and, at both matchings, by $9.4$--$21\times$ over the SVD factorization in 3D. At matched storage the coders run at a looser setting, which encodes faster, so the zfp margin narrows to $0.8$--$1.8\times$ and zfp is the faster scheme in the 2D diffusive field block, at an error of $0.73$.

}%

\section{Conclusion\label{sec:conclusion}}
\noindent We have introduced OSCAR, a physics-based lossy compression method for wave fields in weakly scattering media. 
By exploiting the universal, sample-agnostic concentration of spectral weight on a shell of width $\sim\!1/\ell_s$ in $k$-space, OSCAR reduces storage requirements by a factor $\propto\!k\ell_s\!\times\!n_{\Delta x}^d$ per field realization, cf.~Eq.~\eqref{eq:compression}. 
These predictions are verified in 2D scalar and 3D vector simulations, from quasi-ballistic to diffusive transport, and in a variety of conditions such as disorder generators, source types, polarization, etc.

A key feature distinguishing OSCAR from generic compression methods is the sample-independence of the shell geometry. 
Because all disorder realizations share the same mask, ensemble-averaged second-order quantities can all be computed via convolution entirely in compressed $k$-space, without decompressing the field data for the individual realizations.
This is demonstrated here on sensitivity maps and extends by the same construction to intensity and correlations.
The cost is quadratic in the grid size and competitive with decompression at coarse output resolution over the sizes measured here, cf.\ Sec.~\ref{sec:sm_benchcost}. 
Notably, these second-order observables tolerate substantially higher compression (smaller $\alpha$) than individual field reconstruction.
When used at the cell-averaged resolution of transport applications, their errors fall well below the field-level $\epsilon_{L2}$, as the 3D sensitivity kernels of Fig.~\ref{fig:second_order_3d} demonstrate quantitatively.
Coherent interference between independently compressed fields is faithfully preserved.
The cell-averaged observables admit coarse output discretization that yields an additional $(n_{\Delta x}L_c/\lambda)^d$ reduction.
The error scalings of fields, products, and integrated observables that underlie these results are derived in Appendix~\ref{app:metrics}.

We would like to remark on the domain of applicability of the output coarse-graining. 
OSCAR compression of stored fields, Eq.~\eqref{eq:compression}, applies equally to all domains. The $k\ell_s$ field-storage reduction is set by the universal on-shell concentration and is unaffected by the downstream use of the compressed data. 
What differs across domains is whether the computed quantities, e.g., sensitivity, can be subsequently coarse-grained without losing physically relevant information. 
In conventional DOT, the sensitivity kernel is a product of real, positive intensity Green's functions that vary on the scale of the transport mean free path or longer, so the output can be discretized at that scale. 
Refs.~\cite{2022_Bender_Coherent_Enhancement,2025_Jara_coherent_sensing} extend this result to WFS-DOT.
In coherent-wave domains such as seismic FWI or ground-penetrating radar, the sensitivity kernel carries wavelength-scale Fresnel-zone oscillations that are essential for finite-frequency inversion and for resolving $\lambda$-scale imaging targets. The output must therefore be retained at fine-grid resolution. 
OSCAR provides the same field-storage compression in all cases, whereas the additional output coarse-graining is only warranted in the diffusive limit.

Benchmarks against general-purpose scientific compressors quantify the practical advantage.
The efficiency of these schemes is tied to underdeveloped speckle and degrades as the speckle develops, whereas the OSCAR compression is set by $k\ell_s$ alone.
Tested from quasi-ballistic to diffusive transport, OSCAR stores the fields in less space than SZ3, zfp, and truncated SVD at matched error, with a margin that grows with $k\ell_s$.
Meanwhile at the OSCAR storage, the alternatives return large errors or reconstructions that retain no overlap with the original fields, cf.\ Table~\ref{tab:sm4_audit}.
The complete store-and-read procedure is also $2.5$--$4.2\times$ faster than SZ3 and $1.4$--$2.7\times$ faster than zfp at matched error, and storing a pair of fields costs well under one percent of the solve that produced it, see Sec.~\ref{sec:benchmarks}.

The OSCAR method requires $k\ell_s\!\gg\!1$.
As disorder strength increases toward the localization regime $k\ell_s\!\sim\!1$~\cite{2009_Lagendijk_vanTiggelen_Wiersma_Anderson,2023_Yamilov_AL}, the shell broadens to progressively fill the entire $k$-space and the compression advantage vanishes. 
Within its regime of validity, OSCAR is agnostic to the specific wave equation and discretization scheme, requiring only that the medium support propagating modes with a well-defined dispersion relation and finite scattering mean free path.

The monochromatic formulation presented here extends naturally to time-domain problems. 
A moderately broadband pulse occupies a shell of finite radial extent $\sim\!\Delta\omega/c$ in $k$-space, set (broadened) by the bandwidth as well as by scattering. 
When the bandwidth satisfies $\Delta\omega/\omega\ll1$, the shell remains thin compared to $k$ and OSCAR applies at each temporal frequency independently. 
For broadband sources, the method can be combined with temporal Fourier decomposition, compressing each frequency component onto its respective shell. 

The physical principles behind OSCAR apply also to acoustic, seismic, and other kinds of waves in heterogeneous media where $k\ell_s$ is sufficiently large and where storage of 4D (3D space and frequency or time) wavefields presents an analogous bottleneck~\cite{2023_Wang_Tucker_Compression,2024_Ni_Wavefield_Reconstruction_DAS}.
In seismology, where lossy wavefield compression has been explored using generic methods~\cite{2016_Boehm_wavefield_compression,2022_Kukreja_lossy_checkpoint}, OSCAR offers the additional advantage of enabling sensitivity analysis and ensemble operations directly in compressed space.

By alleviating the storage and post-processing burden that currently limits adoption of full-wave methods, OSCAR lowers the storage barrier to full-wave simulations at scales relevant to biomedical optics, seismic imaging, underwater acoustics, and other fields.
\begin{acknowledgments}
Extensive numerical simulations in this work were carried out on the Mill high-performance computing (HPC) cluster~\cite{2024_Cluster} at Missouri University of Science and Technology.
A part of the computation for this work was performed on the high-performance computing infrastructure operated by Research Support Solutions in the Division of IT at the University of Missouri, Columbia, MO~\cite{2025_Hellbender}. The authors gratefully acknowledge Predrag Lazic for technical assistance and IT support.

The authors declare the following competing interest: A patent application related to the on-shell compression and reconstruction (OSCAR) algorithm described in this paper has been filed by Missouri University of Science and Technology.
\end{acknowledgments}
\section*{Data availability}
The data sets supporting this study are openly available in the Missouri S\&T Scholars' Mine repository~\cite{2026_Jara_OSCAR_deposit}.
The deposit contains one realization of the 2D field ensemble, the co-polarized component of one realization of the 3D vector field, and self-contained Python scripts that perform the complete OSCAR compression and reconstruction cycle on these fields, reproducing the compression ratios and errors reported here.

\appendix
\makeatletter\renewcommand{\thesubsection}{\thesection.\arabic{subsection}}\renewcommand{\p@subsection}{}\makeatother%
\section{Field-error metrics: exact identities and statistics}
\label{app:metrics}\label{sec:sm_metricdefs}
{

This appendix derives the exact and statistical relationships behind the error metrics of Sec.~\ref{sec:2d_field} and the second-order error theory used in Secs.~\ref{sec:secondorder}, \ref{sec:3d_secondorder}, and \ref{sec:benchmarks}.
Their numerical verification, together with the spatial structure of the error, is presented in Supplemental Material Sec.~\ref{sec:sm_metrics}.

OSCAR compression of a stored field $E_0$ involves steps that can be formalized as follows
\begin{equation}
E_c = c\,P E_0,
\qquad
P_{\rm out} = 1 - \frac{\|P E_0\|^2}{\|E_0\|^2},
\label{eq:sm2_chain}
\end{equation}
where $P$ projects the spatial spectrum onto the retained mask (which can be either isotropic or anisotropic), $P_{\rm out}$ is the discarded fraction of the spectral power, and $c$ is a real power-restoring factor applied at reconstruction.
The stored object is the bare projection $PE_0$.
The choice of $c$ adopted in the text is the disorder-averaged constant $c_2 = (1 - \overline{P}_{\rm out})^{-1/2}$, compared against the per-realization and analytic options in Supplemental Material Sec.~\ref{sec:sm_metricverify}.
Norms and inner products are over the analysis domain, $\|E\|^2 = \int |E|^2\,dV$ and $\langle E, F\rangle = \int E^* F\,dV$.
Every error metric below is an analytic function of the single number $P_{\rm out}$, which is the central result of this appendix.

The three field-level metrics $\epsilon_{L2}$, $\epsilon_\varphi$, and $\epsilon_A$ are defined in Eqs.~(\ref{eq:eps_L2})--(\ref{eq:eps_A}) of Sec.~\ref{sec:2d_field}.
The amplitude metric $\epsilon_A$ is a Cauchy--Schwarz test of amplitude-profile proportionality, $\epsilon_A = 0$ if and only if $|E_c| \propto |E_0|$ pointwise, but it is blind to phase, since $E_c \to E_c\,e^{i\varphi(\vec r)}$ with any spatially varying phase leaves it unchanged.
The triangle inequality $\int|E_0||E_c| \ge |\langle E_0,E_c\rangle|$ gives $\epsilon_A \le \epsilon_\varphi$ for any pair of fields, so the amplitude metric always under-reports relative to the phase-sensitive overlap.
The denominators in Eqs.~(\ref{eq:eps_L2})--(\ref{eq:eps_A}) normalize the metrics, not necessarily the reconstruction. 
Indeed, they make $\epsilon_A$ and $\epsilon_\varphi$ invariant under any scalar factor applied to $E_c$, so these two metrics are blind to the choice of $c$ in Eq.~(\ref{eq:sm2_chain}).
Only $\epsilon_{L2}$ depends on $c$, and Supplemental Material Sec.~\ref{sec:sm_metricverify} shows that this dependence is actually weak for the three explored normalization options, making all of them viable.
Table~\ref{tab:sm2_metrics} summarizes the metrics together with the relations derived below.

\begin{table*}[tp]
\centering
{
\caption{Field-level error metrics under the compression steps in Eq.~(\ref{eq:sm2_chain}).
$\mathcal{E}$ and $\mathcal{K}$ are the complete elliptic integrals and $\nu^2 = 1-P_{\rm out}$.
The exact rows require only Parseval orthogonality and hold for each individual field in any dimension, \emph{with no ensemble averaging}.
$^{*}$The $\epsilon_{L2}$ identity becomes exact under the power-restoring factor $c_1$. The $c_2$ normalization adopted in the text preserves the power at the percent level of the realization spread, see Supplemental Material Sec.~\ref{sec:sm_metricverify}.}
\label{tab:sm2_metrics}
\begin{tabular}{lcccc}
\hline
Metric & Phase-sensitive & Exact or statistical & Value & Small-$P_{\rm out}$ limit \\
\hline
$\epsilon_{L2}$ & Yes & Exact identity$^{*}$ & $\sqrt{2(1-\sqrt{1-P_{\rm out}})}$ & $\sqrt{P_{\rm out}}$ \\
$\epsilon_\varphi$ & Yes & Exact identity & $1-\sqrt{1-P_{\rm out}}$ & $P_{\rm out}/2$ \\
$\epsilon_A$ (Eq.~\eqref{eq:eps_A}) & No & Gaussian statistics & $1-\mathcal{E}(\nu)+\tfrac12 P_{\rm out} \mathcal{K}(\nu)$ & $P_{\rm out}/4$ \\
\hline
\end{tabular}
}
\end{table*}

\begin{table*}[t!]
\centering
{
\caption{Error scalings of field, product, and observable classes.
The results are asymptotically exact in the $P_{\rm out} \to 0$ limit.
$V_{\rm coh}$ is the coherence volume and $V$ the system volume.}
\label{tab:sm2_scalings}
\begin{tabular}{lcc}
\hline
Quantity & $L_2$ metric, $\epsilon_{L2}$ & Overlap metric, $\epsilon_\varphi$ \\
\hline
Field $E$ & $\sqrt{P_{\rm out}}$ & $P_{\rm out}/2$ \\
Intensity $|E|^2$ & $\sqrt{P_{\rm out}}$ & $P_{\rm out}/2$ \\
Product $E_1E_2^*$ (pointwise) & $\sqrt{2P_{\rm out}}$ & $P_{\rm out}$ \\
Integrated observables & \multicolumn{2}{c}{$0$ (deterministic) $+\ \mathcal{O}\bigl(P_{\rm out}\sqrt{V_{\rm coh}/V}\bigr)$ fluctuation} \\
\hline
\end{tabular}
}
\end{table*}

\subsection{Exact relationships}
\label{sec:sm_metricexact}

The kept and discarded parts satisfy the relationship $E_0 = PE_0 + (1-P)E_0$.
They have disjoint spectral supports and are therefore orthogonal by construction.
As a direct consequence, the discarded part drops out of the overlap between the original and compressed fields:
\begin{equation}
\langle E_0, E_c\rangle = c\,\langle E_0, PE_0\rangle = c\,\|PE_0\|^2 = \|E_0\|\,\|PE_0\|,
\label{eq:sm2_overlap}
\end{equation}
which is real and positive for any real positive $c$.
The last equality in Eq.~\eqref{eq:sm2_overlap} takes the specific choice $c = c_1 \equiv \|E_0\|/\|PE_0\| = (1 - P_{\rm out})^{-1/2}$, the power-restoring factor built from the norms of the same field, the first of the three options of Supplemental Material Sec.~\ref{sec:sm_metricverify}, and this choice is used throughout this appendix.
The compression steps therefore introduce no global phase.
Using the relationship $\|E_c\| = c_1\,\|PE_0\| = \|E_0\|$,
\begin{equation}
\epsilon_\varphi = 1-\frac{\|PE_0\|}{\|E_0\|} = 1-\sqrt{1-P_{\rm out}} \underset{P_{\rm out}\to 0}{\approx} \frac{P_{\rm out}}{2}.
\label{eq:sm2_identity}
\end{equation}
Expanding the square in $\epsilon_{L2}$ and using Eq.~(\ref{eq:sm2_overlap}),
\begin{equation}
\begin{split}
\epsilon_{L2}^2
&= \frac{\|E_c\|^2 + \|E_0\|^2 - 2\,\mathrm{Re}\,\langle E_0, E_c\rangle}{\|E_0\|^2}\\
&= 2\Bigl(1-\frac{\|PE_0\|}{\|E_0\|}\Bigr) = 2\,\epsilon_\varphi
\;\underset{P_{\rm out}\to 0}{\approx}\; P_{\rm out}.
\end{split}
\label{eq:sm2_L2}
\end{equation}
The relationships of Eqs.~(\ref{eq:sm2_identity}) and (\ref{eq:sm2_L2}) are the first two rows of Table~\ref{tab:sm2_metrics}.

\subsection{Gaussian-speckle statistics and the universal metric ratios}
\label{sec:sm_metricgauss}

Computing the error for the amplitude metric $\epsilon_A$ requires knowledge of the field statistics.
For fully developed speckle the kept and discarded parts are independent circular-Gaussian fields (disjoint spectral supports), so $(E_0, E_c)$ are jointly circular-Gaussian with field correlation $\nu = \sqrt{1-P_{\rm out}}$.
The mean product of correlated Rayleigh amplitudes is the standard bivariate-Rayleigh moment \cite{2007_Goodman_speckle}.
For unit-power fields it reads $\overline{|E_0||E_c|} = (\pi/4)\,{}_2F_1(-\tfrac12,-\tfrac12;1;\nu^2)$, and the hypergeometric identity ${}_2F_1(-\tfrac12,-\tfrac12;1;\nu^2) = (2/\pi)\bigl[2\mathcal{E}(\nu) - (1-\nu^2)\mathcal{K}(\nu)\bigr]$ turns it into the parameter-free prediction
\begin{equation}
\epsilon_A = 1 - \mathcal{E}(\nu) + \tfrac12\,P_{\rm out}\,\mathcal{K}(\nu),
\qquad
\nu^2 = 1-P_{\rm out},
\label{eq:sm2_phipred}
\end{equation}
with $\mathcal{E}$ and $\mathcal{K}$ the complete elliptic integrals.
Equation~\eqref{eq:sm2_phipred} is the result quoted in the fourth column of Table~\ref{tab:sm2_metrics}.
In the small-$P_{\rm out}$ expansion the logarithms of $E$ and $K$ cancel exactly and $\epsilon_A \to P_{\rm out}/4$, while $\epsilon_\varphi \to P_{\rm out}/2$ from Eq.~(\ref{eq:sm2_identity}).
The ratio $\epsilon_\varphi/\epsilon_A$ tends to $2$ in the Gaussian limit, so the amplitude metric under-reports the phase-sensitive error by close to a factor of two, cf.\ Supplemental Material Table~\ref{tab:sm2_verify}.
In the limit of applicability of OSCAR (weak scattering, $k\ls \gg 1$), the Gaussian approximation for the field statistics is accurate for developed speckle~\cite{2007_Goodman_speckle}.

Combining the results of this section with the asymptotic expansion of the retained-power law for the Lorentzian lineshape (Eq.~\eqref{eq:power_ratio} and Supplemental Material Sec.~\ref{sec:sm_rte}), $P_{\rm out} \approx 2/(\pi\alpha)$, all three metrics acquire closed asymptotic forms,
\begin{equation}
\epsilon_{L2} \approx \sqrt{\frac{2}{\pi\alpha}},
\qquad
\epsilon_\varphi \approx \frac{1}{\pi\alpha},
\qquad
\epsilon_A \approx \frac{1}{2\pi\alpha},
\label{eq:sm2_asympt}
\end{equation}
which establishes the empirical $1/\alpha$ law of $\epsilon_\varphi$ in Fig.~\ref{fig:boundary_correction_2d}(a).
In 3D each Cartesian component is compressed as an independent scalar field and Eq.~(\ref{eq:sm2_phipred}) applies per component, exactly for the speckle-dominated components and approximately where the coherent (ballistic) component is large.

\subsection{Products and second-order observables}
\label{sec:sm_metricsecond}

Many physical quantities are quadratic in the field and, therefore, can be statistically analyzed in a similar manner.
To formalize our description we begin by writing $E_i = PE_i + d_i$ for two independently compressed speckle fields and considering a product of the form $Q = E_1 E_2^*$.
Such a construct covers, e.g., the intensity ($E_1 = E_2$) and the transmission-sensitivity kernel $S \propto -\mathrm{Re}[E_{\rm adj} E_{\rm fwd}]$ of Eq.~\eqref{eq:sm4_kernel}.
The compressed-space error decomposes as
\begin{equation}
\Delta Q = PE_1\,d_2^* + d_1\,(PE_2)^* + d_1 d_2^*
\equiv X_1 + X_2 + D,
\label{eq:sm2_decomp}
\end{equation}
two cross terms linear in the discarded amplitudes and one quadratic term.
Three regimes follow, and they provide mathematical justification for our claims in the main text about the properties of the second-order quantities.

For independent speckle fields the terms of Eq.~(\ref{eq:sm2_decomp}) are statistically independent, and at full spatial resolution
\begin{equation}
\frac{\overline{|\Delta Q|^2}}{\overline{|Q|^2}} = p_1 + p_2 - p_1 p_2 \approx 2\,P_{\rm out},
\qquad
\epsilon_S = \sqrt{2}\;\epsilon_{L2},
\label{eq:sm2_product}
\end{equation}
with $p_i$ the discarded fractions.
The raw product error is $\sqrt2$ times the field error because the two cross terms are independent speckle products of equal weight that add in quadrature.

The spatial integrals of the two cross terms vanish identically,
\begin{equation}
\int X_1\,dV = \langle d_2, PE_1\rangle = 0,
\qquad
\int X_2\,dV = 0,
\label{eq:sm2_orthogonality}
\end{equation}
by the same disjoint-support orthogonality as Eq.~(\ref{eq:sm2_overlap}).
Only $\int D\,dV$ survives, a zero-mean fluctuation of relative size $\mathcal{O}(P_{\rm out}\sqrt{V_{\rm coh}/V})$ with $V_{\rm coh}$ the coherence volume.
Consequently, the first-order error vanishes for any observable that is a spatial integral of the product, such as an overlap integral.

The cross terms are concentrated at spatial frequencies at and above the mask edge, cf.\ Supplemental Material Sec.~\ref{sec:sm_cg}, so restricting the output to spatial frequencies $|\vec q| \lesssim \dkm$ suppresses them.
Averaging the product over cells of growing size $L_c$ therefore removes the cross terms, and the remaining error is second order in the discarded amplitude, far below the first-order $\sqrt{P_{\rm out}}$ of the pointwise map.

Table~\ref{tab:sm2_scalings} collects the error scalings of this appendix.

}%

\clearpage
\onecolumngrid
\setcounter{page}{1}
\renewcommand{\thepage}{S\arabic{page}}
\begin{center}
{\Large\bfseries Supplemental Material:\\[4pt] On-shell compression and reconstruction (OSCAR)\\ of monochromatic wave fields in weakly scattering media}\\[10pt]
{Pablo Jara and Alexey Yamilov}
\end{center}
\vspace{6pt}

\begin{quotation}\noindent
This Supplemental Material presents the derivations, definitions, and details on numerical simulations used to study and demonstrate the on-shell compression of scattered wave fields.
Section~\ref{sec:sm_media} characterizes all media, with every transport parameter traced to a specific simulation.
Section~\ref{sec:sm_rte} states the on-shell structure of the wave field, connects it to the radiative transfer equation, and verifies it numerically.
Section~\ref{sec:sm_metrics} resolves the reconstruction error in space and verifies the exact and statistical metric relationships of Appendix~\ref{app:metrics} of the main text numerically in 2D and 3D.
Section~\ref{sec:sm_crossover} demonstrates that reconstruction is robust across the quasi-ballistic to diffusive crossover in both 2D and 3D.
Section~\ref{sec:sm_mask} introduces integral measures of the spectral support, its width and its isotropization, which set the sample-agnostic isotropic mask.
Section~\ref{sec:sm_cg} tests the reconstruction of the sensitivity kernel, directly computed from forward and adjoint solves in both 2D and 3D, quantifies when its coarse graining is valid, and verifies that the compressed representation retains the kernel's full spectral support.
\end{quotation}
\vspace{6pt}

\setcounter{section}{0}
\setcounter{equation}{0}
\setcounter{figure}{0}
\setcounter{table}{0}
\renewcommand{\thesection}{S\arabic{section}}
\renewcommand{\thesubsection}{\thesection.\arabic{subsection}}
\makeatletter
\renewcommand{\p@subsection}{}%
\renewcommand{\p@subsubsection}{}
\def\@hangfrom@section#1#2#3{\@hangfrom{#1#2}\MakeTextUppercase{#3}}%
\let\@sectioncntformat\@undefined
\@removefromreset{equation}{section}%
\makeatother
\renewcommand{\theequation}{S\arabic{equation}}
\renewcommand{\thefigure}{S\arabic{figure}}
\renewcommand{\thetable}{S\arabic{table}}
\renewcommand{\theHsection}{S\arabic{section}}
\renewcommand{\theHsubsection}{S\arabic{section}.\arabic{subsection}}
\renewcommand{\theHequation}{S\arabic{equation}}
\renewcommand{\theHfigure}{S\arabic{figure}}
\renewcommand{\theHtable}{S\arabic{table}}

\section{Media and their characterization}
\label{sec:sm_media}

All numerical models (the media) used in this work are characterized here, and every transport parameter quoted anywhere in the paper or in this Supplemental Material traces to a simulation described in this section, see Tables~\ref{tab:sm3_2d} and \ref{tab:sm3_3d}.
The characterization data of this section use plane-wave illumination for clarity.
Throughout this Supplemental Material an overbar denotes disorder (ensemble) averaging, as in the main text, so the field in any given system can be decomposed into its ensemble average and its fluctuating (speckle) part as $E = \bar E + \delta E$.
Angle brackets are reserved for inner products, $\langle E, F\rangle = \int E^* F\,dV$, and for spatial block means carrying an explicit subscript.

\subsection{Systems}
\label{sec:sm1_systems}

\begin{figure}[t!]
\centering
\includegraphics{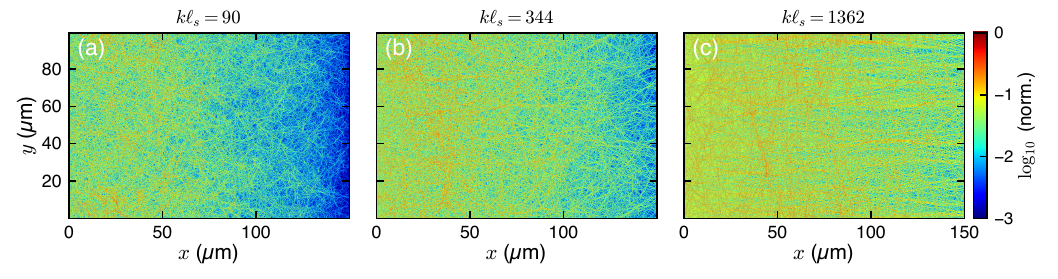}
\caption{Representative 2D fields, intensity $\log_{10} I(x,y)$, one realization at three dilution levels of Table~\ref{tab:sm3_2d}.
(a) $k\ls = 90$, (b) $k\ls = 344$, (c) $k\ls = 1362$.
At $k\ls = 90$ the plane wave decays visibly across the box.
At $k\ls = 1362$ the propagation is quasi-ballistic and the field is attenuated far more slowly.}
\label{fig:sm3_fields2d}
\end{figure}

2D media are ensembles of sub-wavelength air holes (radii $40$--$65$\,nm, minimum separation $50$\,nm) in a background of index $n_{\mathrm{eff}} = 2.85$ at $\lambda_0 = 1.55\um$, TM polarization, with representative fields shown in Fig.~\ref{fig:sm3_fields2d}.
Fields are computed with the direct frequency-domain solver MESTI \cite{sm_mesti} on a grid of twenty points per vacuum wavelength, which is $7.0$ points per wavelength in the background medium.
Three geometries are used.
The remission geometry of the main text has a $250 \times 300\um$ scattering region with perfectly matched layers on all sides and finite-size input on the front face, see Fig.~5 of the main text.
The periodic plane-wave geometry, used for the characterization of this section and benchmarking, has a line source spanning the full width, periodic transverse boundaries, absorbing layers at the two longitudinal faces, and a $99.2 \times 150\um$ box (widened to $198.4\um$, with modified thickness, where noted in Table~\ref{tab:sm3_2d}).
A third geometry, the square plane-wave box, is a $300 \times 300\um$ realization ensemble of the filling-fraction-$0.1$ medium (Level~1) with the same solver settings.
It provides the power-capture and error-measurement ensembles in Figs.~3 and 4 of the main text, with $40$ realizations.
The dilution series varies the hole filling fraction from $0.005$ to $0.1$ at fixed thickness, and two fixed-medium series vary the thickness alone.
The five ensembles of the dilution series are labeled Level~1 through Level~5 in order of increasing dilution, and this label identifies them throughout.
Table~\ref{tab:sm3_2d} lists them in this order, which is also the order of increasing $k\ls$.
The measured on-shell anisotropy is $g = 0.03$--$0.09$, so these are near-isotropic scatterers, and the transport depth $\gamma = L(1 - g)/\ls$, built from the measured $\ls$ and $g$, is used as a defined control parameter throughout.

\begin{figure}[t!]
\centering
\includegraphics{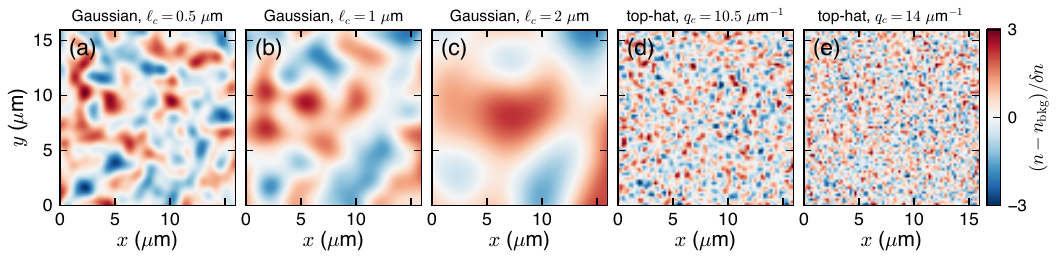}
\caption{Five representative refractive-index (fluctuation) maps $n(x, y, z_0)$, the disorder responsible for wave scattering.
All panels show areas of the same spatial extent and use a common color scale in units of $\delta n$.
Gaussian-correlated disorder with $\ell_c = 0.5$, $1$, and $2\um$ (a--c) and spectral top-hat disorder with $q_c = 10.5$ and $14\um^{-1}$ (d, e), see Table~\ref{tab:sm3_3d}.}
\label{fig:sm3_media}
\end{figure}

\begin{figure}[t!]
\centering
\includegraphics{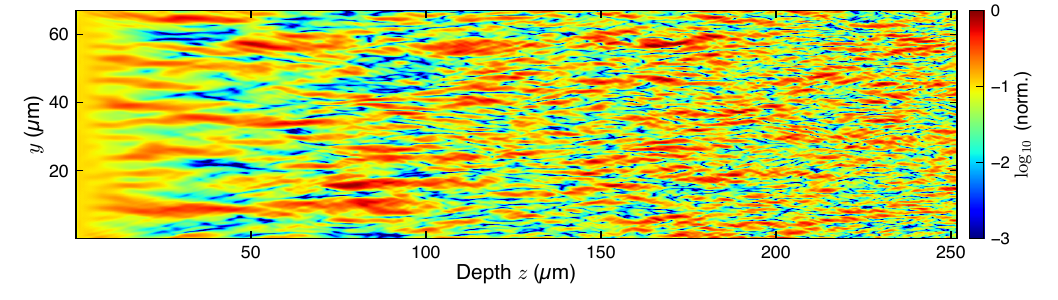}
\caption{Representative 3D intensity cut $\log_{10} I(x_0, y, z)$ taken in the mid-$x$ plane, under the plane-wave illumination used for characterization and benchmarking.
Disorder is Gaussian-correlated with $\ell_c = 1\um$, $\delta n = 0.0144$, $\ls = 27\um$, one realization, see Table~\ref{tab:sm3_3d}.
The ballistic component fades within the first few $\ls$, and the underdeveloped, forward-streaked speckle persists across the remaining depth.}
\label{fig:sm3_fields3d}
\end{figure}

3D media are continuous refractive-index disorder $n(\vec r) = n_{\rm bkg} + \delta n(\vec r)$ with $n_{\rm bkg} = 1.35$ at $\lambda_0 = 0.85\um$, solved with the iterative vector-Maxwell solver WaveSim \cite{sm_wavesim} at three grid points per medium wavelength, periodic transversely with absorbing ends, plane-wave illumination.
Two disorder classes with closed-form spectra are used, see Fig.~\ref{fig:sm3_media}.
Gaussian-correlated disorder of correlation length $\ell_c$ has anisotropy $g = \coth a - 1/a$ with $a = 2(k \ell_c)^2$, obtained by integrating the Born-level phase function, proportional to the generator spectrum evaluated at the momentum transfer, over the scattering angle \cite{sm_ishimaru}.
Spectral top-hat disorder, constant spatial spectrum up to a cutoff $q_c$, has $g = 1 - q_c^2/4k^2$ exactly by the same integration, and its sampled spectrum equals the specification by construction.
Media (spatial distributions of the refractive index) are generated by smoothing or spectrally filtering white Gaussian noise, unit-variance normalized, with the amplitude set by $\delta n$.
The transverse boundaries are periodic and the illumination is a plane wave, so the finite transverse extent acts as a supercell period rather than an aperture.
Statistical fluctuations are computed from the realization-to-realization spread, which stays below $0.3\%$ in $\epsilon_{L2}/\sqrt{\overline{P}_{\rm out}}$ at $\alpha \ge 2$ across seeds and Cartesian components even for the $27\um$-wide deep slabs.
A representative 3D field is shown in Fig.~\ref{fig:sm3_fields3d}.


\subsection{Measured parameters}\label{sec:sm1_params}

The scattering mean free path is obtained from a fit of the decay of the ballistic (co-polarized) coherent amplitude, $|\bar E(z)| \propto e^{-z/2\ls}$, averaged over the transverse plane.
The depth range $5\um < z < \min(4\ls^{\rm ref}, L - 5\um)$ is used in the fit.
Here $\ls^{\rm ref}$ is the design estimate of the mean free path used only to size the window, and the standard error of the ensemble mean is quoted as the uncertainty, except for the $1000$-realization Level~1 ensemble, fitted on the ensemble mean, where the fit-window systematic is quoted instead.
Varying the endpoints of the fit window changes the fitted $\ls$ by $0.03$--$3\%$, a systematic not included in the quoted uncertainties.
In 2D, $\ls$ is fitted this way for every filling fraction of the dilution series, on the $1000$-realization plane-wave Level~1 ensemble ($\ls = 7.8 \pm 0.2\um$) and on the twenty realizations of each of the other levels, and the geometries that share a medium share its $\ls$.
In 3D, every system of Table~\ref{tab:sm3_3d} is fitted individually.
The radial width of the ring is not used in determining $\ls$, because the measured width is broadened (contaminated) by the finite-box window.
The effective wavenumber $k$ is the measured radial peak of the field spectrum, which absorbs the real part of the self-energy, see Eq.~\eqref{eq:dyson} of the main text.
The finite-difference grid additionally imprints a weak fourfold dispersion on the 2D ring, $|\vec k'| = c_0 + c_4\cos 4\theta$ along the ring with $c_4/c_0 \approx 0.8\%$, see Sec.~\ref{sec:corrections} of the main text.
The ensemble fits give $c_0 = 11.477$ and $c_4 = 0.096\um^{-1}$ for the rectangular geometry, and $c_0 = 11.476$ and $c_4 = 0.088\um^{-1}$ for the square ensemble, and $c_0 = 11.524$ and $c_4 = 0.097\um^{-1}$ for the remission geometry.
These constants are determined once from the ensemble and held fixed for every realization.
As a control, refitting $c_0$ per realization changes it by under $10^{-4}$ of its value across the forty square-ensemble realizations and changes the retained power by under $0.3\%$ of its value at $\alpha \ge 4$ and by under $5\%$ at the narrowest masks.
This test confirms that a single mask is appropriate in all realizations at the mask widths used for the results of this work.
Binned in $K(\vec k')$ of Eq.~(\ref{eq:discrete_dispersion}) of the main text, the ring is isotropic, with its peak at $k_p = 11.20\um^{-1}$ at Level~1. The main-text masks are defined there, so their edges follow the measured fourfold shape in $|\vec k'|$. Every quoted $k\ls$ multiplies the ring-center wavenumber, in 2D the angular mean $c_0$ of the ring radius in $|\vec k'|$, by the $\ls$ found from the decay of the field.
The anisotropy $g$ is exact for the top-hat generator and analytic for the Gaussian one, obtained in both cases from the Born-level phase function \cite{sm_ishimaru} evaluated at the measured $k$.
The quoted uncertainty $\delta g = 2(1 - g)\,\delta k/k$ follows from differentiating the top-hat expression $g = 1 - q_c^2/4k^2$, and stays below $10^{-3}$ for every 3D system.
In 2D, $g$ is measured directly as the on-shell angular mean $\overline{\cos\theta}$ \cite{sm_ishimaru} for the scattering angle $\theta$, with the ballistic component excluded.

In tissue-optics terms the 3D systems correspond to reduced scattering coefficients $\mu_s' = (1 - g)/\ls$ from $0.4$ to $165\,\mathrm{cm}^{-1}$ at $\lambda_0 = 0.85\um$, spanning the $5$--$25\,\mathrm{cm}^{-1}$ range reported for soft tissue \cite{sm_jacques}.
A measure of the ring support, the effective shell width used throughout this work for mask selection, is defined next.
Let $\bar{\mathcal{S}}(k')$ denote the spectral density of the field at radial wavenumber $k' = |\vec k'|$, averaged over direction, with the ballistic component excluded, the direction-averaged counterpart of $\rho(k')$ in Eq.~(\ref{eq:lorentzian}) of the main text.
A quantitative measure of the ring width is
\begin{equation}
\ash = \frac{2}{\pi}\,\ls\, \frac{\int \bar{\mathcal{S}}(k')\, dk'}{\max_{k'} \bar{\mathcal{S}}(k')},
\label{eq:sm1_areas}
\end{equation}
a single number obtained by integration.
The normalization is chosen so that a fully developed ring, a Lorentzian of full width $1/\ls$ at half maximum, gives $\ash = 1$.
$\ash$ is the effective thickness of the shell, sharing the normalization with the $\alpha$ parameter of the mask. The $\alpha \gtrsim \ash$ condition ensures adequate coverage of the shell.
At Level~1, $\ash = 0.99$ is close to the fully developed ring, and it grows to $3.74$ at the most dilute level, paralleling the quasi-ballistic broadening measured in 3D, cf.\ Fig.~\ref{fig:sm_shellvol}(c).
This measure is the basis of the mask-support analysis of Sec.~\ref{sec:sm_mask}.

Each system is used in this work as follows.
The remission geometry provides the sensitivity maps of Fig.~5 of the main text, the ring decomposition of Fig.~\ref{fig:sm_decomp}, and the metrics-verification ensemble of Sec.~\ref{sec:sm_metricverify}.
Level~1 provides the fields of Figs.~1 and 2 of the main text and the 2D sensitivity data of Fig.~\ref{fig:sm5_sens2d}(a,c,g).
Levels~1, 3, and 5 are shown in Fig.~\ref{fig:sm3_fields2d}, and the full dilution series provides the ring-support sequence of Eq.~\eqref{eq:sm1_areas} used in Sec.~\ref{sec:sm_mask}.
The Pair medium provides the diffusive block of the 2D sensitivity maps of Fig.~\ref{fig:sm4_maps2d}, the field-storage sequence (Sequence 2) provides its thinner block, and the thickness sequence (Sequence 1) provides the $\epsilon_S(\gamma)$ data of Fig.~\ref{fig:sm5_sens2d}(e).
The field-storage sequence (Sequence 2), together with Level~1 as its $\gamma = 18.6$ member, provides the 2D crossover of Sec.~\ref{sec:sm_crossover}, Figs.~\ref{fig:sm2_rd3d} and \ref{fig:sm2_depth3d}.
The 3D systems of Table~\ref{tab:sm3_3d} support the crossover and mask studies of Secs.~\ref{sec:sm_crossover} and \ref{sec:sm_mask}, Figs.~6--8 of the main text, and the benchmarks of Sec.~\ref{sec:benchmarks} of the main text.

\begin{table}[h]
\caption{2D simulations.
TM, plane wave, periodic width $W = 99.2\um$, $L = 150\um$ unless stated, twenty points per vacuum wavelength $\lambda_0$.
All measured columns are defined in Secs.~\ref{sec:sm1_systems} and \ref{sec:sm1_params}, the $\ls$ uncertainty the standard error of the ensemble mean and the $g$ uncertainty the realization spread, for the Level-1 medium the fit-window systematic of Sec.~\ref{sec:sm1_params}.
Here ff denotes the hole filling fraction and $N_{\rm rlz}$ the archived ensemble size. Individual figures may draw subsets or use larger dedicated ensembles, cf.\ Secs.~\ref{sec:sm_rte_numerics} and \ref{sec:sm_metricverify}.
The last three rows are the fixed-medium constructions, the diffusive medium at ff $= 0.033$ (Pair), the thickness sequence at ff $= 0.05$ (Sequence 1), and the field-storage sequence at the Level~1 statistics (Sequence 2), the latter two varying $\gamma$ by thickness alone.
The remission row is the finite-input demonstration geometry of Fig.~5 of the main text, with light injected on the front face of the $300\um$-deep slab.
Its medium is identical to the Level-1 and square ensembles. Bare numbers in the last column are figures of the main text and S-prefixed numbers figures of this Supplemental Material.}
\label{tab:sm3_2d}
\centering
\small\setlength{\tabcolsep}{2.0pt}
\begin{tabular}{lcccccccp{1.05in}}
\hline\hline
System & ff & $W {\times} L$ ($\um$) & $\ls$ ($\um$) & $k\ls$ & $g$ & $\gamma$ & $N_{\rm rlz}$ & Used in figures \\
\hline
Level 1 & 0.1 & $99.2 {\times} 150$ & $7.8 \pm 0.2$ & 90 & $0.030 \pm 0.006$ & 18.6 & 20 & 1, 2, \ref{fig:sm6_headline}, \ref{fig:sm6_sens}, \ref{fig:sm3_fields2d}(a), \ref{fig:sm2_poutdist}, \ref{fig:sm2_rd3d}, \ref{fig:sm2_depth3d}, \ref{fig:sm5_sens2d}, Tab.~\ref{tab:sm4_audit} \\
Level 2 & 0.05 & $99.2 {\times} 150$ & $13.4 \pm 0.5$ & 156 & $0.061 \pm 0.014$ & 10.5 & 20 & \ref{fig:sm6_headline}, \ref{fig:sm6_sens} \\
Level 3 & 0.02 & $99.2 {\times} 150$ & $29.2 \pm 0.8$ & 344 & $0.089 \pm 0.014$ & 4.68 & 20 & \ref{fig:sm6_headline}, \ref{fig:sm6_sens}, \ref{fig:sm3_fields2d}(b) \\
Level 4 & 0.01 & $99.2 {\times} 150$ & $56.4 \pm 0.8$ & 667 & $0.089 \pm 0.020$ & 2.42 & 20 & \ref{fig:sm6_headline}, \ref{fig:sm6_sens}, Tab.~\ref{tab:sm4_audit} \\
Level 5 & 0.005 & $99.2 {\times} 150$ & $115 \pm 1$ & 1362 & $0.085 \pm 0.023$ & 1.19 & 20 & \ref{fig:sm6_headline}, \ref{fig:sm6_sens}, \ref{fig:sm3_fields2d}(c) \\
Square & 0.1 & $300 {\times} 300$ & $7.8 \pm 0.2$ & 90 & $0.030 \pm 0.006$ & 37.3 & 40 & 3, 4, \ref{fig:sm2_poutdist} \\
Remission & 0.1 & $250 {\times} 300$ & $7.8 \pm 0.2$ & 90 & $0.030 \pm 0.006$ & 37.3 & 20 & 5, \ref{fig:sm_decomp}, \ref{fig:sm2_poutdist}, \ref{fig:sm2_metrics} \\
Pair & 0.033 & $198.4 {\times} 265$ & $18.8 \pm 0.5$ & 221 & $0.064 \pm 0.013$ & 13.2 & 6 & \ref{fig:sm4_maps2d}(e--h) \\
Sequence 1 & 0.05 & $99.2 {\times} (8.8$--$299)$ & $13.4 \pm 0.5$ & 156 & $0.061 \pm 0.014$ & 0.62--20.9 & 10 each & \ref{fig:sm5_sens2d}(e) \\
Sequence 2 & 0.1 & $99.2 {\times} (18.75$--$112.5)$ & $7.8 \pm 0.2$ & 90 & $0.030 \pm 0.006$ & 2.3--14.0 & 10 each & \ref{fig:sm2_rd3d}, \ref{fig:sm2_depth3d}, \ref{fig:sm4_maps2d}(a--d), Tab.~\ref{tab:sm4_audit} \\
\hline\hline
\end{tabular}
\end{table}

\begin{table}[h]
\caption{3D simulations.
All measured columns are defined in Secs.~\ref{sec:sm1_systems} and \ref{sec:sm1_params}, with $\gamma = L(1-g)/\ls$ carrying the uncertainties in quadrature.
$W {\times} L$ is the transverse width and the depth of the interior box, $320^2 \times 1200$ voxels $= 67 \times 67 \times 252\um$ and $128^2 \times 3000$ voxels $= 27 \times 27 \times 630\um$, and $k\ls$ uses $k = 9.982\um^{-1}$.
$N_{\rm rlz}$ denotes the number of realizations.
The Depol.\ column is the depolarized power fraction, realization spread below $0.002$. Entries are absent where only angular histograms were stored. Individual figures may draw subsets, see the last column. Bare numbers in the last column are figures of the main text and S-prefixed numbers figures of this Supplemental Material.}
\label{tab:sm3_3d}
\centering
\footnotesize\setlength{\tabcolsep}{2.5pt}
\begin{tabular}{lccccccccp{1.05in}}
\hline\hline
Generator & $\delta n$ & $W {\times} L$ ($\um$) & $\ls$ ($\um$) & $k\ls$ & $g$ & $\gamma$ & $N_{\rm rlz}$ & Depol. & Used in figures \\
\hline
Top-hat $q_c = 14\um^{-1}$ & 0.014 & $67 {\times} 252$ & $299 \pm 2$ & 2985 & $0.508$ & $0.41 \pm 0.01$ & 10 & 0.161 & 7, \ref{fig:sm6_headline}, \ref{fig:sm6_gamma}, \ref{fig:sm6_sens}, \ref{fig:sm2_rd3d}, \ref{fig:sm2_depth3d}, \ref{fig:sm_shellvol}, \ref{fig:sm5_sens3d}, Tab.~\ref{tab:sm4_audit} \\
Top-hat $q_c = 14\um^{-1}$ & 0.020 & $67 {\times} 252$ & $150 \pm 1$ & 1497 & $0.508$ & $0.83 \pm 0.01$ & 10 & 0.262 & 7, \ref{fig:sm6_headline}, \ref{fig:sm6_gamma}, \ref{fig:sm6_sens}, \ref{fig:sm2_rd3d}, \ref{fig:sm2_depth3d}, \ref{fig:sm_shellvol}, \ref{fig:sm4_maps3d} \\
Top-hat $q_c = 14\um^{-1}$ & 0.030 & $67 {\times} 252$ & $67.4 \pm 0.3$ & 673 & $0.508$ & $1.84 \pm 0.01$ & 10 & 0.409 & 7, \ref{fig:sm6_headline}, \ref{fig:sm6_gamma}, \ref{fig:sm6_sens}, \ref{fig:sm2_rd3d}, \ref{fig:sm2_depth3d}, \ref{fig:sm_shellvol}, \ref{fig:sm5_sens3d} \\
Top-hat $q_c = 14\um^{-1}$ & 0.037 & $67 {\times} 252$ & $44.4 \pm 0.3$ & 443 & $0.508$ & $2.79 \pm 0.02$ & 10 & 0.481 & \ref{fig:sm6_headline}, \ref{fig:sm6_gamma}, \ref{fig:sm6_sens}, \ref{fig:sm5_sens3d} \\
Top-hat $q_c = 14\um^{-1}$ & 0.045 & $67 {\times} 252$ & $30.1 \pm 0.3$ & 300 & $0.508$ & $4.11 \pm 0.04$ & 10 & 0.537 & 6--8, \ref{fig:sm6_headline}, \ref{fig:sm6_gamma}, \ref{fig:sm6_sens}, \ref{fig:sm2_poutdist}, \ref{fig:sm2_rd3d}, \ref{fig:sm2_depth3d}, \ref{fig:sm_shellvol}, \ref{fig:sm5_sens3d} \\
Top-hat $q_c = 10.5\um^{-1}$ & 0.018 & $67 {\times} 252$ & $150 \pm 1$ & 1497 & $0.723$ & $0.47 \pm 0.01$ & 10 & 0.243 & 7, \ref{fig:sm2_rd3d}, \ref{fig:sm2_depth3d}, \ref{fig:sm_shellvol} \\
Top-hat $q_c = 10.5\um^{-1}$ & 0.025 & $67 {\times} 252$ & $78.4 \pm 1.0$ & 783 & $0.723$ & $0.89 \pm 0.01$ & 10 & 0.364 & 7, \ref{fig:sm6_headline}, \ref{fig:sm6_sens}, \ref{fig:sm2_rd3d}, \ref{fig:sm2_depth3d}, \ref{fig:sm_shellvol}, \ref{fig:sm5_sens3d} \\
Top-hat $q_c = 10.5\um^{-1}$ & 0.035 & $67 {\times} 252$ & $40.1 \pm 0.7$ & 400 & $0.723$ & $1.74 \pm 0.03$ & 10 & 0.480 & 7, \ref{fig:sm2_rd3d}, \ref{fig:sm2_depth3d}, \ref{fig:sm_shellvol} \\
Gaussian $\ell_c = 1\um$ & 0.0072 & $67 {\times} 252$ & $104 \pm 8$ & 1038 & $0.995$ & $0.012 \pm 0.001$ & 10 & --- & \ref{fig:sm6_headline}, \ref{fig:sm6_gamma} \\
Gaussian $\ell_c = 1\um$ & 0.0102 & $67 {\times} 252$ & $52.9 \pm 4.6$ & 528 & $0.995$ & $0.024 \pm 0.002$ & 10 & --- & \ref{fig:sm6_headline}, \ref{fig:sm6_gamma} \\
Gaussian $\ell_c = 2\um$ & 0.0102 & $67 {\times} 252$ & $30.4 \pm 4.4$ & 303 & $0.9987$ & $0.011 \pm 0.002$ & 10 & --- & \ref{fig:sm3_media}, Sec.~\ref{sec:sm_mask} \\
Gaussian $\ell_c = 1\um$ & 0.0144 & $67 {\times} 252$ & $27.1 \pm 2.0$ & 271 & $0.995$ & $0.046 \pm 0.003$ & 10 & --- & \ref{fig:sm6_headline}, \ref{fig:sm6_gamma}, \ref{fig:sm3_fields3d} \\
Gaussian $\ell_c = 0.5\um$ & 0.0204 & $67 {\times} 252$ & $25.8 \pm 0.8$ & 258 & $0.980$ & $0.195 \pm 0.006$ & 10 & --- & \ref{fig:sm3_media} \\
Gaussian $\ell_c = 1\um$ & 0.0204 & $67 {\times} 252$ & $12.9 \pm 0.8$ & 129 & $0.995$ & $0.098 \pm 0.006$ & 10 & --- & \ref{fig:sm6_headline}, \ref{fig:sm6_gamma} \\
Gaussian $\ell_c = 0.5\um$ & 0.043 & $67 {\times} 252$ & $5.79 \pm 0.30$ & 58 & $0.980$ & $0.87 \pm 0.05$ & 10 & 0.442 & 7, \ref{fig:sm2_rd3d}, \ref{fig:sm2_depth3d} \\
Top-hat $q_c = 14\um^{-1}$ & 0.045 & $27 {\times} 630$ & $29.9 \pm 0.6$ & 298 & $0.508$ & $10.4 \pm 0.2$ & 10 & 0.604 & 7, \ref{fig:sm6_gamma}, \ref{fig:sm2_poutdist}, \ref{fig:sm2_metrics}, \ref{fig:sm2_rd3d}, \ref{fig:sm2_depth3d}, \ref{fig:sm_shellvol}, \ref{fig:sm4_maps3d}, \ref{fig:sm5_sens3d}, Tab.~\ref{tab:sm4_audit} \\
\hline\hline
\end{tabular}
\end{table}

\FloatBarrier
\section{On-shell structure of the wave field}
\label{sec:sm_rte}

Section~\ref{sec:sm_wigner} connects the on-shell structure to the radiative transfer equation through the Wigner function.
Section~\ref{sec:sm_rte_numerics} numerically verifies the on-shell structure directly on an ensemble of $1000$ disorder realizations, and the measured radial profile follows the parameter-free Lorentzian prediction.
Section~\ref{sec:sm1_estimator} defines the finite-box ring-width estimator used in Fig.~\ref{fig:onshell_schematic} of the main text and quantifies the window broadening it corrects for.

\subsection{Wigner function and the shell-integrated radiative transfer equation (RTE)}
\label{sec:sm_wigner}

In this subsection we reproduce the common derivation of the RTE and point out the connection of the on-shell spectral statement to the specific intensity.
We start with the definition of the Wigner distribution of the field~\cite{sm_rpk1996, sm_akkermans2007},
\begin{equation}
W(\vec R, \vec k') = \int d\vec\rho\;
\overline{E(\vec R + \tfrac{\vec\rho}{2})\, E^*(\vec R - \tfrac{\vec\rho}{2})}\; e^{-i\vec k'\cdot\vec\rho}.
\label{eq:sm_wignerdef}
\end{equation}

\emph{Step 1, exact moment equation.}
Write the Helmholtz equation at $\vec r_1$ and its conjugate at $\vec r_2$, multiply by $E^*(\vec r_2)$ and $E(\vec r_1)$ respectively, and subtract.
With $\nabla_1^2 - \nabla_2^2 = 2\nabla_{\vec R}\cdot\nabla_{\vec\rho}$ and $\nabla_{\vec\rho} \to i\vec k'$ under the Wigner transform,
\begin{equation}
\frac{i}{k}\,\vec k'\cdot\nabla_{\vec R}\, W(\vec R, \vec k')
= -\frac{k_0^2}{2k}\int d\vec\rho\; e^{-i\vec k'\cdot\vec\rho}\,
\overline{\bigl[\delta\varepsilon(\vec r_1) - \delta\varepsilon(\vec r_2)\bigr]\,
E(\vec r_1) E^*(\vec r_2)}.
\label{eq:sm_moment}
\end{equation}
This step is exact for a lossless medium.
All approximations enter through the closure of the right-hand side, where the disorder average couples $W$ to higher correlations of the permittivity fluctuation $\delta\varepsilon$ with the fields.

\emph{Step 2, ladder closure.}
To leading order in the weak disorder ($k\ls \gg 1$), the standard closure \cite{sm_rpk1996, sm_akkermans2007} pairs one $\delta\varepsilon$ with the Born expansion of the adjacent field.
Pairing inside the same field produces the self-energy, a loss term $-W/\ls$ with $\mathrm{Im}\,\Sigma = k/\ls$ evaluated at the peak.
Pairing across the two fields produces the vertex, a gain term that re-emits the scattered weight through the pair of average Green's functions, with $|\bar G(\vec k')|^2 = (\pi\ls/k)\, A(\vec k')$.
Here $A(\vec k') \equiv -\mathrm{Im}\,\bar G(\vec k')/\pi$ is the spectral function of the average Green's function of Eq.~\eqref{eq:green_avg} of the main text.
Its radial profile is the Lorentzian of Eq.~\eqref{eq:lorentzian} of the main text, of full width $1/\ls$, sharply peaked on the shell $|\vec k'| = k$ for $k\ls \gg 1$.
Explicitly, both pairings carry the same ladder vertex $U(\vec k', \vec k'') = k_0^4\, \tilde C_\varepsilon(\vec k' - \vec k'')$, with $\tilde C_\varepsilon$ the spectrum of the permittivity correlation.
Pairing within the same field yields the self-energy,
\begin{equation}
\Sigma(\vec k') = -\int \frac{d\vec k''}{(2\pi)^d}\; U(\vec k', \vec k'')\, \bar G(\vec k'').
\label{eq:sm_born}
\end{equation}
Its imaginary part turns the corresponding piece of the right-hand side of Eq.~(\ref{eq:sm_moment}) into the loss term $-(\mathrm{Im}\,\Sigma/k)\, W = -W/\ls$, after division of Eq.~(\ref{eq:sm_moment}) by $i$ and with $k'/k \to 1$.
The cross-field pairing dresses the same vertex with the pair of average Green's functions and re-emits the scattered weight, so Eq.~(\ref{eq:sm_moment}) closes into
\begin{equation}
\hat k'\cdot\nabla_{\vec R}\, W(\vec R, \vec k')
= -\frac{1}{\ls}\, W(\vec R, \vec k')
+ \frac{|\bar G(\vec k')|^2}{\ls} \int \frac{d\vec k''}{(2\pi)^d}\; U(\vec k', \vec k'')\, W(\vec R, \vec k'').
\label{eq:sm_prekinetic}
\end{equation}
To connect Eq.~(\ref{eq:sm_prekinetic}) to the on-shell form, the exact spectral identity $-\mathrm{Im}\,\bar G(\vec k') = |\bar G(\vec k')|^2\, \mathrm{Im}\,\Sigma(k') = \pi A(k')$ is used, so the gain prefactor is $\pi A(k')/k$. 
This gives the same spectral function $A$ that carries the radial lineshape.
Evaluating $U$ under the sharply peaked $A$ with both momenta on the shell reduces it to a function of the scattering angle alone, normalized to the phase function $p$, while the optical theorem applied to Eq.~(\ref{eq:sm_born}) fixes the overall gain rate to the same $1/\ls$ as in the loss term.
We obtain
\begin{equation}
\hat k'\cdot\nabla_{\vec R}\, W(\vec R, \vec k')
= -\frac{1}{\ls}\, W(\vec R, \vec k')
+ \frac{1}{\ls}\, A(k') \int d\hat k''\; p(\hat k'\cdot\hat k'')\; I(\vec R, \hat k''),
\label{eq:sm_kinetic}
\end{equation}
where $I$ is the radially integrated angular density and the normalizations are
\begin{equation}
I(\vec R, \hat k) \equiv \int_0^\infty W(\vec R, k''\hat k)\, d\sigma(k''),
\qquad
\int_0^\infty A(k'')\, d\sigma(k'') = 1,
\qquad
\int p(\hat k'\cdot\hat k'')\, d\hat k'' = 1,
\label{eq:sm_norms}
\end{equation}
with $d\sigma \propto k''^{\,d-1} dk''$ the radial measure in $d$ dimensions.
The geometric prefactors are set by the two unit normalizations and by flux conservation.
Three on-shell evaluations are made in writing Eqs.~(\ref{eq:sm_prekinetic}) and (\ref{eq:sm_kinetic}), each valid to $O\bigl((k\ls)^{-1}\bigr)$ because $A$ is sharply peaked: (i) the prefactor of the gradient term is set to $k'/k \to 1$, (ii) the rate $\mathrm{Im}\,\Sigma$ is taken constant across the peak, and (iii) the disorder vertex is evaluated with both momenta on the shell, which reduces it to the phase function $p$ of angles alone.

\emph{Step 3, radial integration.}
Integrate Eq.~(\ref{eq:sm_kinetic}) over $d\sigma(k')$ term by term.
The gradient and loss terms integrate directly to $\hat k'\cdot\nabla_{\vec R} I$ and $-I/\ls$.
In the gain term the entire radial dependence is concentrated in the prefactor $A(k')$, which integrates to unity by Eq.~(\ref{eq:sm_norms}).
\emph{The radial profile drops out of every term}, leaving the familiar RTE for the angular-spatial density~\cite{sm_ishimaru, sm_akkermans2007},
\begin{equation}
\hat k'\cdot\nabla_{\vec R}\, I(\vec R, \hat k')
= -\frac{1}{\ls}\, I(\vec R, \hat k')
+ \frac{1}{\ls} \int p(\hat k'\cdot\hat k'')\, I(\vec R, \hat k'')\, d\hat k''.
\label{eq:sm_rte}
\end{equation}

\emph{Step 4, self-consistency of factorization.}
Conversely, the ansatz
\begin{equation}
W(\vec R, \vec k') \simeq A(|\vec k'|)\; I(\vec R, \hat k')
\label{eq:sm_ansatz}
\end{equation}
solves Eq.~(\ref{eq:sm_kinetic}) identically whenever $I$ obeys Eq.~(\ref{eq:sm_rte}).
The gradient and loss terms act on the ansatz multiplicatively, while the gain is proportional to $A(k')$ by construction.
The factorization is therefore not an additional assumption on top of the ladder approximation.
It is a component part of the ladder solution.
The RTE spreads $I$ in space and isotropizes it in direction, which is the familiar widening of the specific intensity with propagation.
The radial profile $A(|\vec k'|)$ is not acted upon and stays pinned at full width $1/\ls$.
In real space the ansatz of Eq.~(\ref{eq:sm_ansatz}) is the factorization of the field correlation function in the center-of-mass and relative coordinates $\vec R = (\vec r_1 + \vec r_2)/2$ and $\vec\rho = \vec r_1 - \vec r_2$~\cite{sm_shapiro1986},
\begin{equation}
\overline{E(\vec R + \tfrac{\vec\rho}{2})\, E^*(\vec R - \tfrac{\vec\rho}{2})}
= \overline{I}(\vec R)\, g(\rho),
\label{eq:sm_factorization}
\end{equation}
with the short-range field correlation~\cite{sm_shapiro1986, sm_yamilov2008}
\begin{equation}
g(\rho) =
\begin{cases}
\dfrac{\sin k\rho}{k\rho}\; e^{-\rho/2\ls}, & \text{3D},\\[2ex]
J_0(k\rho)\; e^{-\rho/2\ls}, & \text{2D}.
\end{cases}
\label{eq:sm_C1}
\end{equation}
Here $\overline{I}(\vec R)$ is the slowly varying intensity envelope, and the Fourier transform of $g(\rho)$ in $\vec\rho$ is the radial Lorentzian of Eq.~\eqref{eq:lorentzian} of the main text.
This fact justifies the radial integration of Step 3, and it holds at leading order in $1/k\ls$.

The above considerations show that the RTE of Eq.~(\ref{eq:sm_rte}) should be viewed as `shell-integrated' rather than on-shell.
Therefore the on-shell structure that OSCAR exploits is not in contradiction with the RTE description.
More precisely, the specific intensity $I(\vec R, \hat k)$ is defined in Eq.~(\ref{eq:sm_norms}) as the radial integral of $W$.
Scattering redistributes spectral weight from one direction $\hat k'$ to another at the same $|\vec k'| = k$, so the entire transport dynamics unfolds inside the stored band, and one depth-independent mask accommodates every stage of the propagation process the RTE is designed to describe.
Importantly, OSCAR operates on fields in a specific system, while preserving the wave transport information that the RTE describes only in terms of the statistically averaged specific intensity.

\subsection{Numerical verification}
\label{sec:sm_rte_numerics}

\begin{figure}[h]
\includegraphics{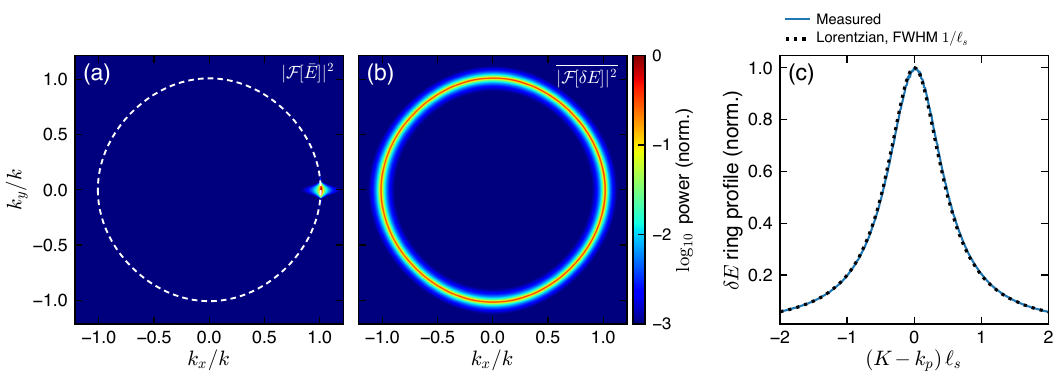}
\caption{Decomposition of the wave field over $1000$ disorder realizations (2D system of the main text).
(a) Spatial spectrum of the coherent field $\bar E$, the ballistic streak.
No ring appears, because the speckle averages out of $\bar E$.
The finite-ensemble residual of the incoherent power, suppressed by $1/N = 10^{-3}$, falls below the color scale.
The white dashed line, a guide to the eye, traces the measured ring of (b).
(b) Spatial spectrum of the speckle field $\delta E$, the ring of Eq.~\eqref{eq:lorentzian} of the main text.
Both spectra are normalized by their common maximum and share the logarithmic color scale.
Axes are normalized by the ring radius $k$.
(c) Radial profile of the speckle ring in $K(\vec k')$ of Eq.~\eqref{eq:discrete_dispersion} of the main text, around the ring radius $k_p$, in units of $1/\ls$. Dotted: the Lorentzian of full width $1/\ls$, with no free parameters.
The measured full width at half maximum is $1.00/\ls$.}
\label{fig:sm_decomp}
\end{figure}

Here we numerically verify the predictions of Sec.~\ref{sec:sm_wigner} directly on an ensemble of $1000$ disorder realizations in the remission geometry of Sec.~III\,A of the main text with the Level~1 statistics of Table~\ref{tab:sm3_2d}, $250 \times 300\um$ with absorbing boundaries and one of the five stored modes of the finite-size input of the main text per realization.
The ensemble-averaged $\bar E$ field, often referred to as ballistic or coherent, does not average out to zero because of its fixed phase relationship to the realization-independent incident field.
The remainder $\delta E = E - \bar E$ is the speckle or `random' field, which inherently depends on each disorder realization.
Figure~\ref{fig:sm_decomp}(a,b) shows spatial spectra of the two field components.
The coherent part carries $2.8\%$ of the total power in this $38\,\ls$-deep sample, and its spectrum is the forward fan expected of a decaying collimated excitation, see Fig.~\ref{fig:sm_decomp}(a).
The speckle part carries the remaining $97.2\%$, concentrated on the ring, see Fig.~\ref{fig:sm_decomp}(b).
The measured ring follows $k(\theta) = 11.524 + 0.097 \cos 4\theta\um^{-1}$.
The fourfold modulation is the leading angular anisotropy of the five-point discretized Laplacian.
Expanding its dispersion relation, $(4/\Delta x^2)\sum_{i}\sin^2(k_i\Delta x/2) = \mathrm{const}$, to next order in $k\Delta x$ gives a $\cos4\theta$ term of amplitude $k(k\Delta x)^2/96$, which evaluates to $0.096\um^{-1}$ at $\Delta x = \lambda_0/20$, in agreement with the measured value.
The mean radius combines the self-energy shift of Eq.~\eqref{eq:dyson} of the main text with the isotropic part of the same grid correction.
Read in $K(\vec k')$ of Eq.~\eqref{eq:discrete_dispersion} of the main text, in which the ring is a circle, the radial profile of the speckle ring follows the parameter-free prediction, the Lorentzian of Eq.~\eqref{eq:lorentzian} of the main text with full width $1/\ls$ from the independently measured $\ls = 7.8 \pm 0.2\um$, see Fig.~\ref{fig:sm_decomp}(c).
The measured full width at half maximum is $1.00/\ls$, within $1\%$ of the intrinsic width. The medium is bounded by absorbing layers on all four sides, so the box of depth $L = 300\um$ and width $W = 250\um$ imposes a $\mathrm{sinc}^2$ window of full width $0.886 \times 2\pi/L = 0.14/\ls$ on $k_x$ and $0.17/\ls$ on $k_y$.
The depth independence predicted by Step 4 is confirmed in Fig.~\ref{fig:onshell_schematic}(b) of the main text.

\subsection{The ring-width estimator behind Fig.~\ref{fig:onshell_schematic} of the main text}
\label{sec:sm1_estimator}

Figure~\ref{fig:onshell_schematic} of the main text estimates the ring width from finite boxes cut one at a time from individual realizations, and this subsection defines that single-box procedure, including its measured accuracy and its treatment of the box size.

A square box of side $W$ centered at $(x_0, y_0)$ is cut from the field (the sharp box is the window) and zero-padded onto a fixed $2048^2$ grid, and then Fourier transformed.
A box-size-dependent transform length makes the reading oscillate across the $W$ sequence because of the changing spectral lattice, and padding every box to the same grid removes the artifact.
The power spectrum is binned in $K(\vec k')$ of Eq.~\eqref{eq:discrete_dispersion} of the main text, in which the ring is a circle, so no correction for the lattice dispersion is needed.
The ballistic sector $|\theta| < 30^{\circ}$ is excluded so that it does not contaminate the halo contribution.
The azimuthally averaged radial profile is accumulated in bins of $0.01\um^{-1}$ over $\pm 1.2\um^{-1}$ in the radial offset $q$ from the ring center.
The spectrum of a single finite box also carries a diffuse background, produced by the speckle fluctuations and by the sidelobe leakage of the sharp-box window from the strong ring.
This background is a property of the single-box measurement, not of the medium. The ensemble spectrum of Sec.~\ref{sec:sm_rte_numerics} carries no off-shell background.
The background is estimated as the mean spectral level at $1.0\um^{-1} < |q| < 1.2\um^{-1}$ and subtracted, so the width reading refers to the ring alone.
The peak position is estimated from parabolic interpolation, and the full width is read at half of the background-subtracted maximum with the outermost crossings linearly interpolated.
Panels (b) and (c) of Fig.~\ref{fig:onshell_schematic} show the mean and standard deviation of this reading computed from $1000$ realizations (and five transverse box positions in (b)).
The dashed curve combines the ring and window widths in quadrature,
\begin{equation}
\dksf(W) = \sqrt{(1/\ls)^2 + (c_W/W)^2},
\qquad
c_W = 0.886 \times 2\pi = 5.57,
\label{eq:sm1_quadrature}
\end{equation}
where $0.886$ is the full width at half maximum of $\mathrm{sinc}^2$, so $c_W/W$ is the width of the spectral window of a sharp box of side $W$.
The formula is exact in the two limits where one width dominates the other.
In between it systematically underestimates the combined width, because the Lorentzian wings make the true width of the convolution closer to the sum of the two widths than to the square-root combination of Eq.~\eqref{eq:sm1_quadrature}.
The offset between the dashed curve and the data in Fig.~\ref{fig:onshell_schematic}(c) at the larger boxes is attributed to this effect, and the smallest box, where the window dominates, falls slightly below the curve because the half-maximum reading of a window-dominated profile is biased low.

\FloatBarrier
\section{\texorpdfstring{Field-error metrics: spatial structure and numerical verification}{Field-error metrics: spatial structure and numerical verification}}
\label{sec:sm_metrics}

This section examines the reconstruction error in detail and verifies the error-metric relationships numerically.
The metrics are defined in Eqs.~(\ref{eq:eps_L2})--(\ref{eq:eps_A}) of the main text, and the exact and statistical relationships among them are derived in Appendix~\ref{app:metrics} of the main text.
We use the notation of that appendix, with $E_c = c\,PE_0$ the reconstruction and $P_{\rm out}$ the discarded fraction of the spectral power, cf.\ Eq.~\eqref{eq:sm2_chain} of the main text.
Three options for the power-restoring factor $c$ are compared in Sec.~\ref{sec:sm_metricverify} below.

\subsection{Spatially resolved error: the boundary effect}
\label{sec:sm_metriczones}

The relationships above are spatially uniform only under periodic boundary conditions in every direction.
In practice, open boundaries contribute excess error, as shown throughout this work.
This section introduces the mathematical apparatus used to spatially resolve and quantify this error.

\emph{The boundary (Gibbs) zone and the trim.}
A sharp band mask of width $\dkm$ results in a sinc-modulated real-space kernel with envelope $\propto 1/(\dkm\,x)$.
Convolution with this kernel corrupts the field within $\sim 1/\dkm$ of any sharp edge of the domain.
Zero padding does not alleviate the problem, since it replaces the periodic wrap (step) by a field-to-zero step that generates oscillations of a similar magnitude (measured edge-layer errors $0.93$ padded vs $0.90$ unpadded).
The depth-resolved error profiles of Fig.~\ref{fig:sm2_depth3d} below show this confinement directly.
We introduce the following measure of the excess error as a function of a boundary trim of width $t$: $\epsilon_\varphi(t)$ is the metric of Eq.~(\ref{eq:eps_varphi}) of the main text evaluated after a layer of width $t$ is removed from every open boundary of the analysis domain, and $\epsilon_\varphi(\infty)$ is its interior value.
The kernel envelope sets the decay of this measure.
Squaring the $1/(\dkm\,x)$ envelope and integrating beyond the trim gives a tail $\propto 1/(\dkm^2 t)$, and weighting by the intensity at the boundary and normalizing by the domain power gives
\begin{equation}
\epsilon_\varphi(t)-\epsilon_\varphi(\infty) \sim
\frac{1}{\dkm^2\, t}\,\frac{\oint_{\partial\Omega} I\,dS}{\int_{\Omega} I\,dV},
\qquad t \gtrsim 1/\dkm,
\label{eq:sm2_trim}
\end{equation}
so only the boundaries the field actually reaches contribute.
This $1/t$ decay is the dashed guide of Fig.~4(b) of the main text, and the excess saturates toward its finite $t \to 0$ value once the trim falls below the corrupted-layer width.
Interior quantities throughout this work use the trim $t = 3/\dkm$, except at the longitudinal faces of the 3D slabs, treated below.
The mask is isotropic, but the corrupted depth at a given face is set by the spectral width of the stored field in the direction normal to that face.
Near the longitudinal faces of a 3D slab that width is the forward width of the shell, broadened by the slab window, so the 3D trim at those faces uses the measured $\dksf$, cf.~Sec.~\ref{sec:sm_mask}.

\subsection{\texorpdfstring{Power normalization schemes and numerical verification}{Power normalization schemes and numerical verification}}\label{sec:sm_metricverify}

As shown in Appendix~\ref{app:metrics} of the main text, the projection $PE$ removes the $P_{\rm out}$ fraction of the power in the original field.
Here, we consider three specific schemes for restoring it in a reconstruction of the form $c\,PE$, where the real factor $c$ is:
\begin{equation}
c_1 = \frac{\|E\|}{\|PE\|}
\;\;\text{(per realization)},
\qquad
c_2 = \bigl(1 - \overline{P}_{\rm out}\bigr)^{-1/2}
\;\;\text{(disorder averaged)},
\qquad
c_3 = P_{\rm in}(\alpha)^{-1/2}
\;\;\text{(analytic)}.
\label{eq:sm2_schemes}
\end{equation}
$c_1$ is computed from the norms of the specific instance of the field, $c_2$ is computed from the disorder-averaged discarded fraction measured on the ensemble, and $c_3$ is computed from the analytic retained-power relationship $P_{\rm in} = (2/\pi)\arctan\alpha$ in Eq.~\eqref{eq:power_ratio} of the main text, with no input from the simulations.
Below we compare these schemes on the same ensembles, and select one convention ($c_2$) used throughout this work.

\begin{table}[h]
\centering
\caption{Summary of the numerical verification of the error-metrics relationships in Sec.~\ref{sec:sm_metrics}.
Equations numbered (A) are in Appendix A of the main text and those numbered (S) in this Supplemental Material.
The geometries, boundary conditions, and illumination of the four data sets are specified in the text.
All predictions are parameter free.}
\label{tab:sm2_verify}
\begin{tabular}{llll}
\hline
Prediction & Where derived & Measured & Data set \\
\hline
$\arg\langle E_0,E_c\rangle = 0$ & Eq.~(\ref{eq:sm2_overlap}) & $3\times10^{-9}$ rad (single precision) & 2D, $N_{\rm rlz}=20$ \\
$\epsilon_\varphi = 1-\sqrt{1-P_{\rm out}}$ & Eq.~(\ref{eq:sm2_identity}) & Holds to $6$ digits & 2D, $N_{\rm rlz}=20$ \\
$\epsilon_A$ elliptic formula & Eq.~(\ref{eq:sm2_phipred}) & Deviation at or below $10^{-3}$ at every $\alpha$ & 2D, $N_{\rm rlz}=20$ \\
$\epsilon_\varphi/\epsilon_A \to 2$ & Eq.~(\ref{eq:sm2_phipred}) & $1.8$--$2.5$, within $10\%$ of $2$ at $\alpha \ge 5$ & 2D, $N_{\rm rlz}=20$ \\
$\epsilon_{L2} = \sqrt{P_{\rm out}}$, 3D, all $(\gamma, g)$ & Eq.~(\ref{eq:sm2_L2}) & Ratio $0.81$--$1.07$ (mean $0.93$) & 3D crossover \\
$\epsilon_{L2} = \sqrt{P_{\rm out}}$, 3D, $\gamma = 10.4$ & Eq.~(\ref{eq:sm2_L2}) & Ratio $0.95$--$1.08$, six masks & 3D deep slab \\
Depth-flat interior error & Eq.~(\ref{eq:sm2_depthres}) & Flat to $0.4$--$10\%$ outside boundary layers & 3D crossover \\
$\epsilon_S = \sqrt2\,\epsilon_{L2}$ pointwise & Eq.~(\ref{eq:sm2_product}) & $\to \sqrt2$ at $\alpha \gtrsim 8$, depth independent & 2D, $100$ rlz \\
$\int X_i\,dV = 0$ & Eq.~(\ref{eq:sm2_orthogonality}) & $<6\times10^{-16}$, every realization & 2D, $100$ rlz \\
Cross terms carry the raw error & Eq.~(\ref{eq:sm2_decomp}) & $92$--$97\%$ of error power at $\alpha = 5$--$12$ & 2D, $100$ rlz \\
Integrated cross product & Eq.~(\ref{eq:sm2_orthogonality}) & $3.2\%$ where pointwise is $72\%$ & 2D mode pair \\
Boundary-layer confinement, $1/t$ tail & Eq.~(\ref{eq:sm2_trim}) & Observed, Fig.~4(b) of the main text & 2D, $N_{\rm rlz}=40$ \\
Kernel invariance to excitation mode & --- & The $25$-pair set and a single pair agree to within $3\%$ & 2D, $100$ rlz \\
\hline
\end{tabular}
\end{table}

\begin{figure}[t!]
\centering
\includegraphics[width=0.88\textwidth]{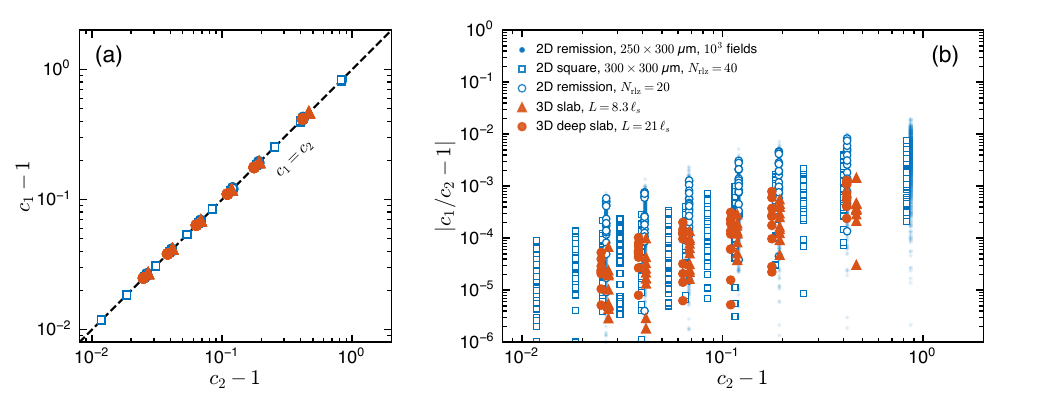}
\caption{The per-realization power-restoring factor $c_1 = (1-P_{\rm out})^{-1/2}$ vs the ensemble constant $c_2 = (1-\overline{P}_{\rm out})^{-1/2}$ of Eq.~(\ref{eq:sm2_schemes}), with every ensemble and every mask width shown together.
(a) All realizations fall on the line $c_1 = c_2$ (dashed).
(b) The relative deviation $|c_1/c_2 - 1|$ stays below $2\times10^{-3}$ in 3D, at or below the $10^{-2}$ level for the square ensemble and the twenty-realization remission ensemble, and reaches $2\times10^{-2}$ only at the narrowest mask of the $10^3$-field remission ensemble.
The scatter shrinks with the field volume, cf.\ the $300\times300\um$ square ensemble and the $250\times300\um$ remission ensemble of the same medium.}
\label{fig:sm2_poutdist}
\end{figure}

Four independent data sets test every relation above, as summarized in Table~\ref{tab:sm2_verify}.
The first is the $N_{\rm rlz} = 20$ realization ensemble of the 2D remission system of the main text ($250 \times 300\um$, finite-size input, $k\ls \approx 90$, the remission row of Table~\ref{tab:sm3_2d}).
The second is a $1000$-realization 2D ensemble in the remission geometry of Sec.~III\,A of the main text with the Level~1 statistics of Table~\ref{tab:sm3_2d} ($250 \times 300\um$, absorbing boundaries, five input modes per realization), of which $100$ realizations also carry the five output-mode fields for the product tests and forward/backward pairs for the sensitivity tests.
The third comprises the three-dimensional (3D) crossover calculations of Table~\ref{tab:sm3_3d} (periodic transverse boundaries, plane-wave illumination), eight $(\gamma, g)$ systems with ten realizations each spanning quasi-ballistic to diffusive transport ($\gamma = 0.41$--$4.11$, $k\ls = 58$--$2985$), extended by the deep-slab calculations to $\gamma = 10.4$.
The fourth is the $40$-realization square ensemble of Sec.~III\,A of the main text, which provides the trim calibration of Fig.~4(b).
The per-realization factor $c_1$ is compared with the ensemble constant $c_2$ in Fig.~\ref{fig:sm2_poutdist}, for every ensemble and mask width.
The agreement justifies the fixed $c_2$ normalization used throughout.

\begin{figure}[t!]
\centering
\includegraphics[width=0.88\textwidth]{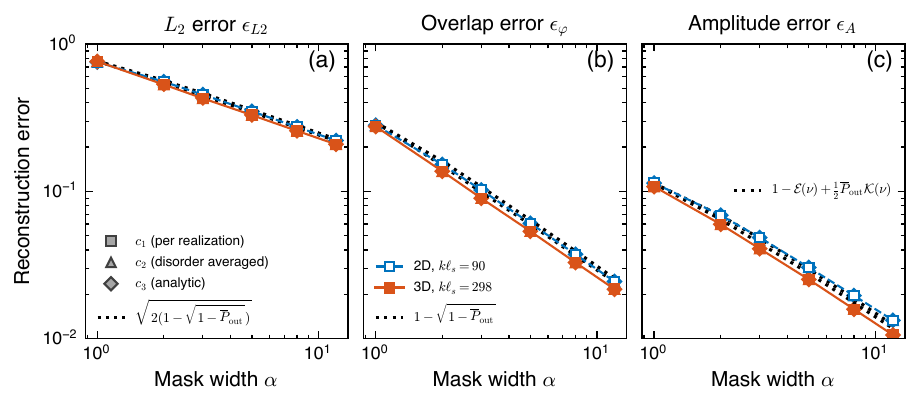}
\caption{Field-error metrics vs the normalized mask width $\alpha = \dkm\,\ls$, in both dimensionalities.
Color denotes the dimensionality: blue open symbols and dashed lines, the 2D $N_{\rm rlz} = 20$ ensemble ($k\ls = 90$, dispersion-shaped mask). Orange filled symbols and solid lines are the 3D deep slab ($k\ls = 298$, $L = 21\,\ls$, top-hat disorder, isotropic masks, co-polarized component, ten realizations).
Marker shape denotes the renormalization: square, triangle, and diamond for the $c_1$, $c_2$, and $c_3$ schemes of Eq.~(\ref{eq:sm2_schemes}).
The three schemes coincide within the marker size at every $\alpha$. Error bars are realization spreads, below $0.9\%$ of the mean over the twenty 2D realizations and $0.25\%$ over the ten 3D ones, and covered by the symbols.
The panels follow the metric order of Table~\ref{tab:sm2_metrics} of the main text: (a) the relative $L_2$ error, (b) the overlap error $\epsilon_\varphi$, (c) the amplitude metric $\epsilon_A$.
Dotted lines: the parameter-free predictions of Table~\ref{tab:sm2_metrics} of the main text, evaluated at each ensemble's measured $\overline{P}_{\rm out}$.}
\label{fig:sm2_metrics}
\end{figure}

Figure~\ref{fig:sm2_metrics} compares the three schemes of Eq.~(\ref{eq:sm2_schemes}) on the 2D ensemble and the 3D deep slab.
In the $L_2$ panel the three options are indistinguishable.
In the overlap and amplitude panels the three nested markers coincide identically, because both metrics are invariant under any scalar factor applied to the reconstruction.
The 2D and 3D curves differ only through their retained-power laws $\overline{P}_{\rm out}(\alpha)$, since each metric is a function of $\overline{P}_{\rm out}$ alone.


\section{Field compression tests: from quasi-ballistic to diffusive transport}\label{sec:sm_crossover}

This section presents a comprehensive set of tests of field compression across a range of systems spanning the crossover from the quasi-ballistic to the diffusive regime of wave transport.
The system parameters are given in Tables~\ref{tab:sm3_2d} and \ref{tab:sm3_3d}.
In 2D they span $\gamma = 2.3$--$18.6$, and in 3D they span $\gamma = 0.41$--$10.4$ and two disorder classes with $k\ls = 58$--$2985$.

\begin{figure}[t!]
\centering
\includegraphics{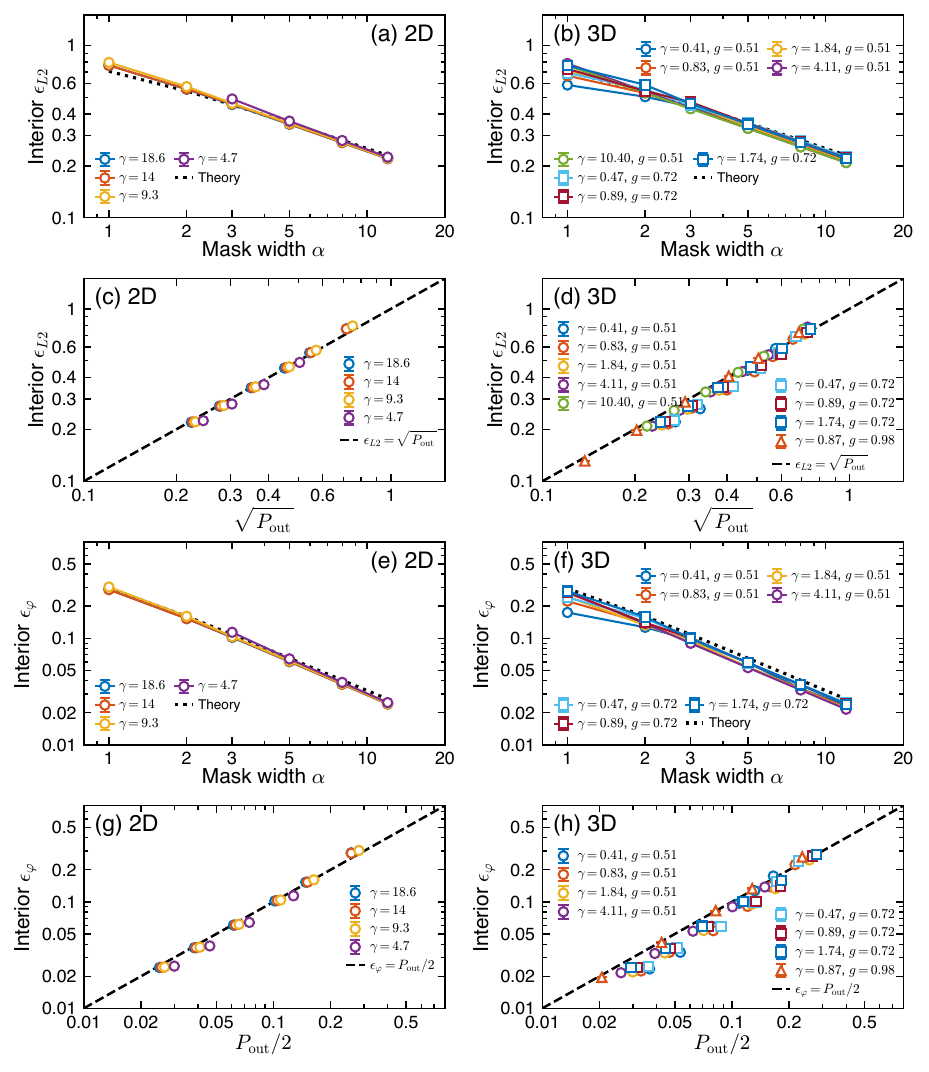}
\caption{Reconstruction fidelity across the crossover into the diffusive regime, in 2D (left column, the fixed-medium thickness sequence (Sequence 2) of the Level~1 medium of Table~\ref{tab:sm3_2d}, $k\ls \approx 90$, $N_{\rm rlz} = 10$ per thickness) and 3D (right column, the systems of Table~\ref{tab:sm3_3d}, co-polarized component, ten realizations per system).
Error bars are smaller than the markers.
(a, b) Interior $\epsilon_{L2}$ vs the mask width $\alpha = \dkm\,\ls$.
The dotted curve in each panel is the universal prediction, obtained by evaluating Eq.~\eqref{eq:power_ratio} of the main text at each $\alpha$ and inserting the result into Eq.~\eqref{eq:metric_identities} of the main text.
The measured per-system predictions $\sqrt{\overline{P}_{\rm out}}$ spread system to system by $4$--$8\%$ in both dimensionalities due to the finite-box spectral window.
2D points for systems where the trim $t = 3/\dkm$ leaves no interior (all of $\gamma = 2.3$ and $\alpha \le 2$ of $\gamma = 4.7$) are not shown.
(c, d) The same data vs $\sqrt{P_{\rm out}}$, collapsing onto the diagonal in both dimensionalities.
(e--h) The same two tests for the overlap error $\epsilon_\varphi$ vs its small-$P_{\rm out}$ prediction $\epsilon_\varphi = P_{\rm out}/2$, in both dimensionalities.
The $g = 0.98$ medium appears only in the collapse panels (d, h), and the deep slab, for which $\epsilon_\varphi$ was not stored, only in (b, d).}
\label{fig:sm2_rd3d}
\end{figure}

\begin{figure}[t!]
\centering
\includegraphics{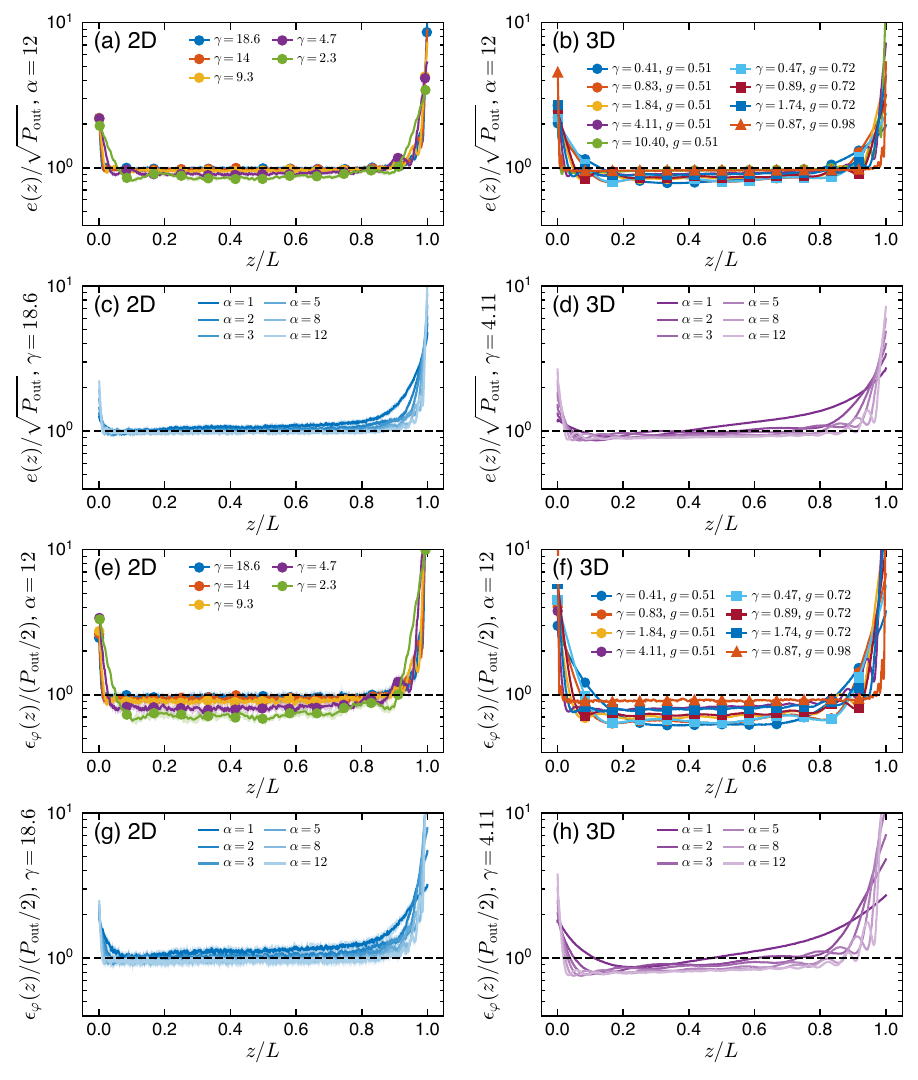}
\caption{Depth-resolved reconstruction error $e(z) = \epsilon_{L2}(z)$, Eq.~(\ref{eq:sm2_depthres}), in 2D (left column, the fixed-medium thickness sequence (Sequence 2), $\gamma = 2.3$--$18.6$ at $k\ls \approx 90$) and 3D (right column, the systems of Table~\ref{tab:sm3_3d}, co-polarized component).
(a, b) The local error normalized by $\sqrt{P_{\rm out}}$ for the $\alpha = 12$ mask, vs depth normalized to the thickness.
(c, d) The same for the mask sequence $\alpha = 1$--$12$ in the diffusive system ($\gamma = 18.6$ in 2D and $\gamma = 4.11$ in 3D).
(e--h) The phase-sensitive error $\epsilon_\varphi$ compared to the $P_{\rm out}/2$ prediction.
The interior is flat in every panel. For $e(z)$ it sits at the predicted level in the diffusive systems and below it in the thinnest 2D systems, while $\epsilon_\varphi$ sits below the small-$P_{\rm out}$ prediction throughout, by up to a quarter, in both dimensionalities. The deviations at the open faces are boundary layers of thickness $\propto 1/\dkm$. Shaded bands are the realization spreads, plus and minus one standard deviation across the realizations of each condition, ten in 2D and in the 3D $e(z)$ panels and three to ten in the 3D $\epsilon_\varphi$ panels. They stay below $2\%$ of the mean in 3D, except $3.4\%$ for the $g = 0.98$ medium, and within $5$--$13\%$ in 2D. The deep slab, for which $\epsilon_\varphi$ was not stored, appears in (b) only.}
\label{fig:sm2_depth3d}
\end{figure}

The interior error $\epsilon_{L2}$ vs the dimensionless mask width $\alpha$ is measured for all 2D and 3D systems at six mask widths, see Fig.~\ref{fig:sm2_rd3d}.
The 2D systems form a sequence of simulations of media with the same type of disorder (Level~1, $k\ls \approx 90$) and increasing thickness $L$, see Table~\ref{tab:sm3_2d}.
These correspond to $\gamma = L(1-g)/\ls = 2.3$--$18.6$, with ten realizations per thickness computed for statistical averaging.
The 3D systems are the crossover set of Table~\ref{tab:sm3_3d}, eight $(\gamma, g)$ systems with ten realizations each, spanning two disorder classes with $k\ls = 58$--$2985$ and extended by the deep-slab calculations to $\gamma = 10.4$.
The curves are nearly indistinguishable across the tested range in both dimensionalities.
Replotted vs $\sqrt{P_{\rm out}}$, the entire 2D and 3D data set collapses onto the relationship $\epsilon_{L2} = \sqrt{P_{\rm out}}$ of Eq.~(\ref{eq:sm2_L2}) of the main text, with the measured ratio between $0.81$ and $1.07$ (mean $0.93$) over the eight crossover systems and all masks, and between $0.92$ and $1.07$ in 2D.
At $\gamma = 2.3$, the system is too small and the trim $t = 3/\dkm$ leaves no interior at any tested mask width, so this system appears only in the depth profiles of Fig.~\ref{fig:sm2_depth3d}(a,e). Those profiles lie below the global-$P_{\rm out}$ level, consistent with the discarded power concentrating at the faces at small $\gamma$.
This result confirms that the compression error is indeed set by the discarded spectral power and does not depend on the transport regime.

At small mask widths the universal dotted prediction of Figs.~\ref{fig:sm2_rd3d}(a,b,e,f) underestimates the measured error, most visibly in 2D.
The deviation arises because the nominal $\alpha$ overstates the mask width in units of the measured shell width.
Indeed, $\alpha = \dkm\,\ls \equiv \dkm/\dks$, whereas in a finite-size system $\dksf > \dks$, cf.\ Sec.~\ref{sec:sm1_estimator}.
Therefore, the effective $\alpha$ is in fact smaller, which increases the theoretical prediction for the error and explains the discrepancy.
Restoration of the agreement can be seen directly in panels (c,d,g,h), since the same data collapse onto the predictions when plotted vs the measured $P_{\rm out}$.
Figure~\ref{fig:sm2_depth3d} presents the depth-resolved error
\begin{equation}
\epsilon_{L2}(z) = \frac{\|c_2 P E - E\|_z}{\|E\|_z},
\label{eq:sm2_depthres}
\end{equation}
where $\|\cdot\|_z$ denotes the $L_2$ norm of Eq.~(\ref{eq:eps_L2}) of the main text restricted to the transverse plane at depth $z$. The depth-resolved overlap error $\epsilon_\varphi(z)$ is defined analogously.
Normalized by $\sqrt{P_{\rm out}}$, this error stays at or below one throughout the bulk of the system, with deviations confined to boundary layers of thickness $\propto 1/\dkm$, which can be reduced by increasing the parameter $\alpha$ (wider mask).
The depolarized power fraction, defined in Table~\ref{tab:sm3_3d}, rises to $0.60$ at $\gamma = 10.4$, so the relationship is checked per polarization component rather than assumed.
At $\alpha = 3$ across the crossover grid, $\epsilon_{L2}/\sqrt{P_{\rm out}} = 0.88$--$1.01$ for the co-polarized component, $0.92$--$1.01$ for the cross-polarized transverse one, and $0.92$--$1.02$ for the longitudinal one.
The $y$- and $z$-polarized field components exhibit statistically the same level of errors as the co-polarized component reported in Figs.~\ref{fig:sm2_rd3d}(b,d,f,h) and \ref{fig:sm2_depth3d}(b,d,f,h), and are not shown here.

\FloatBarrier
\section{Mask selection}\label{sec:sm_mask}

In this section we determine the required mask width from direct measurements across the quasi-ballistic to diffusive crossover and conclude that the isotropic mask is the optimum default choice, while the specific choice of its dimensionless width $\alpha = \dkm\,\ls$ determines the reconstruction fidelity.

\begin{figure}[t!]
\centering
\includegraphics{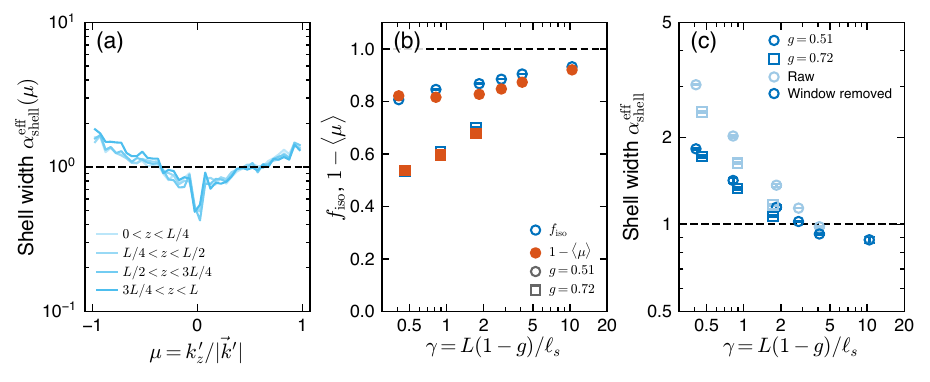}
\caption{Integral measures of the shell support.
(a) The per-angle shell width $\ash(\mu)$, cf.\ Eq.~(\ref{eq:sm_veff}), in the 3D system with $\gamma = 10.4$, cf.\ Table~\ref{tab:sm3_3d}, with the field windowed into four equal depth intervals.
Panels (b,c) report data for 3D systems of increasing transport depth ($\gamma = 0.41$--$10.4$, Table~\ref{tab:sm3_3d}), spanning the quasi-ballistic to diffusive crossover.
(b) The isotropy fraction $f_{\rm iso}$ and $1 - \langle\mu\rangle$, cf.\ Eq.~(\ref{eq:sm_fiso}), with circles the $g = 0.51$ systems and squares $g = 0.72$, both trending toward unity as the shell isotropizes with increasing $\gamma$.
(c) The effective shell width $\ash$ of Eq.~(\ref{eq:sm1_areas}).
The dashed line marks the fully developed shell, $\ash = 1$.
Pale symbols are the raw $\ash$ and full-color symbols have the finite-$L$ window broadening removed, see text.
Sample-to-sample standard deviation is smaller than the markers.}
\label{fig:sm_shellvol}
\end{figure}

We first introduce an integral measure of the shell width.
Let $\mathcal{S}(k', \mu)$ be the spectral density at radial wavenumber $k' = |\vec k'|$ and direction $\mu = k_z'/k'$, with the ballistic component excluded, which for plane-wave excitation is concentrated along the incidence direction $\mu=1$.
$\bar{\mathcal{S}}(k') = \tfrac{1}{2}\int_{-1}^{1} \mathcal{S}(k', \mu)\, d\mu$ is its angular average.
The spectral densities are computed for each disorder realization, and Fig.~\ref{fig:sm_shellvol} reports the means over realizations.
The effective shell width $\ash$ is the equivalent width of $\bar{\mathcal{S}}(k')$ introduced in Eq.~(\ref{eq:sm1_areas}), and its per-angle version is
\begin{equation}
\ash(\mu) = \frac{2}{\pi}\,\ls\, \frac{\int \mathcal{S}(k', \mu)\, dk'}{\max_{k'} \mathcal{S}(k', \mu)},
\label{eq:sm_veff}
\end{equation}
per-direction widths obtained by integration.

We quantify the anisotropy based on the angular spectral density $\chi(\mu) = \int \mathcal{S}(k', \mu)\, dk'$.
When $\chi(\mu)$ is expanded over the Legendre polynomials $P_l(\mu)$, the isotropy fraction is the share of its power carried by the isotropic $l = 0$ term,
\begin{equation}
f_{\rm iso} = \frac{\chi_0^2}{\sum_{l} \chi_l^2},
\qquad
\chi_l = \sqrt{\tfrac{2l+1}{2}} \int_{-1}^{1} \chi(\mu)\, P_l(\mu)\, d\mu,
\qquad
\langle\mu\rangle = \frac{\int_{-1}^{1} \chi(\mu)\,\mu\, d\mu}{\int_{-1}^{1} \chi(\mu)\, d\mu},
\label{eq:sm_fiso}
\end{equation}
which equals one for an isotropic shell.
The first moment $\langle\mu\rangle$ also measures the net forward bias of the scattered power, the field-level analog of the single-scattering anisotropy $g$. 
We use it as a complementary parameter to test shell isotropization.

Figure~\ref{fig:sm_shellvol}(a) resolves the shell width in depth in the thickest system, $\gamma = 10.4$.
The four quarters of the depth give the same $\ash(\mu)$, so in a long enough system the shell width does not depend on depth, as asserted in Sec.~\ref{sec:sm_rte}.
Figure~\ref{fig:sm_shellvol}(b) shows the isotropization with increasing $\gamma$.
The isotropy fraction of Eq.~(\ref{eq:sm_fiso}) approaches unity and the mean cosine approaches zero as $\gamma$ increases beyond unity.
In systems with $\gamma < 1$, the entire anisotropic shell can be captured by choosing $\alpha \gtrsim \ash$, cf.\ Fig.~\ref{fig:sm_shellvol}(c).
Figure~\ref{fig:sm_shellvol}(c) shows $\ash(\gamma)$ for the top-hat systems of the crossover set.
In shorter systems the finite-$L$ broadening of Sec.~\ref{sec:sm_media} inflates the raw width (pale symbols).
Removing this finite-size effect with the known window model (full color) shows that the convergence $\ash \to 1$ happens even faster than the raw data suggest.
An isotropic mask with $\alpha \gtrsim \ash$ is therefore sufficient to cover the measured support in every regime we tested.
The forward-cone mask applied to the quasi-ballistic media in Sec.~\ref{sec:sm_benchres} of the main text is chosen in the same way, from the ensemble rather than from the individual realization.
In the most forward-scattering medium of Table~\ref{tab:sm3_3d}, the Gaussian medium with $\ell_c = 2\um$, the direction $\mu_{\rm min}$ enclosing $99.5\%$ of the scattered shell power lies between $0.925$ and $0.930$ in each of the five realizations tested. This is a stricter test than the $\ell_c = 1\um$ medium of Fig.~\ref{fig:sm6_gamma} of the main text.
A cone edge chosen at or below this interval therefore covers every realization, and the cone built from the ensemble value retains no less than the targeted $99.5\%$ of the ring power in each of them.
When the support has not been measured in advance, a deliberately generous edge, set from a conservative estimate of the scattering anisotropy, remains sample independent and forgoes only part of the compression gain.

\FloatBarrier
\section{Compression tests of second-order quantities and their coarse graining}\label{sec:sm_cg}

\subsection{Reconstruction of the sensitivity kernel across the crossover}\label{sec:sm_cg_kernel}

We consider the transmission-sensitivity kernel, Eq.~(\ref{eq:sm4_kernel}) of the main text, which captures the interference of two distinct but correlated fields in the same system, and test the fidelity of its reconstruction in both 2D and 3D systems in different transport regimes.

\begin{figure}[t!]
\centering
\includegraphics{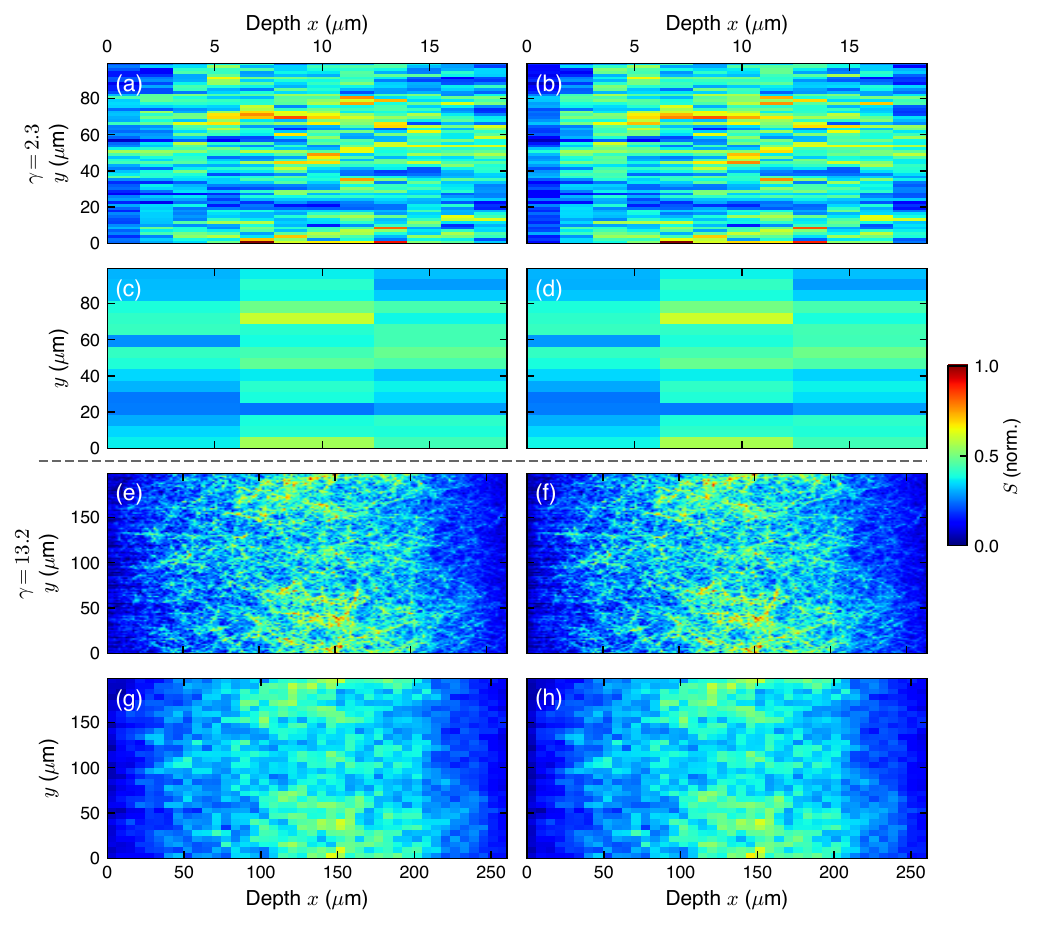}
\caption{Single-realization 2D sensitivity maps in two transport regimes.
Panels (a)--(d) show the $\gamma = 2.3$ member of the fixed-medium thickness sequence of Table~\ref{tab:sm3_2d} (Sequence 2), with $k\ls \approx 90$ and $L = 18.75\um$.
Panels (e)--(h) show the diffusive system, the Pair medium of Table~\ref{tab:sm3_2d}, with filling fraction $0.033$, $\ls = 18.8\um$, $k\ls \approx 221$, $L = 265\um$, and $\gamma = 13.2$.
The maps are computed via Eq.~(\ref{eq:sm4_kernel}) using an isotropic mask with the same $\alpha = 12$ for both regimes.
Maps computed using raw fields are on the left and those using compressed fields (via convolution) on the right.
(a,b,e,f) are coarse-grained at $L_c = \lambda_0$ cells and (c,d,g,h) at $4\lambda_0$.
All four panels of a (top or bottom) block are displayed in the same units, with a single color scale spanning zero to the maximum of the direct (no-compression) computation.
The two coarsening cells match the arrow-marked cell sizes in Fig.~\ref{fig:sm5_sens2d}(a).}
\label{fig:sm4_maps2d}
\end{figure}

\begin{figure}[t!]
\centering
\includegraphics{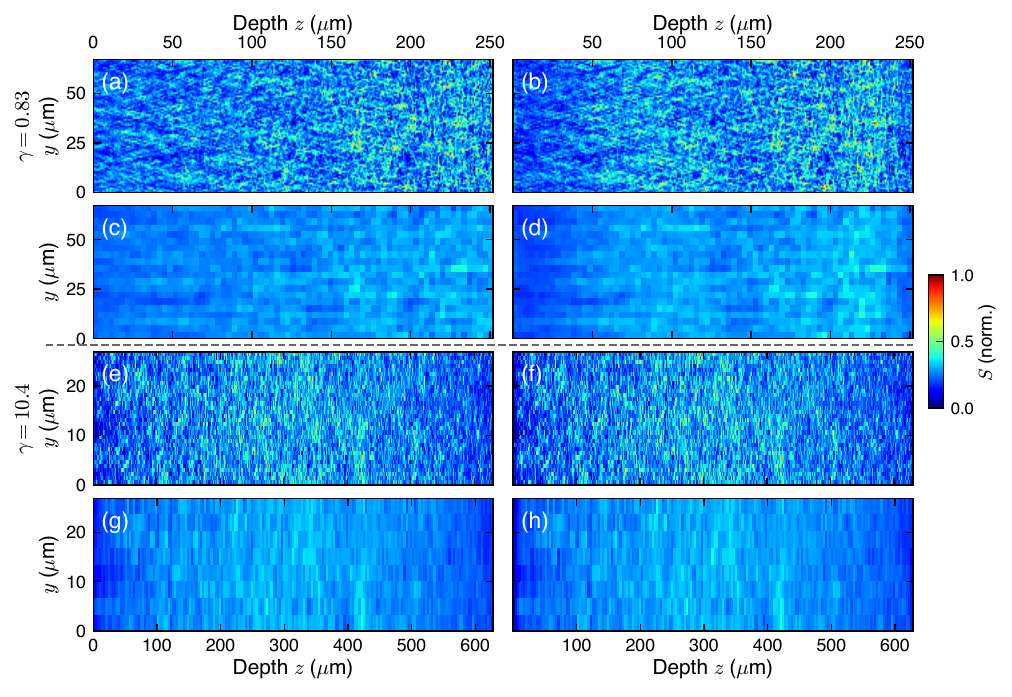}
\caption{Single-realization (section of) 3D sensitivity maps $S(x_0,y,z)$ of Eq.~(\ref{eq:sm4_kernel}) in two transport regimes, shown in the mid-$x$ plane.
Panels (a)--(d) show the quasi-ballistic system with $\gamma = 0.83$ and $k\ls = 1497$, the $\ls = 150\um$ system of Table~\ref{tab:sm3_3d}.
Panels (e)--(h) show the diffusive deep-slab system of Table~\ref{tab:sm3_3d}, with $\gamma = 10.4$ and $k\ls = 298$.
The maps are computed via Eq.~(\ref{eq:sm4_kernel}) using an isotropic mask with the same $\alpha = 12$ for both regimes.
Maps computed using raw fields are on the left and those using compressed fields (via convolution) on the right.
(a,b,e,f) are coarse-grained at $L_c = \lambda_0$ cells and (c,d,g,h) at $4\lambda_0$.
All four panels of a (top or bottom) block are displayed in the same units, with a single color scale spanning zero to the maximum of the direct (no-compression) computation.
The two coarsening cells match the arrow-marked cell sizes in Fig.~\ref{fig:sm5_sens2d}(b).}
\label{fig:sm4_maps3d}
\end{figure}

The forward and adjoint fields are constructed as in Sec.~\ref{sec:3d_secondorder} of the main text.
This single adjoint solve replaces $N_{\rm ch}$ per-channel solves, where $N_{\rm ch}$ is the number of propagating channels of the exit surface.
In 2D, $N_{\rm ch} = kW/\pi \approx 363$ and $743$ for the two blocks of Fig.~\ref{fig:sm4_maps2d}.
In 3D, $N_{\rm ch} = k^2 A/4\pi \approx 3.6\times10^4$ per polarization for the exit plane of area $A$ of the quasi-ballistic system in Fig.~\ref{fig:sm4_maps3d}, and $5.7\times10^3$ for the deep slab.

The compressed sensitivity kernel in Eq.~(\ref{eq:sm4_kernel}) is evaluated using the fields compressed with the same isotropic mask of dimensionless width $\alpha = 12$ in both regimes and in both dimensionalities.
Power-restoring renormalization is accomplished by the disorder-averaged constant $c_2$ of Sec.~\ref{sec:sm_metricverify}.
Figures~\ref{fig:sm4_maps2d} and \ref{fig:sm4_maps3d} show the object itself in both dimensionalities, one quasi-ballistic and one diffusive block each.
The figures compare sensitivity maps computed directly (panels in the left column) with those obtained via convolution of the compressed fields, cf.\ Eq.~(\ref{eq:intensity_convolution}) of the main text (panels in the right column).

Panels (a,b,e,f) in Figs.~\ref{fig:sm4_maps2d} and \ref{fig:sm4_maps3d} show accurate reproduction of the wavelength-scale and the long-range features in the sensitivity map. 
The reconstruction errors are quantified in Fig.~\ref{fig:sm5_sens2d}.
The fidelity of the reconstruction can be traced to the fact that the retained spectral support of the sensitivity extends to $|\vec q| = 2k$ and beyond, cf.\ Fig.~\ref{fig:sm5_sens2d}(g,h).


\subsection{Coarse graining}\label{sec:sm_cg_cells}

Sensitivities enter observables through integrals over extended perturbations.
The relevant error is therefore the coarse-grained one, Eq.~(\ref{eq:sm5_epss}) of the main text.
Coarse graining reduces the wavelength-scale error with increasing cell size until it levels off at the coarsest cells.

Figure~\ref{fig:sm5_sens2d}(a,b) shows the measured $\epsilon_S(L_c)$ for masks with $\alpha = 1$--$12$ in 2D and 3D.
Figure~\ref{fig:sm5_sens2d}(c,d) plots the same errors vs the compression ratio $C$, the number of grid samples divided by the number of retained shell coefficients, cf.\ Eq.~(\ref{eq:compression}) of the main text.
In the 3D $\gamma = 4.11$ system at $\alpha = 3$ (compression ratio $C = 210$), the error falls from $0.52$ at the native sampling to $0.063$ after coarse graining.
At $\alpha = 12$ (compression ratio $C = 52$) the coarse-grained error is $0.011$.
Figure~\ref{fig:sm5_sens2d}(e,f) shows the native and coarse-grained errors at $\alpha = 12$ as system thickness $\gamma$ is increased across the crossover into the diffusive regime of wave transport.
The native error depends only weakly on $\gamma$, while the coarse-grained error falls across the crossover, cf.\ Fig.~\ref{fig:sm5_sens2d}(e,f).
The gain of coarse graining grows with the sample thickness at fixed cell size.
In 2D it rises from $49$ to $132$ between $\gamma = 4.9$ and $20.9$ at the same cell, and in 3D from $23$ to $51$ between $\gamma = 4.11$ and $10.4$ in the same medium, cf.\ Fig.~\ref{fig:sm5_sens2d}(e,f).
The thinner systems reach smaller gains, in part because their cells are limited by the box size in 2D and are a smaller fraction of $\ls$ in 3D.

Three conditions are required for coarse graining to be effective.
First, the mask must cover the support of the field, $\alpha\gtrsim\ash$, cf.\ Sec.~\ref{sec:sm_mask}.
Second, the stored box must be large enough to resolve the mask in $k$-space, $L/\ls \gtrsim 2\pi/\alpha$ with $L$ the depth.
The 3D system with $k\ls \approx 3000$ is the closest to this limit, $L/\ls = 0.84$, and its retained power at $\alpha = 12$ is $0.92$ instead of the $0.93$--$0.95$ of the other 3D systems.
Third, the cell size $L_c$ must exceed the correlation length of the error, $\ls/\alpha = 1/\dkm$, because the compression error of the product is concentrated at spatial frequencies at and above the mask edge.

\subsection{Spectral support of the sensitivity kernel}\label{sec:sm_specsupport}

\begin{figure}[p]
\centering
\includegraphics{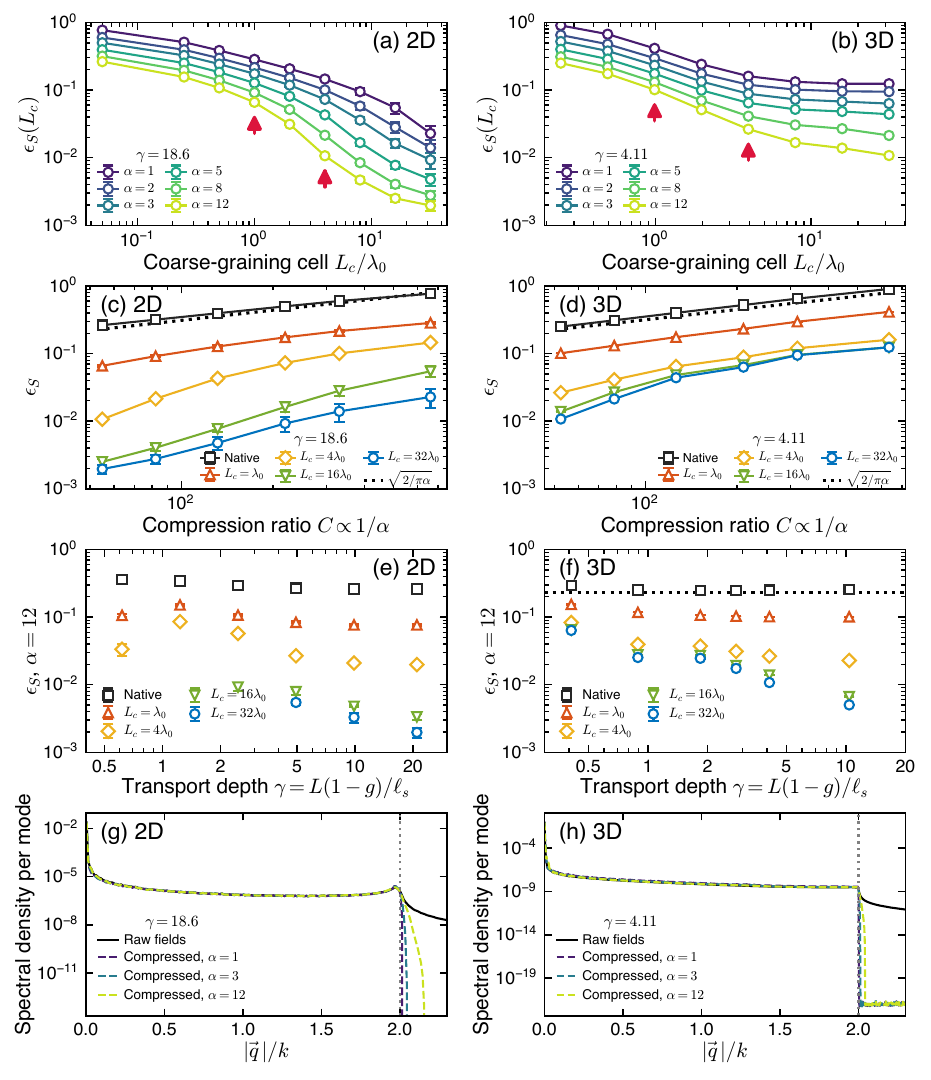}
\caption{Compression error of the sensitivity kernel defined in Eq.~(\ref{eq:sm4_kernel}) of the main text, for isotropic masks and the disorder-averaged $c_2$ normalization, cf.\ Sec.~\ref{sec:sm_metricverify}.
Panels in the left column show 2D data and panels in the right column show 3D data.
(a, b) $\epsilon_S(L_c)$ of Eq.~(\ref{eq:sm5_epss}) of the main text for masks with $\alpha = 1$--$12$.
The 2D system is the plane-wave Level~1 medium of Table~\ref{tab:sm3_2d}, with $k\ls \approx 90$ and $\gamma = 18.6$, mean $\pm$ std over twenty realizations.
The 3D system is the $\gamma = 4.11$ member of Table~\ref{tab:sm3_3d}, mean $\pm$ std over ten realizations.
Arrows mark the coarse-graining cells corresponding to the maps depicted in Figs.~\ref{fig:sm4_maps2d} and \ref{fig:sm4_maps3d}.
(c, d) $\epsilon_S$ at the native sampling and after coarse graining, vs the compression ratio $C$, cf.\ Eq.~(\ref{eq:compression}) of the main text, for the same systems.
The dotted guide traces the field-level law $\sqrt{2/(\pi\alpha)}$.
(e, f) The native and coarse-grained errors at $\alpha = 12$, for cell sizes up to $L_c = 32\lambda_0$, as functions of the transport depth $\gamma$.
The 2D data use the fixed-medium thickness sequence of Table~\ref{tab:sm3_2d} (Sequence 1) with ten realizations per thickness.
The 3D data are five top-hat systems spanning $\gamma = 0.41$--$4.11$ of Table~\ref{tab:sm3_3d}, extended by the deep slab with $\gamma = 10.4$.
Block averaging retains only whole cells, so the large-cell entries are absent for the thinnest systems in 2D (small $L$).
(g, h) Single-realization radial spectrum of $S$ for the systems of panels (a, b) computed from the raw and compressed fields, in 3D from sinc-interpolated fields, see text, cf.\ Eq.~(\ref{eq:sm5_pspec}).
In both 2D and 3D, the sensitivity kernel obtained from the compressed fields retains its support up to the dotted line, which marks $|\vec q| = 2k$.}
\label{fig:sm5_sens2d}
\label{fig:sm5_sens3d}
\end{figure}

The support of the product of two shell-confined fields in Eq.~(\ref{eq:sm4_kernel}) is not limited to low spatial frequencies.
The convolution of two shells extends up to $|\vec q| = 2(k+\dkm/2)$, which is greater than $2k$ irrespective of the mask width $\dkm$.
Therefore, the kernel of Eq.~(\ref{eq:sm4_kernel}) carries structure down to half the wavelength by construction.
To verify that this structure survives compression, we measure the radial spectral density
\begin{equation}
P_S(q) = \bigl\langle |\tilde S(\vec q)|^2 \bigr\rangle_{|\vec q| = q},
\label{eq:sm5_pspec}
\end{equation}
where $\tilde S(\vec q)$ is the Fourier transform of $S(\vec r)$ and the average runs over the Fourier modes with $|\vec q| = q$.
Figure~\ref{fig:sm5_sens2d}(g,h) shows $P_S(q)$ computed from the raw and from the compressed fields in 2D and 3D.
In 3D the support edge $2k$ exceeds the grid Nyquist limit $\pi/\Delta x = 1.5k$, so the fields are sinc-interpolated onto a twofold finer grid before the product is formed.
This step is exact for the grid-band-limited fields and makes the spectra alias-free up to $3k$.
In 2D the Nyquist limit is $3.5k$ and no interpolation is needed.
The two spectra coincide across the occupied band at every mask width shown.
The compressed spectrum terminates at its mask-limited edge $2(k + \dkm/2)$ and vanishes beyond it.
In 2D the spectrum carries the ring--ring contact peak at exactly $2k$, which is fully reproduced in the compressed representation.
Multiplication of the shell-projected fields in real space equals the convolution of their shell coefficients, verified to $6\times10^{-33}$ in 3D and to $3\times10^{-31}$ in 2D, in double precision, as the fraction of the product's spectral power beyond the support edge.
The per-mode spectral density of Fig.~\ref{fig:sm5_sens2d}(g,h) peaks toward $q \to 0$.
However, since the number of Fourier modes per radial bin grows as $q^{d-1}$, the total power has a significant contribution from the large-$q$ phase space as well.
Indeed, we compute that the band between $q = k$ and $2k$ carries a large share of the kernel power in the diffusive systems of Fig.~\ref{fig:sm5_sens2d}, $39\%$ at $\gamma = 18.6$ in 2D and $49\%$ at $\gamma = 4.11$ in 3D, and it is fully captured within the OSCAR procedure, reproducing its wavelength-scale content, see Fig.~\ref{fig:sm5_sens2d}(g,h) and Figs.~\ref{fig:sm4_maps2d} and \ref{fig:sm4_maps3d}.

\makeatletter
\setcounter{NAT@ctr}{0}
\let\NAT@origwrout\NAT@wrout
\renewcommand\NAT@wrout[5]{\NAT@origwrout{S#1}{#2}{#3}{#4}{#5}}
\def\@biblabel#1{[S#1]}
\makeatother


\begin{thebibliography}{53}%
\makeatletter
\providecommand \@ifxundefined [1]{%
 \@ifx{#1\undefined}
}%
\providecommand \@ifnum [1]{%
 \ifnum #1\expandafter \@firstoftwo
 \else \expandafter \@secondoftwo
 \fi
}%
\providecommand \@ifx [1]{%
 \ifx #1\expandafter \@firstoftwo
 \else \expandafter \@secondoftwo
 \fi
}%
\providecommand \natexlab [1]{#1}%
\providecommand \enquote  [1]{``#1''}%
\providecommand \bibnamefont  [1]{#1}%
\providecommand \bibfnamefont [1]{#1}%
\providecommand \citenamefont [1]{#1}%
\providecommand \href@noop [0]{\@secondoftwo}%
\providecommand \href [0]{\begingroup \@sanitize@url \@href}%
\providecommand \@href[1]{\@@startlink{#1}\@@href}%
\providecommand \@@href[1]{\endgroup#1\@@endlink}%
\providecommand \@sanitize@url [0]{\catcode `\\12\catcode `\$12\catcode
  `\&12\catcode `\#12\catcode `\^12\catcode `\_12\catcode `\%12\relax}%
\providecommand \@@startlink[1]{}%
\providecommand \@@endlink[0]{}%
\providecommand \url  [0]{\begingroup\@sanitize@url \@url }%
\providecommand \@url [1]{\endgroup\@href {#1}{\urlprefix }}%
\providecommand \urlprefix  [0]{URL }%
\providecommand \Eprint [0]{\href }%
\providecommand \doibase [0]{https://doi.org/}%
\providecommand \selectlanguage [0]{\@gobble}%
\providecommand \bibinfo  [0]{\@secondoftwo}%
\providecommand \bibfield  [0]{\@secondoftwo}%
\providecommand \translation [1]{[#1]}%
\providecommand \BibitemOpen [0]{}%
\providecommand \bibitemStop [0]{}%
\providecommand \bibitemNoStop [0]{.\EOS\space}%
\providecommand \EOS [0]{\spacefactor3000\relax}%
\providecommand \BibitemShut  [1]{\csname bibitem#1\endcsname}%
\let\auto@bib@innerbib\@empty
\bibitem [{\citenamefont {Notaro{\v{s}}}\ \emph {et~al.}(2023)\citenamefont
  {Notaro{\v{s}}}, \citenamefont {Andriulli},\ and\ \citenamefont
  {Bagci}}]{2023_Notarus_Computational_EM}%
  \BibitemOpen
  \bibfield  {author} {\bibinfo {author} {\bibfnamefont {B.~M.}\ \bibnamefont
  {Notaro{\v{s}}}}, \bibinfo {author} {\bibfnamefont {F.~P.}\ \bibnamefont
  {Andriulli}},\ and\ \bibinfo {author} {\bibfnamefont {H.}~\bibnamefont
  {Bagci}},\ }\bibfield  {title} {\bibinfo {title} {Guest editorial: Frontiers
  in computational electromagnetics},\ }\href
  {https://doi.org/10.1109/TAP.2023.3337967} {\bibfield  {journal} {\bibinfo
  {journal} {IEEE Trans. Antennas Propag.}\ }\textbf {\bibinfo {volume} {71}},\
  \bibinfo {pages} {9175} (\bibinfo {year} {2023})}\BibitemShut {NoStop}%
\bibitem [{\citenamefont {Mosk}\ \emph {et~al.}(2012)\citenamefont {Mosk},
  \citenamefont {Lagendijk}, \citenamefont {Lerosey},\ and\ \citenamefont
  {Fink}}]{2012_Mosk_review}%
  \BibitemOpen
  \bibfield  {author} {\bibinfo {author} {\bibfnamefont {A.~P.}\ \bibnamefont
  {Mosk}}, \bibinfo {author} {\bibfnamefont {A.}~\bibnamefont {Lagendijk}},
  \bibinfo {author} {\bibfnamefont {G.}~\bibnamefont {Lerosey}},\ and\ \bibinfo
  {author} {\bibfnamefont {M.}~\bibnamefont {Fink}},\ }\bibfield  {title}
  {\bibinfo {title} {Controlling waves in space and time for imaging and
  focusing in complex media},\ }\href {https://doi.org/10.1038/nphoton.2012.88}
  {\bibfield  {journal} {\bibinfo  {journal} {Nat. Photonics}\ }\textbf
  {\bibinfo {volume} {6}},\ \bibinfo {pages} {283} (\bibinfo {year}
  {2012})}\BibitemShut {NoStop}%
\bibitem [{\citenamefont {Park}\ \emph {et~al.}(2018)\citenamefont {Park},
  \citenamefont {Yu}, \citenamefont {Lee}, \citenamefont {Lai},\ and\
  \citenamefont {Park}}]{2018_Park_perspective}%
  \BibitemOpen
  \bibfield  {author} {\bibinfo {author} {\bibfnamefont {J.-H.}\ \bibnamefont
  {Park}}, \bibinfo {author} {\bibfnamefont {Z.}~\bibnamefont {Yu}}, \bibinfo
  {author} {\bibfnamefont {K.}~\bibnamefont {Lee}}, \bibinfo {author}
  {\bibfnamefont {P.}~\bibnamefont {Lai}},\ and\ \bibinfo {author}
  {\bibfnamefont {Y.}~\bibnamefont {Park}},\ }\bibfield  {title} {\bibinfo
  {title} {Perspective: Wavefront shaping techniques for controlling multiple
  light scattering in biological tissues: Toward in vivo applications},\ }\href
  {https://doi.org/10.1063/1.5033917} {\bibfield  {journal} {\bibinfo
  {journal} {APL Photonics}\ }\textbf {\bibinfo {volume} {3}},\ \bibinfo
  {pages} {100901} (\bibinfo {year} {2018})}\BibitemShut {NoStop}%
\bibitem [{\citenamefont {Gigan}\ \emph {et~al.}(2022)\citenamefont {Gigan},
  \citenamefont {Katz}, \citenamefont {de~Aguiar}, \citenamefont {Andresen},
  \citenamefont {Aubry}, \citenamefont {Bertolotti}, \citenamefont {Bossy},
  \citenamefont {Bouchet}, \citenamefont {Brake}, \citenamefont {Brasselet},
  \citenamefont {Bromberg}, \citenamefont {Cao}, \citenamefont {Chaigne},
  \citenamefont {Cheng}, \citenamefont {Choi}, \citenamefont {Čižmár},
  \citenamefont {Cui}, \citenamefont {Curtis}, \citenamefont {Defienne},
  \citenamefont {Hofer}, \citenamefont {Horisaki}, \citenamefont {Horstmeyer},
  \citenamefont {Ji}, \citenamefont {LaViolette}, \citenamefont {Mertz},
  \citenamefont {Moser}, \citenamefont {Mosk}, \citenamefont {Pégard},
  \citenamefont {Piestun}, \citenamefont {Popoff}, \citenamefont {Phillips},
  \citenamefont {Psaltis}, \citenamefont {Rahmani}, \citenamefont {Rigneault},
  \citenamefont {Rotter}, \citenamefont {Tian}, \citenamefont {Vellekoop},
  \citenamefont {Waller}, \citenamefont {Wang}, \citenamefont {Weber},
  \citenamefont {Xiao}, \citenamefont {Xu}, \citenamefont {Yamilov},
  \citenamefont {Yang},\ and\ \citenamefont {Y{\i}lmaz}}]{2022_Gigan_roadmap}%
  \BibitemOpen
  \bibfield  {author} {\bibinfo {author} {\bibfnamefont {S.}~\bibnamefont
  {Gigan}}, \bibinfo {author} {\bibfnamefont {O.}~\bibnamefont {Katz}},
  \bibinfo {author} {\bibfnamefont {H.~B.}\ \bibnamefont {de~Aguiar}}, \bibinfo
  {author} {\bibfnamefont {E.~R.}\ \bibnamefont {Andresen}}, \bibinfo {author}
  {\bibfnamefont {A.}~\bibnamefont {Aubry}}, \bibinfo {author} {\bibfnamefont
  {J.}~\bibnamefont {Bertolotti}}, \bibinfo {author} {\bibfnamefont
  {E.}~\bibnamefont {Bossy}}, \bibinfo {author} {\bibfnamefont
  {D.}~\bibnamefont {Bouchet}}, \bibinfo {author} {\bibfnamefont
  {J.}~\bibnamefont {Brake}}, \bibinfo {author} {\bibfnamefont
  {S.}~\bibnamefont {Brasselet}}, \bibinfo {author} {\bibfnamefont
  {Y.}~\bibnamefont {Bromberg}}, \bibinfo {author} {\bibfnamefont
  {H.}~\bibnamefont {Cao}}, \bibinfo {author} {\bibfnamefont {T.}~\bibnamefont
  {Chaigne}}, \bibinfo {author} {\bibfnamefont {Z.}~\bibnamefont {Cheng}},
  \bibinfo {author} {\bibfnamefont {W.}~\bibnamefont {Choi}}, \bibinfo {author}
  {\bibfnamefont {T.}~\bibnamefont {Čižmár}}, \bibinfo {author}
  {\bibfnamefont {M.}~\bibnamefont {Cui}}, \bibinfo {author} {\bibfnamefont
  {V.~R.}\ \bibnamefont {Curtis}}, \bibinfo {author} {\bibfnamefont
  {H.}~\bibnamefont {Defienne}}, \bibinfo {author} {\bibfnamefont
  {M.}~\bibnamefont {Hofer}}, \bibinfo {author} {\bibfnamefont
  {R.}~\bibnamefont {Horisaki}}, \bibinfo {author} {\bibfnamefont
  {R.}~\bibnamefont {Horstmeyer}}, \bibinfo {author} {\bibfnamefont
  {N.}~\bibnamefont {Ji}}, \bibinfo {author} {\bibfnamefont {A.~K.}\
  \bibnamefont {LaViolette}}, \bibinfo {author} {\bibfnamefont
  {J.}~\bibnamefont {Mertz}}, \bibinfo {author} {\bibfnamefont
  {C.}~\bibnamefont {Moser}}, \bibinfo {author} {\bibfnamefont {A.~P.}\
  \bibnamefont {Mosk}}, \bibinfo {author} {\bibfnamefont {N.~C.}\ \bibnamefont
  {Pégard}}, \bibinfo {author} {\bibfnamefont {R.}~\bibnamefont {Piestun}},
  \bibinfo {author} {\bibfnamefont {S.}~\bibnamefont {Popoff}}, \bibinfo
  {author} {\bibfnamefont {D.~B.}\ \bibnamefont {Phillips}}, \bibinfo {author}
  {\bibfnamefont {D.}~\bibnamefont {Psaltis}}, \bibinfo {author} {\bibfnamefont
  {B.}~\bibnamefont {Rahmani}}, \bibinfo {author} {\bibfnamefont
  {H.}~\bibnamefont {Rigneault}}, \bibinfo {author} {\bibfnamefont
  {S.}~\bibnamefont {Rotter}}, \bibinfo {author} {\bibfnamefont
  {L.}~\bibnamefont {Tian}}, \bibinfo {author} {\bibfnamefont {I.~M.}\
  \bibnamefont {Vellekoop}}, \bibinfo {author} {\bibfnamefont {L.}~\bibnamefont
  {Waller}}, \bibinfo {author} {\bibfnamefont {L.}~\bibnamefont {Wang}},
  \bibinfo {author} {\bibfnamefont {T.}~\bibnamefont {Weber}}, \bibinfo
  {author} {\bibfnamefont {S.}~\bibnamefont {Xiao}}, \bibinfo {author}
  {\bibfnamefont {C.}~\bibnamefont {Xu}}, \bibinfo {author} {\bibfnamefont
  {A.}~\bibnamefont {Yamilov}}, \bibinfo {author} {\bibfnamefont
  {C.}~\bibnamefont {Yang}},\ and\ \bibinfo {author} {\bibfnamefont
  {H.}~\bibnamefont {Y{\i}lmaz}},\ }\bibfield  {title} {\bibinfo {title}
  {Roadmap on wavefront shaping and deep imaging in complex media},\ }\href
  {https://doi.org/10.1088/2515-7647/ac76f9} {\bibfield  {journal} {\bibinfo
  {journal} {J. Phys.: Photonics}\ }\textbf {\bibinfo {volume} {4}},\ \bibinfo
  {pages} {042501} (\bibinfo {year} {2022})}\BibitemShut {NoStop}%
\bibitem [{\citenamefont {Cao}\ \emph {et~al.}(2022)\citenamefont {Cao},
  \citenamefont {Mosk},\ and\ \citenamefont
  {Rotter}}]{2022_Cao_Mosk_Rotter_review}%
  \BibitemOpen
  \bibfield  {author} {\bibinfo {author} {\bibfnamefont {H.}~\bibnamefont
  {Cao}}, \bibinfo {author} {\bibfnamefont {A.~P.}\ \bibnamefont {Mosk}},\ and\
  \bibinfo {author} {\bibfnamefont {S.}~\bibnamefont {Rotter}},\ }\bibfield
  {title} {\bibinfo {title} {Shaping the propagation of light in complex
  media},\ }\href {https://doi.org/10.1038/s41567-022-01677-x} {\bibfield
  {journal} {\bibinfo  {journal} {Nat. Phys.}\ }\textbf {\bibinfo {volume}
  {18}},\ \bibinfo {pages} {994} (\bibinfo {year} {2022})}\BibitemShut
  {NoStop}%
\bibitem [{\citenamefont {Fercher}\ \emph {et~al.}(2003)\citenamefont
  {Fercher}, \citenamefont {Drexler}, \citenamefont {Hitzenberger},\ and\
  \citenamefont {Lasser}}]{2003_Fercher_OCT}%
  \BibitemOpen
  \bibfield  {author} {\bibinfo {author} {\bibfnamefont {A.~F.}\ \bibnamefont
  {Fercher}}, \bibinfo {author} {\bibfnamefont {W.}~\bibnamefont {Drexler}},
  \bibinfo {author} {\bibfnamefont {C.~K.}\ \bibnamefont {Hitzenberger}},\ and\
  \bibinfo {author} {\bibfnamefont {T.}~\bibnamefont {Lasser}},\ }\bibfield
  {title} {\bibinfo {title} {Optical coherence tomography---principles and
  applications},\ }\href {https://doi.org/10.1088/0034-4885/66/2/204}
  {\bibfield  {journal} {\bibinfo  {journal} {Rep. Prog. Phys.}\ }\textbf
  {\bibinfo {volume} {66}},\ \bibinfo {pages} {239} (\bibinfo {year}
  {2003})}\BibitemShut {NoStop}%
\bibitem [{\citenamefont {Kim}(2010)}]{2010_Kim_holographic}%
  \BibitemOpen
  \bibfield  {author} {\bibinfo {author} {\bibfnamefont {M.~K.}\ \bibnamefont
  {Kim}},\ }\bibfield  {title} {\bibinfo {title} {Principles and techniques of
  digital holographic microscopy},\ }\href {https://doi.org/10.1117/6.0000006}
  {\bibfield  {journal} {\bibinfo  {journal} {SPIE Rev.}\ }\textbf {\bibinfo
  {volume} {1}},\ \bibinfo {pages} {018005} (\bibinfo {year}
  {2010})}\BibitemShut {NoStop}%
\bibitem [{\citenamefont {Kraszewski}\ and\ \citenamefont
  {Pluci\'{n}ski}(2016)}]{2016_Kraszewski_coherent}%
  \BibitemOpen
  \bibfield  {author} {\bibinfo {author} {\bibfnamefont {M.}~\bibnamefont
  {Kraszewski}}\ and\ \bibinfo {author} {\bibfnamefont {J.}~\bibnamefont
  {Pluci\'{n}ski}},\ }\bibfield  {title} {\bibinfo {title} {Coherent-wave
  {M}onte {C}arlo method for simulating light propagation in tissue},\ }in\
  \href {https://doi.org/10.1117/12.2213213} {\emph {\bibinfo {booktitle}
  {Proc. SPIE}}},\ Vol.\ \bibinfo {volume} {9706}\ (\bibinfo {year} {2016})\
  p.\ \bibinfo {pages} {970611}\BibitemShut {NoStop}%
\bibitem [{\citenamefont {McCoy}\ \emph {et~al.}(2021)\citenamefont {McCoy},
  \citenamefont {Shneidman}, \citenamefont {Davis},\ and\ \citenamefont
  {Aizenberg}}]{2021_McCoy_FDTD}%
  \BibitemOpen
  \bibfield  {author} {\bibinfo {author} {\bibfnamefont {D.~E.}\ \bibnamefont
  {McCoy}}, \bibinfo {author} {\bibfnamefont {A.~V.}\ \bibnamefont
  {Shneidman}}, \bibinfo {author} {\bibfnamefont {A.~L.}\ \bibnamefont
  {Davis}},\ and\ \bibinfo {author} {\bibfnamefont {J.}~\bibnamefont
  {Aizenberg}},\ }\bibfield  {title} {\bibinfo {title} {Finite-difference
  time-domain ({FDTD}) optical simulations: A primer for the life sciences and
  bio-inspired engineering},\ }\href
  {https://doi.org/10.1016/j.micron.2021.103160} {\bibfield  {journal}
  {\bibinfo  {journal} {Micron}\ }\textbf {\bibinfo {volume} {151}},\ \bibinfo
  {pages} {103160} (\bibinfo {year} {2021})}\BibitemShut {NoStop}%
\bibitem [{\citenamefont {Virieux}\ and\ \citenamefont
  {Operto}(2009)}]{2009_Virieux_FWI}%
  \BibitemOpen
  \bibfield  {author} {\bibinfo {author} {\bibfnamefont {J.}~\bibnamefont
  {Virieux}}\ and\ \bibinfo {author} {\bibfnamefont {S.}~\bibnamefont
  {Operto}},\ }\bibfield  {title} {\bibinfo {title} {An overview of
  full-waveform inversion in exploration geophysics},\ }\href
  {https://doi.org/10.1190/1.3238367} {\bibfield  {journal} {\bibinfo
  {journal} {Geophysics}\ }\textbf {\bibinfo {volume} {74}},\ \bibinfo {pages}
  {WCC1} (\bibinfo {year} {2009})}\BibitemShut {NoStop}%
\bibitem [{\citenamefont {Liu}\ and\ \citenamefont
  {Gu}(2012)}]{2012_Liu_seismic}%
  \BibitemOpen
  \bibfield  {author} {\bibinfo {author} {\bibfnamefont {Q.}~\bibnamefont
  {Liu}}\ and\ \bibinfo {author} {\bibfnamefont {Y.~J.}\ \bibnamefont {Gu}},\
  }\bibfield  {title} {\bibinfo {title} {Seismic imaging: From classical to
  adjoint tomography},\ }\href {https://doi.org/10.1016/j.tecto.2012.07.006}
  {\bibfield  {journal} {\bibinfo  {journal} {Tectonophysics}\ }\textbf
  {\bibinfo {volume} {566--567}},\ \bibinfo {pages} {31} (\bibinfo {year}
  {2012})}\BibitemShut {NoStop}%
\bibitem [{\citenamefont {Jensen}\ \emph {et~al.}(2011)\citenamefont {Jensen},
  \citenamefont {Kuperman}, \citenamefont {Porter},\ and\ \citenamefont
  {Schmidt}}]{2011_Jensen_ocean}%
  \BibitemOpen
  \bibfield  {author} {\bibinfo {author} {\bibfnamefont {F.~B.}\ \bibnamefont
  {Jensen}}, \bibinfo {author} {\bibfnamefont {W.~A.}\ \bibnamefont
  {Kuperman}}, \bibinfo {author} {\bibfnamefont {M.~B.}\ \bibnamefont
  {Porter}},\ and\ \bibinfo {author} {\bibfnamefont {H.}~\bibnamefont
  {Schmidt}},\ }\href {https://doi.org/10.1007/978-1-4419-8678-8} {\emph
  {\bibinfo {title} {Computational Ocean Acoustics}}},\ \bibinfo {edition}
  {2nd}\ ed.\ (\bibinfo  {publisher} {Springer},\ \bibinfo {year}
  {2011})\BibitemShut {NoStop}%
\bibitem [{\citenamefont {Lin}\ \emph {et~al.}(2012)\citenamefont {Lin},
  \citenamefont {Collis},\ and\ \citenamefont {Duda}}]{2012_Lin_parabolic}%
  \BibitemOpen
  \bibfield  {author} {\bibinfo {author} {\bibfnamefont {Y.-T.}\ \bibnamefont
  {Lin}}, \bibinfo {author} {\bibfnamefont {J.~M.}\ \bibnamefont {Collis}},\
  and\ \bibinfo {author} {\bibfnamefont {T.~F.}\ \bibnamefont {Duda}},\
  }\bibfield  {title} {\bibinfo {title} {A three-dimensional parabolic equation
  model of sound propagation using higher-order operator splitting and
  {Pad\'{e}} approximants},\ }\href {https://doi.org/10.1121/1.4754421}
  {\bibfield  {journal} {\bibinfo  {journal} {J. Acoust. Soc. Am.}\ }\textbf
  {\bibinfo {volume} {132}},\ \bibinfo {pages} {EL364} (\bibinfo {year}
  {2012})}\BibitemShut {NoStop}%
\bibitem [{\citenamefont {Graham}\ \emph {et~al.}(2012)\citenamefont {Graham},
  \citenamefont {Hou}, \citenamefont {Lakkis},\ and\ \citenamefont
  {Scheichl}}]{2012_Graham_book}%
  \BibitemOpen
  \bibinfo {editor} {\bibfnamefont {I.}~\bibnamefont {Graham}}, \bibinfo
  {editor} {\bibfnamefont {T.}~\bibnamefont {Hou}}, \bibinfo {editor}
  {\bibfnamefont {O.}~\bibnamefont {Lakkis}},\ and\ \bibinfo {editor}
  {\bibfnamefont {R.}~\bibnamefont {Scheichl}},\ eds.,\ \href
  {https://doi.org/10.1007/978-3-642-22061-6} {\emph {\bibinfo {title}
  {Numerical analysis of multiscale problems}}},\ \bibinfo {series} {Lecture
  Notes in Computational Science and Engineering}, Vol.~\bibinfo {volume} {83}\
  (\bibinfo  {publisher} {Springer},\ \bibinfo {address} {Berlin, Heidelberg},\
  \bibinfo {year} {2012})\BibitemShut {NoStop}%
\bibitem [{\citenamefont {Mache}\ and\ \citenamefont
  {Vellekoop}(2026)}]{2026_Mache_domain_decomposition}%
  \BibitemOpen
  \bibfield  {author} {\bibinfo {author} {\bibfnamefont {S.}~\bibnamefont
  {Mache}}\ and\ \bibinfo {author} {\bibfnamefont {I.~M.}\ \bibnamefont
  {Vellekoop}},\ }\bibfield  {title} {\bibinfo {title} {Domain decomposition of
  the modified {B}orn series approach for large-scale wave propagation
  simulations},\ }\href {https://doi.org/10.1016/j.jcp.2025.114619} {\bibfield
  {journal} {\bibinfo  {journal} {J. Comput. Phys.}\ }\textbf {\bibinfo
  {volume} {550}},\ \bibinfo {pages} {114619} (\bibinfo {year}
  {2026})}\BibitemShut {NoStop}%
\bibitem [{\citenamefont {Yee}(1966)}]{1966_Yee_FDTD}%
  \BibitemOpen
  \bibfield  {author} {\bibinfo {author} {\bibfnamefont {K.~S.}\ \bibnamefont
  {Yee}},\ }\bibfield  {title} {\bibinfo {title} {Numerical solution of initial
  boundary value problems involving {M}axwell's equations in isotropic media},\
  }\href {https://doi.org/10.1109/TAP.1966.1138693} {\bibfield  {journal}
  {\bibinfo  {journal} {IEEE Trans. Antennas Propag.}\ }\textbf {\bibinfo
  {volume} {14}},\ \bibinfo {pages} {302} (\bibinfo {year} {1966})}\BibitemShut
  {NoStop}%
\bibitem [{\citenamefont {Taflove}\ and\ \citenamefont
  {Hagness}(2005)}]{2005_Taflove_FDTD}%
  \BibitemOpen
  \bibfield  {author} {\bibinfo {author} {\bibfnamefont {A.}~\bibnamefont
  {Taflove}}\ and\ \bibinfo {author} {\bibfnamefont {S.~C.}\ \bibnamefont
  {Hagness}},\ }\href@noop {} {\emph {\bibinfo {title} {Computational
  Electrodynamics: The Finite-Difference Time-Domain Method}}},\ \bibinfo
  {edition} {3rd}\ ed.\ (\bibinfo  {publisher} {Artech House},\ \bibinfo
  {address} {Boston, MA},\ \bibinfo {year} {2005})\BibitemShut {NoStop}%
\bibitem [{\citenamefont {Ball}\ and\ \citenamefont
  {Li}(2024)}]{2024_Ball_microscopy}%
  \BibitemOpen
  \bibfield  {author} {\bibinfo {author} {\bibfnamefont {J.~M.}\ \bibnamefont
  {Ball}}\ and\ \bibinfo {author} {\bibfnamefont {W.}~\bibnamefont {Li}},\
  }\bibfield  {title} {\bibinfo {title} {Using high-resolution microscopy data
  to generate realistic structures for electromagnetic {FDTD} simulations from
  complex biological models},\ }\href
  {https://doi.org/10.1038/s41596-023-00947-z} {\bibfield  {journal} {\bibinfo
  {journal} {Nat. Protoc.}\ }\textbf {\bibinfo {volume} {19}},\ \bibinfo
  {pages} {1348} (\bibinfo {year} {2024})}\BibitemShut {NoStop}%
\bibitem [{\citenamefont {Krischer}\ \emph {et~al.}(2015)\citenamefont
  {Krischer}, \citenamefont {Fichtner}, \citenamefont {Zukauskaite},\ and\
  \citenamefont {Igel}}]{2015_Krischer_LASIF}%
  \BibitemOpen
  \bibfield  {author} {\bibinfo {author} {\bibfnamefont {L.}~\bibnamefont
  {Krischer}}, \bibinfo {author} {\bibfnamefont {A.}~\bibnamefont {Fichtner}},
  \bibinfo {author} {\bibfnamefont {S.}~\bibnamefont {Zukauskaite}},\ and\
  \bibinfo {author} {\bibfnamefont {H.}~\bibnamefont {Igel}},\ }\bibfield
  {title} {\bibinfo {title} {Large-scale seismic inversion framework},\ }\href
  {https://doi.org/10.1785/0220140248} {\bibfield  {journal} {\bibinfo
  {journal} {Seismol. Res. Lett.}\ }\textbf {\bibinfo {volume} {86}},\ \bibinfo
  {pages} {1198} (\bibinfo {year} {2015})}\BibitemShut {NoStop}%
\bibitem [{\citenamefont {Zhao}\ \emph {et~al.}(2024)\citenamefont {Zhao},
  \citenamefont {Curtis},\ and\ \citenamefont {Zhang}}]{2024_Zhao_Bayesian}%
  \BibitemOpen
  \bibfield  {author} {\bibinfo {author} {\bibfnamefont {X.}~\bibnamefont
  {Zhao}}, \bibinfo {author} {\bibfnamefont {A.}~\bibnamefont {Curtis}},\ and\
  \bibinfo {author} {\bibfnamefont {X.}~\bibnamefont {Zhang}},\ }\bibfield
  {title} {\bibinfo {title} {Physically structured variational inference for
  {Bayesian} full waveform inversion},\ }\href
  {https://doi.org/10.1029/2024JB029557} {\bibfield  {journal} {\bibinfo
  {journal} {J. Geophys. Res.: Solid Earth}\ }\textbf {\bibinfo {volume}
  {129}},\ \bibinfo {pages} {e2024JB029557} (\bibinfo {year}
  {2024})}\BibitemShut {NoStop}%
\bibitem [{\citenamefont {Oliveira}\ \emph {et~al.}(2021)\citenamefont
  {Oliveira}, \citenamefont {Lin},\ and\ \citenamefont
  {Porter}}]{2021_Oliveira_shallow_acoustics}%
  \BibitemOpen
  \bibfield  {author} {\bibinfo {author} {\bibfnamefont {T.~C.~A.}\
  \bibnamefont {Oliveira}}, \bibinfo {author} {\bibfnamefont {Y.-T.}\
  \bibnamefont {Lin}},\ and\ \bibinfo {author} {\bibfnamefont {M.~B.}\
  \bibnamefont {Porter}},\ }\bibfield  {title} {\bibinfo {title} {Underwater
  sound propagation modeling in a complex shallow water environment},\ }\href
  {https://doi.org/10.3389/fmars.2021.751327} {\bibfield  {journal} {\bibinfo
  {journal} {Front. Mar. Sci.}\ }\textbf {\bibinfo {volume} {8}},\ \bibinfo
  {pages} {751327} (\bibinfo {year} {2021})}\BibitemShut {NoStop}%
\bibitem [{\citenamefont {Lee}\ and\ \citenamefont
  {Schultz}(1995)}]{1995_Lee_3D_acoustics}%
  \BibitemOpen
  \bibfield  {author} {\bibinfo {author} {\bibfnamefont {D.}~\bibnamefont
  {Lee}}\ and\ \bibinfo {author} {\bibfnamefont {M.~H.}\ \bibnamefont
  {Schultz}},\ }\href {https://doi.org/10.1142/2789} {\emph {\bibinfo {title}
  {Numerical Ocean Acoustic Propagation in Three Dimensions}}}\ (\bibinfo
  {publisher} {World Scientific},\ \bibinfo {address} {Singapore},\ \bibinfo
  {year} {1995})\BibitemShut {NoStop}%
\bibitem [{\citenamefont {Akkermans}\ and\ \citenamefont
  {Montambaux}(2007)}]{2007_Akkermans_Mesoscopic_Physics}%
  \BibitemOpen
  \bibfield  {author} {\bibinfo {author} {\bibfnamefont {E.}~\bibnamefont
  {Akkermans}}\ and\ \bibinfo {author} {\bibfnamefont {G.}~\bibnamefont
  {Montambaux}},\ }\href {https://doi.org/10.1017/CBO9780511618833} {\emph
  {\bibinfo {title} {Mesoscopic Physics of Electrons and Photons}}}\ (\bibinfo
  {publisher} {Cambridge University Press},\ \bibinfo {year}
  {2007})\BibitemShut {NoStop}%
\bibitem [{\citenamefont {Wang}\ \emph {et~al.}(2023)\citenamefont {Wang},
  \citenamefont {Zhang},\ and\ \citenamefont
  {Lei}}]{2023_Wang_Tucker_Compression}%
  \BibitemOpen
  \bibfield  {author} {\bibinfo {author} {\bibfnamefont {W.}~\bibnamefont
  {Wang}}, \bibinfo {author} {\bibfnamefont {W.}~\bibnamefont {Zhang}},\ and\
  \bibinfo {author} {\bibfnamefont {T.}~\bibnamefont {Lei}},\ }\bibfield
  {title} {\bibinfo {title} {Compression of seismic forward modeling wavefield
  using {TuckerMPI}},\ }\href {https://doi.org/10.1016/j.cageo.2023.105298}
  {\bibfield  {journal} {\bibinfo  {journal} {Comput. Geosci.}\ }\textbf
  {\bibinfo {volume} {172}},\ \bibinfo {pages} {105298} (\bibinfo {year}
  {2023})}\BibitemShut {NoStop}%
\bibitem [{\citenamefont {Ni}\ \emph {et~al.}(2024)\citenamefont {Ni},
  \citenamefont {Denolle}, \citenamefont {Shi}, \citenamefont {Lipovsky},
  \citenamefont {Pan},\ and\ \citenamefont
  {Kutz}}]{2024_Ni_Wavefield_Reconstruction_DAS}%
  \BibitemOpen
  \bibfield  {author} {\bibinfo {author} {\bibfnamefont {Y.}~\bibnamefont
  {Ni}}, \bibinfo {author} {\bibfnamefont {M.~A.}\ \bibnamefont {Denolle}},
  \bibinfo {author} {\bibfnamefont {Q.}~\bibnamefont {Shi}}, \bibinfo {author}
  {\bibfnamefont {B.~P.}\ \bibnamefont {Lipovsky}}, \bibinfo {author}
  {\bibfnamefont {S.}~\bibnamefont {Pan}},\ and\ \bibinfo {author}
  {\bibfnamefont {J.~N.}\ \bibnamefont {Kutz}},\ }\bibfield  {title} {\bibinfo
  {title} {Wavefield reconstruction of distributed acoustic sensing: Lossy
  compression, wavefield separation, and edge computing},\ }\href
  {https://doi.org/10.1029/2024JH000247} {\bibfield  {journal} {\bibinfo
  {journal} {J. Geophys. Res.: Mach. Learn. Comput.}\ }\textbf {\bibinfo
  {volume} {1}},\ \bibinfo {pages} {e2024JH000247} (\bibinfo {year}
  {2024})}\BibitemShut {NoStop}%
\bibitem [{\citenamefont {Muir}\ and\ \citenamefont
  {Zhan}(2021)}]{2021_Muir_wavefield}%
  \BibitemOpen
  \bibfield  {author} {\bibinfo {author} {\bibfnamefont {J.~B.}\ \bibnamefont
  {Muir}}\ and\ \bibinfo {author} {\bibfnamefont {Z.}~\bibnamefont {Zhan}},\
  }\bibfield  {title} {\bibinfo {title} {Seismic wavefield reconstruction using
  a pre-conditioned wavelet--curvelet compressive sensing approach},\ }\href
  {https://doi.org/10.1093/gji/ggab222} {\bibfield  {journal} {\bibinfo
  {journal} {Geophys. J. Int.}\ }\textbf {\bibinfo {volume} {227}},\ \bibinfo
  {pages} {303} (\bibinfo {year} {2021})}\BibitemShut {NoStop}%
\bibitem [{\citenamefont {Boehm}\ \emph {et~al.}(2016)\citenamefont {Boehm},
  \citenamefont {Hanzich}, \citenamefont {de~la Puente},\ and\ \citenamefont
  {Fichtner}}]{2016_Boehm_wavefield_compression}%
  \BibitemOpen
  \bibfield  {author} {\bibinfo {author} {\bibfnamefont {C.}~\bibnamefont
  {Boehm}}, \bibinfo {author} {\bibfnamefont {M.}~\bibnamefont {Hanzich}},
  \bibinfo {author} {\bibfnamefont {J.}~\bibnamefont {de~la Puente}},\ and\
  \bibinfo {author} {\bibfnamefont {A.}~\bibnamefont {Fichtner}},\ }\bibfield
  {title} {\bibinfo {title} {Wavefield compression for adjoint methods in
  full-waveform inversion},\ }\href {https://doi.org/10.1190/geo2015-0653.1}
  {\bibfield  {journal} {\bibinfo  {journal} {Geophysics}\ }\textbf {\bibinfo
  {volume} {81}},\ \bibinfo {pages} {R385} (\bibinfo {year}
  {2016})}\BibitemShut {NoStop}%
\bibitem [{\citenamefont {Kukreja}\ \emph {et~al.}(2022)\citenamefont
  {Kukreja}, \citenamefont {H{\"u}ckelheim}, \citenamefont {Louboutin},
  \citenamefont {Washbourne}, \citenamefont {Kelly},\ and\ \citenamefont
  {Gorman}}]{2022_Kukreja_lossy_checkpoint}%
  \BibitemOpen
  \bibfield  {author} {\bibinfo {author} {\bibfnamefont {N.}~\bibnamefont
  {Kukreja}}, \bibinfo {author} {\bibfnamefont {J.}~\bibnamefont
  {H{\"u}ckelheim}}, \bibinfo {author} {\bibfnamefont {M.}~\bibnamefont
  {Louboutin}}, \bibinfo {author} {\bibfnamefont {J.}~\bibnamefont
  {Washbourne}}, \bibinfo {author} {\bibfnamefont {P.~H.~J.}\ \bibnamefont
  {Kelly}},\ and\ \bibinfo {author} {\bibfnamefont {G.~J.}\ \bibnamefont
  {Gorman}},\ }\bibfield  {title} {\bibinfo {title} {Lossy checkpoint
  compression in full waveform inversion: a case study with {ZFP}v0.5.5 and the
  overthrust model},\ }\href {https://doi.org/10.5194/gmd-15-3815-2022}
  {\bibfield  {journal} {\bibinfo  {journal} {Geosci. Model Dev.}\ }\textbf
  {\bibinfo {volume} {15}},\ \bibinfo {pages} {3815} (\bibinfo {year}
  {2022})}\BibitemShut {NoStop}%
\bibitem [{\citenamefont {Liang}\ \emph {et~al.}(2023)\citenamefont {Liang},
  \citenamefont {Zhao}, \citenamefont {Di}, \citenamefont {Li}, \citenamefont
  {Underwood}, \citenamefont {Gok}, \citenamefont {Tian}, \citenamefont {Deng},
  \citenamefont {Calhoun}, \citenamefont {Tao}, \citenamefont {Chen},\ and\
  \citenamefont {Cappello}}]{sm_sz3}%
  \BibitemOpen
  \bibfield  {author} {\bibinfo {author} {\bibfnamefont {X.}~\bibnamefont
  {Liang}}, \bibinfo {author} {\bibfnamefont {K.}~\bibnamefont {Zhao}},
  \bibinfo {author} {\bibfnamefont {S.}~\bibnamefont {Di}}, \bibinfo {author}
  {\bibfnamefont {S.}~\bibnamefont {Li}}, \bibinfo {author} {\bibfnamefont
  {R.}~\bibnamefont {Underwood}}, \bibinfo {author} {\bibfnamefont {A.~M.}\
  \bibnamefont {Gok}}, \bibinfo {author} {\bibfnamefont {J.}~\bibnamefont
  {Tian}}, \bibinfo {author} {\bibfnamefont {J.}~\bibnamefont {Deng}}, \bibinfo
  {author} {\bibfnamefont {J.~C.}\ \bibnamefont {Calhoun}}, \bibinfo {author}
  {\bibfnamefont {D.}~\bibnamefont {Tao}}, \bibinfo {author} {\bibfnamefont
  {Z.}~\bibnamefont {Chen}},\ and\ \bibinfo {author} {\bibfnamefont
  {F.}~\bibnamefont {Cappello}},\ }\bibfield  {title} {\bibinfo {title} {{SZ3}:
  A modular framework for composing prediction-based error-bounded lossy
  compressors},\ }\href@noop {} {\bibfield  {journal} {\bibinfo  {journal}
  {IEEE Trans. Big Data}\ }\textbf {\bibinfo {volume} {9}},\ \bibinfo {pages}
  {485} (\bibinfo {year} {2023})},\ \bibinfo {note} {code at
  \url{https://github.com/szcompressor/SZ3}}%
\BibitemShut {NoStop}%
\bibitem [{\citenamefont {Zhao}\ \emph {et~al.}(2021)\citenamefont {Zhao},
  \citenamefont {Di}, \citenamefont {Dmitriev}, \citenamefont {Tonellot},
  \citenamefont {Chen},\ and\ \citenamefont {Cappello}}]{sm_sz3b}%
  \BibitemOpen
  \bibfield  {author} {\bibinfo {author} {\bibfnamefont {K.}~\bibnamefont
  {Zhao}}, \bibinfo {author} {\bibfnamefont {S.}~\bibnamefont {Di}}, \bibinfo
  {author} {\bibfnamefont {M.}~\bibnamefont {Dmitriev}}, \bibinfo {author}
  {\bibfnamefont {T.-L.~D.}\ \bibnamefont {Tonellot}}, \bibinfo {author}
  {\bibfnamefont {Z.}~\bibnamefont {Chen}},\ and\ \bibinfo {author}
  {\bibfnamefont {F.}~\bibnamefont {Cappello}},\ }\bibfield  {title} {\bibinfo
  {title} {Optimizing error-bounded lossy compression for scientific data by
  dynamic spline interpolation},\ }in\ \href@noop {} {\emph {\bibinfo
  {booktitle} {Proc. IEEE 37th Int. Conf. on Data Engineering (ICDE)}}}\
  (\bibinfo  {publisher} {IEEE},\ \bibinfo {year} {2021})\ p.\ \bibinfo {pages}
  {1643}%
\BibitemShut {NoStop}%
\bibitem [{\citenamefont {Lindstrom}(2014)}]{sm_zfp}%
  \BibitemOpen
  \bibfield  {author} {\bibinfo {author} {\bibfnamefont {P.}~\bibnamefont
  {Lindstrom}},\ }\bibfield  {title} {\bibinfo {title} {Fixed-rate compressed
  floating-point arrays},\ }\href@noop {} {\bibfield  {journal} {\bibinfo
  {journal} {IEEE Trans. Vis. Comput. Graph.}\ }\textbf {\bibinfo {volume}
  {20}},\ \bibinfo {pages} {2674} (\bibinfo {year} {2014})},\ \bibinfo {note}
  {code at \url{https://github.com/LLNL/zfp}}%
\BibitemShut {NoStop}%
\bibitem [{\citenamefont {Golub}\ and\ \citenamefont
  {Van~Loan}(2013)}]{sm_svd}%
  \BibitemOpen
  \bibfield  {author} {\bibinfo {author} {\bibfnamefont {G.~H.}\ \bibnamefont
  {Golub}}\ and\ \bibinfo {author} {\bibfnamefont {C.~F.}\ \bibnamefont
  {Van~Loan}},\ }\href@noop {} {\emph {\bibinfo {title} {Matrix
  Computations}}},\ \bibinfo {edition} {4th}\ ed.\ (\bibinfo  {publisher}
  {Johns Hopkins University Press},\ \bibinfo {address} {Baltimore},\ \bibinfo
  {year} {2013})\ \bibinfo {note} {ch.~2.4}%
\BibitemShut {NoStop}%
\bibitem [{\citenamefont {Ishimaru}(1978)}]{1978_Ishimaru_Wave_Propagation}%
  \BibitemOpen
  \bibfield  {author} {\bibinfo {author} {\bibfnamefont {A.}~\bibnamefont
  {Ishimaru}},\ }\href@noop {} {\emph {\bibinfo {title} {Wave Propagation and
  Scattering in Random Media}}}\ (\bibinfo  {publisher} {Academic Press},\
  \bibinfo {address} {New York},\ \bibinfo {year} {1978})\BibitemShut {NoStop}%
\bibitem [{\citenamefont {Sheng}(2006)}]{2006_Sheng_Wave_Scattering}%
  \BibitemOpen
  \bibfield  {author} {\bibinfo {author} {\bibfnamefont {P.}~\bibnamefont
  {Sheng}},\ }\href {https://doi.org/10.1007/3-540-29156-3} {\emph {\bibinfo
  {title} {Introduction to Wave Scattering, Localization and Mesoscopic
  Phenomena}}},\ \bibinfo {edition} {2nd}\ ed.,\ \bibinfo {series} {Springer
  Series in Materials Science}, Vol.~\bibinfo {volume} {88}\ (\bibinfo
  {publisher} {Springer},\ \bibinfo {year} {2006})\BibitemShut {NoStop}%
\bibitem [{\citenamefont {Lagendijk}\ and\ \citenamefont {van
  Tiggelen}(1996)}]{1996_Lagendijk_vanTiggelen_review}%
  \BibitemOpen
  \bibfield  {author} {\bibinfo {author} {\bibfnamefont {A.}~\bibnamefont
  {Lagendijk}}\ and\ \bibinfo {author} {\bibfnamefont {B.~A.}\ \bibnamefont
  {van Tiggelen}},\ }\bibfield  {title} {\bibinfo {title} {Resonant multiple
  scattering of light},\ }\href {https://doi.org/10.1016/0370-1573(95)00065-8}
  {\bibfield  {journal} {\bibinfo  {journal} {Phys. Rep.}\ }\textbf {\bibinfo
  {volume} {270}},\ \bibinfo {pages} {143} (\bibinfo {year}
  {1996})}\BibitemShut {NoStop}%
\bibitem [{\citenamefont {Jacques}(2013)}]{2013_Jacques_optical}%
  \BibitemOpen
  \bibfield  {author} {\bibinfo {author} {\bibfnamefont {S.~L.}\ \bibnamefont
  {Jacques}},\ }\bibfield  {title} {\bibinfo {title} {Optical properties of
  biological tissues: a review},\ }\href
  {https://doi.org/10.1088/0031-9155/58/11/R37} {\bibfield  {journal} {\bibinfo
   {journal} {Phys. Med. Biol.}\ }\textbf {\bibinfo {volume} {58}},\ \bibinfo
  {pages} {R37} (\bibinfo {year} {2013})}\BibitemShut {NoStop}%
\bibitem [{\citenamefont {van Rossum}\ and\ \citenamefont
  {Nieuwenhuizen}(1999)}]{1999_vanRossum_diffuse_waves}%
  \BibitemOpen
  \bibfield  {author} {\bibinfo {author} {\bibfnamefont {M.~C.~W.}\
  \bibnamefont {van Rossum}}\ and\ \bibinfo {author} {\bibfnamefont {T.~M.}\
  \bibnamefont {Nieuwenhuizen}},\ }\bibfield  {title} {\bibinfo {title}
  {Multiple scattering of classical waves: microscopy, mesoscopy, and
  diffusion},\ }\href {https://doi.org/10.1103/RevModPhys.71.313} {\bibfield
  {journal} {\bibinfo  {journal} {Rev. Mod. Phys.}\ }\textbf {\bibinfo {volume}
  {71}},\ \bibinfo {pages} {313} (\bibinfo {year} {1999})}\BibitemShut
  {NoStop}%
\bibitem [{\citenamefont {Jara}\ \emph {et~al.}(2023)\citenamefont {Jara},
  \citenamefont {Lin}, \citenamefont {Hsu}, \citenamefont {Cao},\ and\
  \citenamefont {Yamilov}}]{2023_Jara_Simulation_Coherent_Remission}%
  \BibitemOpen
  \bibfield  {author} {\bibinfo {author} {\bibfnamefont {P.}~\bibnamefont
  {Jara}}, \bibinfo {author} {\bibfnamefont {H.~C.}\ \bibnamefont {Lin}},
  \bibinfo {author} {\bibfnamefont {C.~W.}\ \bibnamefont {Hsu}}, \bibinfo
  {author} {\bibfnamefont {H.}~\bibnamefont {Cao}},\ and\ \bibinfo {author}
  {\bibfnamefont {A.}~\bibnamefont {Yamilov}},\ }\bibfield  {title} {\bibinfo
  {title} {Simulation of coherent remission in planar disordered medium},\ }in\
  \href {https://doi.org/10.23919/ACES57841.2023.10114726} {\emph {\bibinfo
  {booktitle} {2023 Int. Appl. Comput. Electromagn. Soc. Symp. (ACES)}}}\
  (\bibinfo  {publisher} {IEEE},\ \bibinfo {year} {2023})\ pp.\ \bibinfo
  {pages} {1--2}\BibitemShut {NoStop}%
\bibitem [{\citenamefont {Bender}\ \emph {et~al.}(2022)\citenamefont {Bender},
  \citenamefont {Goetschy}, \citenamefont {Hsu}, \citenamefont {Y{\i}lmaz},
  \citenamefont {Jara}, \citenamefont {Yamilov},\ and\ \citenamefont
  {Cao}}]{2022_Bender_Coherent_Enhancement}%
  \BibitemOpen
  \bibfield  {author} {\bibinfo {author} {\bibfnamefont {N.}~\bibnamefont
  {Bender}}, \bibinfo {author} {\bibfnamefont {A.}~\bibnamefont {Goetschy}},
  \bibinfo {author} {\bibfnamefont {C.~W.}\ \bibnamefont {Hsu}}, \bibinfo
  {author} {\bibfnamefont {H.}~\bibnamefont {Y{\i}lmaz}}, \bibinfo {author}
  {\bibfnamefont {P.}~\bibnamefont {Jara}}, \bibinfo {author} {\bibfnamefont
  {A.}~\bibnamefont {Yamilov}},\ and\ \bibinfo {author} {\bibfnamefont
  {H.}~\bibnamefont {Cao}},\ }\bibfield  {title} {\bibinfo {title} {Coherent
  enhancement of optical remission in diffusive media},\ }\href
  {https://doi.org/10.1073/pnas.2207089119} {\bibfield  {journal} {\bibinfo
  {journal} {Proc. Natl. Acad. Sci. U.S.A.}\ }\textbf {\bibinfo {volume}
  {119}},\ \bibinfo {pages} {e2207089119} (\bibinfo {year} {2022})}\BibitemShut
  {NoStop}%
\bibitem [{\citenamefont {Jara}\ \emph {et~al.}(2025)\citenamefont {Jara},
  \citenamefont {Goetschy}, \citenamefont {Cao},\ and\ \citenamefont
  {Yamilov}}]{2025_Jara_coherent_sensing}%
  \BibitemOpen
  \bibfield  {author} {\bibinfo {author} {\bibfnamefont {P.}~\bibnamefont
  {Jara}}, \bibinfo {author} {\bibfnamefont {A.}~\bibnamefont {Goetschy}},
  \bibinfo {author} {\bibfnamefont {H.}~\bibnamefont {Cao}},\ and\ \bibinfo
  {author} {\bibfnamefont {A.}~\bibnamefont {Yamilov}},\ }\bibfield  {title}
  {\bibinfo {title} {Harnessing coherent-wave control for sensing
  applications},\ }\href {https://doi.org/10.1103/wwbs-pftg} {\bibfield
  {journal} {\bibinfo  {journal} {Phys. Rev. Appl.}\ }\textbf {\bibinfo
  {volume} {24}},\ \bibinfo {pages} {054027} (\bibinfo {year}
  {2025})}\BibitemShut {NoStop}%
\bibitem [{\citenamefont {Arridge}(1999)}]{1999_Arridge_optical_tomography}%
  \BibitemOpen
  \bibfield  {author} {\bibinfo {author} {\bibfnamefont {S.~R.}\ \bibnamefont
  {Arridge}},\ }\bibfield  {title} {\bibinfo {title} {Optical tomography in
  medical imaging},\ }\href {https://doi.org/10.1088/0266-5611/15/2/022}
  {\bibfield  {journal} {\bibinfo  {journal} {Inverse Probl.}\ }\textbf
  {\bibinfo {volume} {15}},\ \bibinfo {pages} {R41} (\bibinfo {year}
  {1999})}\BibitemShut {NoStop}%
\bibitem [{\citenamefont {Arridge}\ and\ \citenamefont
  {Schotland}(2009)}]{2009_Arridge_Schotland_review}%
  \BibitemOpen
  \bibfield  {author} {\bibinfo {author} {\bibfnamefont {S.~R.}\ \bibnamefont
  {Arridge}}\ and\ \bibinfo {author} {\bibfnamefont {J.~C.}\ \bibnamefont
  {Schotland}},\ }\bibfield  {title} {\bibinfo {title} {Optical tomography:
  forward and inverse problems},\ }\href
  {https://doi.org/10.1088/0266-5611/25/12/123010} {\bibfield  {journal}
  {\bibinfo  {journal} {Inverse Probl.}\ }\textbf {\bibinfo {volume} {25}},\
  \bibinfo {pages} {123010} (\bibinfo {year} {2009})}\BibitemShut {NoStop}%
\bibitem [{\citenamefont {Durduran}\ \emph {et~al.}(2010)\citenamefont
  {Durduran}, \citenamefont {Choe}, \citenamefont {Baker},\ and\ \citenamefont
  {Yodh}}]{2010_Durduran_DOT}%
  \BibitemOpen
  \bibfield  {author} {\bibinfo {author} {\bibfnamefont {T.}~\bibnamefont
  {Durduran}}, \bibinfo {author} {\bibfnamefont {R.}~\bibnamefont {Choe}},
  \bibinfo {author} {\bibfnamefont {W.~B.}\ \bibnamefont {Baker}},\ and\
  \bibinfo {author} {\bibfnamefont {A.~G.}\ \bibnamefont {Yodh}},\ }\bibfield
  {title} {\bibinfo {title} {Diffuse optics for tissue monitoring and
  tomography},\ }\href {https://doi.org/10.1088/0034-4885/73/7/076701}
  {\bibfield  {journal} {\bibinfo  {journal} {Rep. Prog. Phys.}\ }\textbf
  {\bibinfo {volume} {73}},\ \bibinfo {pages} {076701} (\bibinfo {year}
  {2010})}\BibitemShut {NoStop}%
\bibitem [{\citenamefont {Sarma}\ \emph {et~al.}(2016)\citenamefont {Sarma},
  \citenamefont {Yamilov}, \citenamefont {Petrenko}, \citenamefont {Bromberg},\
  and\ \citenamefont {Cao}}]{2016_Sarma_Open_Channels}%
  \BibitemOpen
  \bibfield  {author} {\bibinfo {author} {\bibfnamefont {R.}~\bibnamefont
  {Sarma}}, \bibinfo {author} {\bibfnamefont {A.~G.}\ \bibnamefont {Yamilov}},
  \bibinfo {author} {\bibfnamefont {S.}~\bibnamefont {Petrenko}}, \bibinfo
  {author} {\bibfnamefont {Y.}~\bibnamefont {Bromberg}},\ and\ \bibinfo
  {author} {\bibfnamefont {H.}~\bibnamefont {Cao}},\ }\bibfield  {title}
  {\bibinfo {title} {Control of energy density inside a disordered medium by
  coupling to open or closed channels},\ }\href
  {https://doi.org/10.1103/PhysRevLett.117.086803} {\bibfield  {journal}
  {\bibinfo  {journal} {Phys. Rev. Lett.}\ }\textbf {\bibinfo {volume} {117}},\
  \bibinfo {pages} {086803} (\bibinfo {year} {2016})}\BibitemShut {NoStop}%
\bibitem [{\citenamefont {Bender}\ \emph {et~al.}(2020)\citenamefont {Bender},
  \citenamefont {Yamilov}, \citenamefont {Y{\i}lmaz},\ and\ \citenamefont
  {Cao}}]{2020_Bender_Eigenchannels}%
  \BibitemOpen
  \bibfield  {author} {\bibinfo {author} {\bibfnamefont {N.}~\bibnamefont
  {Bender}}, \bibinfo {author} {\bibfnamefont {A.}~\bibnamefont {Yamilov}},
  \bibinfo {author} {\bibfnamefont {H.}~\bibnamefont {Y{\i}lmaz}},\ and\
  \bibinfo {author} {\bibfnamefont {H.}~\bibnamefont {Cao}},\ }\bibfield
  {title} {\bibinfo {title} {Fluctuations and correlations of transmission
  eigenchannels in diffusive media},\ }\href
  {https://doi.org/10.1103/PhysRevLett.125.165901} {\bibfield  {journal}
  {\bibinfo  {journal} {Phys. Rev. Lett.}\ }\textbf {\bibinfo {volume} {125}},\
  \bibinfo {pages} {165901} (\bibinfo {year} {2020})}\BibitemShut {NoStop}%
\bibitem [{\citenamefont {Lin}\ \emph {et~al.}(2022)\citenamefont {Lin},
  \citenamefont {Wang},\ and\ \citenamefont {Hsu}}]{2022_Lin_MESTI_Augmented}%
  \BibitemOpen
  \bibfield  {author} {\bibinfo {author} {\bibfnamefont {H.-C.}\ \bibnamefont
  {Lin}}, \bibinfo {author} {\bibfnamefont {Z.}~\bibnamefont {Wang}},\ and\
  \bibinfo {author} {\bibfnamefont {C.~W.}\ \bibnamefont {Hsu}},\ }\bibfield
  {title} {\bibinfo {title} {Fast multi-source nanophotonic simulations using
  augmented partial factorization},\ }\href
  {https://doi.org/10.1038/s43588-022-00370-6} {\bibfield  {journal} {\bibinfo
  {journal} {Nat. Comput. Sci.}\ }\textbf {\bibinfo {volume} {2}},\ \bibinfo
  {pages} {815} (\bibinfo {year} {2022})}\BibitemShut {NoStop}%
\bibitem [{202(2024)}]{2024_Cluster}%
  \BibitemOpen
  \href {https://doi.org/10.71674/PH64-N397} {\bibinfo {title} {The {M}ill
  {HPC} {C}luster}},\ \bibinfo {howpublished} {{M}issouri {U}niversity of
  {S}cience and {T}echnology, \url{https://scholarsmine.mst.edu/the-mill/1/}}
  (\bibinfo {year} {2024})\BibitemShut {NoStop}%
\bibitem [{\citenamefont {Osnabrugge}\ \emph {et~al.}(2016)\citenamefont
  {Osnabrugge}, \citenamefont {Leedumrongwatthanakun},\ and\ \citenamefont
  {Vellekoop}}]{2016_Osnabrugge_wavesim}%
  \BibitemOpen
  \bibfield  {author} {\bibinfo {author} {\bibfnamefont {G.}~\bibnamefont
  {Osnabrugge}}, \bibinfo {author} {\bibfnamefont {S.}~\bibnamefont
  {Leedumrongwatthanakun}},\ and\ \bibinfo {author} {\bibfnamefont {I.~M.}\
  \bibnamefont {Vellekoop}},\ }\bibfield  {title} {\bibinfo {title} {A
  convergent {Born} series for solving the inhomogeneous {Helmholtz} equation
  in arbitrarily large media},\ }\href
  {https://doi.org/10.1016/j.jcp.2016.06.034} {\bibfield  {journal} {\bibinfo
  {journal} {J. Comput. Phys.}\ }\textbf {\bibinfo {volume} {322}},\ \bibinfo
  {pages} {113} (\bibinfo {year} {2016})}\BibitemShut {NoStop}%
\bibitem [{\citenamefont {Goodman}(2007)}]{2007_Goodman_speckle}%
  \BibitemOpen
  \bibfield  {author} {\bibinfo {author} {\bibfnamefont {J.~W.}\ \bibnamefont
  {Goodman}},\ }\href@noop {} {\emph {\bibinfo {title} {Speckle Phenomena in
  Optics: Theory and Applications}}}\ (\bibinfo  {publisher} {Roberts and
  Company},\ \bibinfo {address} {Englewood, CO},\ \bibinfo {year}
  {2007})\BibitemShut {NoStop}%
\bibitem [{\citenamefont {Anderson}\ \emph {et~al.}(1999)\citenamefont
  {Anderson} \emph {et~al.}}]{sm_lapack}%
  \BibitemOpen
  \bibfield  {author} {\bibinfo {author} {\bibfnamefont {E.}~\bibnamefont
  {Anderson}} \emph {et~al.},\ }\href@noop {} {\emph {\bibinfo {title}
  {{LAPACK} Users' Guide}}},\ \bibinfo {edition} {3rd}\ ed.\ (\bibinfo
  {publisher} {SIAM},\ \bibinfo {address} {Philadelphia},\ \bibinfo {year}
  {1999})%
\BibitemShut {NoStop}%
\bibitem [{\citenamefont {Lagendijk}\ \emph {et~al.}(2009)\citenamefont
  {Lagendijk}, \citenamefont {van Tiggelen},\ and\ \citenamefont
  {Wiersma}}]{2009_Lagendijk_vanTiggelen_Wiersma_Anderson}%
  \BibitemOpen
  \bibfield  {author} {\bibinfo {author} {\bibfnamefont {A.}~\bibnamefont
  {Lagendijk}}, \bibinfo {author} {\bibfnamefont {B.}~\bibnamefont {van
  Tiggelen}},\ and\ \bibinfo {author} {\bibfnamefont {D.~S.}\ \bibnamefont
  {Wiersma}},\ }\bibfield  {title} {\bibinfo {title} {Fifty years of {A}nderson
  localization},\ }\href {https://doi.org/10.1063/1.3206091} {\bibfield
  {journal} {\bibinfo  {journal} {Phys. Today}\ }\textbf {\bibinfo {volume}
  {62}},\ \bibinfo {pages} {24} (\bibinfo {year} {2009})}\BibitemShut {NoStop}%
\bibitem [{\citenamefont {Yamilov}\ \emph {et~al.}(2023)\citenamefont
  {Yamilov}, \citenamefont {Skipetrov}, \citenamefont {Hughes}, \citenamefont
  {Minkov}, \citenamefont {Yu},\ and\ \citenamefont {Cao}}]{2023_Yamilov_AL}%
  \BibitemOpen
  \bibfield  {author} {\bibinfo {author} {\bibfnamefont {A.}~\bibnamefont
  {Yamilov}}, \bibinfo {author} {\bibfnamefont {S.~E.}\ \bibnamefont
  {Skipetrov}}, \bibinfo {author} {\bibfnamefont {T.~W.}\ \bibnamefont
  {Hughes}}, \bibinfo {author} {\bibfnamefont {M.}~\bibnamefont {Minkov}},
  \bibinfo {author} {\bibfnamefont {Z.}~\bibnamefont {Yu}},\ and\ \bibinfo
  {author} {\bibfnamefont {H.}~\bibnamefont {Cao}},\ }\bibfield  {title}
  {\bibinfo {title} {Anderson localization of electromagnetic waves in three
  dimensions},\ }\href {https://doi.org/10.1038/s41567-023-02091-7} {\bibfield
  {journal} {\bibinfo  {journal} {Nat. Phys.}\ }\textbf {\bibinfo {volume}
  {19}},\ \bibinfo {pages} {1308} (\bibinfo {year} {2023})}\BibitemShut
  {NoStop}%
\bibitem [{202(2025)}]{2025_Hellbender}%
  \BibitemOpen
  \href {https://doi.org/10.32469/10355/97710} {\bibinfo {title} {The
  {H}ellbender {HPC} {C}luster}},\ \bibinfo {howpublished} {{R}esearch
  {S}upport {S}olutions, {D}ivision of {IT}, {U}niversity of {M}issouri,
  {C}olumbia, {MO}} (\bibinfo {year} {2025})%
\BibitemShut {NoStop}%
\bibitem [{\citenamefont {Jara}\ and\ \citenamefont {Yamilov}(2026)}]{2026_Jara_OSCAR_deposit}%
  \BibitemOpen
  \bibfield  {author} {\bibinfo {author} {\bibfnamefont {P.}~\bibnamefont
  {Jara}}\ and\ \bibinfo {author} {\bibfnamefont {A.}~\bibnamefont
  {Yamilov}},\ }\bibfield  {title} {\bibinfo {title} {On-shell compression and
  reconstruction ({OSCAR}) of monochromatic wave fields in weakly scattering
  media, research data},\ }\bibinfo {howpublished}
  {\url{https://scholarsmine.mst.edu/research_data/19}, {M}issouri {S}\&{T}
  {S}cholars' {M}ine} (\bibinfo {year} {2026})%
\BibitemShut {NoStop}%
\end{thebibliography}

\begin{thebibliography}{S9}
\bibitem{sm_mesti} H.-C. Lin, Z. Wang, and C. W. Hsu, Nat. Comput. Sci. \textbf{2}, 815 (2022).
\bibitem{sm_wavesim} G. Osnabrugge, S. Leedumrongwatthanakun, and I. M. Vellekoop, J. Comput. Phys. \textbf{322}, 113 (2016).
\bibitem{sm_ishimaru} A. Ishimaru, \emph{Wave Propagation and Scattering in Random Media} (Academic Press, New York, 1978).
\bibitem{sm_jacques} S. L. Jacques, Phys. Med. Biol. \textbf{58}, R37 (2013).
\bibitem{sm_rpk1996} L. Ryzhik, G. Papanicolaou, and J. B. Keller, Wave Motion \textbf{24}, 327 (1996).
\bibitem{sm_akkermans2007} E. Akkermans and G. Montambaux, \emph{Mesoscopic Physics of Electrons and Photons} (Cambridge University Press, 2007), Ch.~4.
\bibitem{sm_shapiro1986} B. Shapiro, Phys. Rev. Lett. \textbf{57}, 2168 (1986).
\bibitem{sm_yamilov2008} A. Yamilov, Phys. Rev. B \textbf{78}, 045104 (2008).
\end{thebibliography}
\end{document}